\documentclass[aps,prb,floatfix,twocolumn,showpacs]{revtex4}
\usepackage{bm}
\usepackage{amsmath}
\usepackage{amssymb}
\usepackage{graphicx}
\usepackage{color}
\usepackage{epstopdf}
\usepackage{epsfig}

\begin{document}

\newcommand{\ve}{\varepsilon}
\newcommand{\vt}{\vartheta}
\newcommand{\de}{\delta}
\newcommand{\De}{\Delta}
\newcommand{\vp}{\varphi}
\newcommand{\ka}{\kappa}
\newcommand{\la}{\lambda}
\newcommand{\p}{\phi}
\newcommand{\ps}{\psi}
\newcommand{\ro}{\rho}
\newcommand{\si}{\sigma}
\newcommand{\ze}{\zeta}
\newcommand{\te}{\theta}
\newcommand{\up}{\upsilon}
\newcommand{\om}{\omega}
\newcommand{\ga}{\gamma}
\newcommand{\al}{\alpha}
\newcommand{\be}{\beta}
\newcommand{\vs}{\varsigma}

\title{Anderson Localization in Metamaterials and Other Complex
Media}
\author{Sergey A. Gredeskul$^{1}$, Yuri S. Kivshar$^{2}$, Ara A. Asatryan$%
^{3}$, Konstantin Y. Bliokh$^{4,5}$,\\Yuri P. Bliokh$^{6}$, Valentin D. Freilikher$^{7}$, Ilya V.
Shadrivov$^{2}$}
\affiliation{$^1$ Ben Gurion University of the Negev, 84105 Beer-Sheva, Israel}
\affiliation{$^2$ Nonlinear Physics Centre, Research School of Physics and Engineering,
The Australian National University, Canberra, ACT 0200, Australia}
\affiliation{$^3$Department of Mathematical Sciences, University of Technology, Sydney,
NSW 2007, Australia}
\affiliation{$^4$Advanced Science Institute, RIKEN, Wako-shi, Saitama 351-0198, Japan}
\affiliation{$^5$A. Usikov Institute of Radiophysics and Electronics, Kharkov 61085,
Ukraine}
\affiliation{$^6$Department of Physics, Technion - Israel Institute of Technology, 32100
Haifa, Israel}
\affiliation{$^7$Department of Physics, Bar-Ilan University, Raman-Gan, 52900, Israel}
\date{\today}

\begin{abstract}
We review some recent (mostly ours) results on the Anderson localization of
light and electron waves in complex disordered systems, including: (i)
left-handed metamaterials, (ii) magneto-active optical structures, (iii)
graphene superlattices, and (iv) nonlinear dielectric media. First, we
demonstrate that left-handed metamaterials can significantly suppress
localization of light and lead to an anomalously enhanced transmission. This
suppression is essential at the long-wavelength limit in the case of normal
incidence, at specific angles of oblique incidence (Brewster anomaly), and
in the vicinity of the zero-$\varepsilon $ or zero-$\mu $ frequencies for
dispersive metamaterials. Remarkably, in disordered samples comprised of
alternating normal and left-handed metamaterials, the reciprocal Lyapunov
exponent and reciprocal transmittance increment can differ from each other.
Second, we study magneto-active multilayered structures, which exhibit
nonreciprocal localization of light depending on the direction of
propagation and on the polarization. At resonant frequencies or
realizations, such nonreciprocity results in effectively unidirectional
transport of light. Third, we discuss the analogy between the wave
propagation through multilayered samples with metamaterials and the charge
transport in graphene, which enables a simple physical explanation of
unusual conductive properties of disordered graphene superlatices. We
predict disorder-induced resonances of the transmission coefficient at
oblique incidence of the Dirac quasiparticles. Finally, we demonstrate that
an interplay of nonlinearity and disorder in dielectric media can lead to
bistability of individual localized states excited inside the medium at
resonant frequencies. This results in nonreciprocity of the wave
transmission and unidirectional transport of light.
\end{abstract}

\pacs{42.25.Dd, 72.15.Rn, 78.67.Pt, 78.20.Ls, 72.80.Vp, 42.65 Pc}
\maketitle
\tableofcontents


\section{Introduction}

\label{sec:intro}

Anderson localization is one of the most fundamental phenomena in the
physics of disordered systems. Being predicted in the seminal paper~\cite{And} %
 for spin excitations, and then extended to
electrons and other one-particle excitations in solids~\cite{LGP,Imry} and
classical waves~\cite{John,Ping91,Ping07,Sok}, it became a paradigm of the
modern physics~\cite{AL}. The study of this phenomenon remains a
hot topic throughout its more than 50-years history. It is
constantly  stimulated by
new experimental results, including the most recent observations in
microwaves~\cite{Genack1,we-PRL-1,Ref-exp-2}, optics~\cite%
{optics,Segev,Silberberg}, and Bose-Einstein condensates~\cite{BEC}.

Being a universal wave phenomenon, Anderson localization has natural
implications in novel exotic wave systems, such as photonic crystals, meta- and magnetooptical materials,  graphene superlattices. Indeed, left-handed metamaterials, nonlinear and magnetooptical
materials, and graphene~\ \cite%
{Veselago,Pendry,Shalaev,Katsnelson2,Geim,periodic-1,Zvezdin} are involved
in design and engineering of various multilayered structures operating in a broad spectral range, from optical to microwave frequencies. Random wave
scattering and localization naturally appear in such systems, either due to technological imperfections or owing to the
intentially designed random lattices. Importantly, exotic properties of the
constituent materials essentially require consideration of the interplay of
the Anderson localization with various additional effects: absorption and
gain~\cite{FrePusYur,we-PRL-1,gain,Paasschens,Asatryan98}, polarization and
spin \cite{Sipe,BF,anti1,anti2}, nonlinearity \cite%
{nonlin-loc1,nonlin-loc2,Silberberg,Segev,Shadrivov}, and magnetooptical
phenomena \cite{ErbacherLenkeMaret,InoueFujii,Bliokh-mag}. In this review,
we describe novel remarkable features of Anderson localization of waves in
multilayered structures composed of non-conventional materials with
unique intrinsic properties.

We start our review with Sec.~\ref{sec:multilay} which introduces the basic
concepts and general formalism describing the wave propagation, scattering,
and localization in in random-layered media. Anderson localization
originates from the interference of multiply scattered waves, manifesting
itself most profoundly in one-dimensional (1D) systems where all states
become localized~\cite{Mott,Furst}. Due to one-dimensional geometry, such
systems are well analyzed~\cite{LGP,FG,IKM}, including the mathematical
level of rigorousness of the results~\cite{PF,GMP}. We describe the exact
transfer-matrix approach to the wave propagation and scattering in layered
media. The main spatial scale of localization, i.e., \textit{localization
length}, can be defined in two ways: (i) via the Lyapunov exponent of the
random system and (ii) via the decrement of the wave transmission dependent
on the system. In usual Anderson-localization problems, these two
localization lengths coincide with each other.

In Section~\ref{sec:meta} we consider transmission and localization
properties of the multilayered H-stacks comprised of normal materials with
right-handed $R-$layers and mixed M-stacks, including also left-handed $L-$%
layers with negative refractive index~\cite{Veselago}. The opposite signs of
the phase and group velocities in metamaterials lead to partial or complete
cancellation of the phase accumulation in multilayered M-stacks. We show
that this cancellation suppresses the interference of multiple scattering
waves and the localization itself~\cite%
{Asatryan07,Asatryan10a,Asatryan10b,Asatryan12}. Using the weak scattering
approximation (WSA)~\cite{Asatryan07,Asatryan10a}, we give detailed
analytical and numerical description of transmission and localization
properties of both M- and H-stacks and reveal a number of intriguing
results. Namely: (i) in the long wave limit localization lengths defined via
the Lyapunov exponent and transmission decrement differ from each other in
M-stacks, (ii) there exist two ballistic regimes in the H-stacks, (iii)
essential suppression of localization at special angles in the case of
oblique incidence (Brewster anomaly) and in the vicinity of special
frequencies (zero-$\varepsilon$ or zero-$\mu$ frequencies) is observed.
Finally, in Section~\ref{subsec:enlight} we discuss an anomalous enhancement~%
\cite{Asatryan07} of wave transmission in minimally disordered
alternated M-stacks of metamaterials, where the layer thicknesses are equal
and only dielectric permittivities (or only magnetic permeabilities) vary.

Section~\ref{sec:novel} is devoted to the study of novel localization
features in novel materials. We start with discussion of localization of
light propagating through magneto-active multilayered structures, with
either Faraday or Cotton-Muton (Voigt) geometries (Section ~\ref%
{subsec:nonreciprocal}). We show that magnetooptical effects can
significantly affect the phase relations, resulting in nonreciprocal
localization depending on direction of the wave propagation and polarization
of light. At resonant frequencies corresponding to the excitation of
localized states inside the sample, a nonreciprocal shift of the the
resonance results in effectively unidirectional transmission of light\cite%
{Bliokh-mag}. In Section ~\ref{subsec:graf}, conducting properties of a
graphene layer subject to stratified electric field are considered. The
close analogy between charge transport in such system and wave transmission
through multilayered stack~\cite{BFSN} underpins remarkable conductive
properties of disordered graphene\cite{Cheianov}. We predict
disorder-induced resonances of the transmission coefficient at oblique
incidence of electron waves. Finally, in Section ~\ref{subsec:nonlin}, we
examine the interplay between nonlinearity and disorder in resonant
transmission through a random-leyered dielectric medium~\cite{Shadrivov}.
Owing to effective energy localization and pumping, even weak Kerr
nonlinearity can play a crucial role leading to bistability of Anderson
localized states inside the medium. Akin to the magneto-optical structures,
this brings about unidirectional transmission of light.


\section{Random Multilayered Structures}

\label{sec:multilay} \numberwithin{equation}{section}


\subsection{Transmission Length and Lyapunov Exponent}

\label{subsec:transm} 
As it was mentioned above, 1D Anderson localization results is exponential
decay of the transmission coefficient with the length $L$ of the sample. For
multilayered systems, it worth to use the total number of layers $N$ and
mean layer thickness $L/N$. In what follows we use dimensionless variables
measuring all lengths in mean layer thickness units while the time
dependence is chosen in the form $e^{-i\omega t}$. For simplicity throughout
all this review we mainly consider the lossless stacks. The detailed results
concerning to the case of stacks with losses can be found in original works.

Introduce the dimensionless transmission length $l_N$ on a realization

\begin{equation*}
\frac{1}{l_N}=-\frac{\ln |T_{N}|}{N} =-\frac{\mathrm{Re}\ln T_{N}}{N}
\label{LL-1}
\end{equation*}%
and "averaged" $N$-dependent dimensionless transmission length $l_T\equiv
l_T(N)$ of a multilayered $N-$layered stack

\begin{equation}
\frac{1}{l_T}=-\left\langle \frac{\ln |T_{N}|}{N}\right\rangle
=-\left\langle \frac{\mathrm{Re}\ln T_{N}}{N}\right\rangle.  \label{LL-0}
\end{equation}%
Here $T_N$ is the stack amplitude transmission coefficient related to its
transmittivity $\mathcal{T}_N$ by equality $\mathcal{T}_N=|T_N|^2.$ Due to
self-averaging of $\ln |T_{N}|/N$, both these lengths $l_T$ and $l_N$ tend
to the same limit

\begin{equation}
\lim_{N\rightarrow \infty } l_T=\lim_{N\rightarrow \infty} l_N=l,
\label{LL-3}
\end{equation}%
as the number $N$ of layers tends to infinity. Following\cite{Baluni} we
recall $l$ as localization length. This localization length is related
directly to the transmission properties. Its reciprocal value is nothing but
decrement of the stack transmission coefficient.

Transmission coefficient entering these equations is naturally expressed in
terms of the total $T$-matrix of the stack written in the running wave
basis. Consider transmission of the plane wave incident normally from the
left to the stack comprised of even number $N$ of layers and embedded into
free space. In the simplest case, the wave is described in terms of two
component vector of, say, an electric field $e.$ Within a uniform medium
with dielectric permittivity $\varepsilon$ and magnetic permeability $\mu,$
the field $e$ has the form

\begin{equation}  \label{field}
e(z)=e^{+}e^{ikz}+e^{-}e^{-ikz}, \ \ \ \ k=\frac{\omega}{c}\sqrt{%
\varepsilon\mu},
\end{equation}
with $z$-axis directed to the right (here and below all lengths of the
problem are dimensionless and measured in the mean layer thickness).

If the components of vector ${\vec e}$ are normalized by such a way that the
energy flux of the wave (\ref{field}) is $|e^{+}|^{2}-|e^{-}|^{2},$ then the
amplitudes

\begin{equation}  \label{amplitudes}
\vec e_{L,R}=\left(
\begin{array}{c}
e^{+}_{L,R} \\
\\
e^{-}_{L,R}%
\end{array}
\right)
\end{equation}
of the field from both sides out of the $N-$layer stack are related by its
transfer matrix ${\hat T}(N)$

\begin{eqnarray}  \label{T-stack-1}
{\vec e}|_{L}={\hat T}(N){\vec e}|_{R},
\end{eqnarray}
which is expressed via transmission and reflection coefficients of the stack
as

\begin{eqnarray}  \label{T-stack-2}
{\hat T}(N)= \left\Vert
\begin{array}{ccc}
\displaystyle{\frac{1}{T_{N}}} &  & \displaystyle{\frac{R^{*}_{N}}{T^{*}_{N}}%
} \\
&  &  \\
\displaystyle{\frac{R_{N}}{T_{N}}} &  & \displaystyle{\frac{1}{T^{*}_{N}}}%
\end{array}
\right\Vert,
\end{eqnarray}
where asterisk stands for the complex conjugation.

The methods of calculation of transmission coefficient

\begin{equation}  \label{TransmCoeff}
T(N)=\left(\hat{T}_{11}\right)^{-1}
\end{equation}
are discussed in the next Subsection.

In what follows, we consider stacks composed of weak scattering layers with
reflection coefficients of each layer much smaller than $1.$ In spite of
this, for a sufficiently long stack the transmission coefficient is
exponentially small $\left\vert T_{N}\right\vert\sim \exp {(-\kappa N)}$
with decrement coinciding with reciprocal localization length $%
\kappa=l^{-1}_{T}$ (localized regime). However a short stack comprising a
comparatively small number of layers is almost transparent $|R_{N}|^{2}\ll 1$
(ballistic regime). Here the transmission length takes the form

\begin{equation}
l_T\approx b=\frac{\langle |R_{N}|^{2}\rangle }{2N},  \label{LL-21}
\end{equation}%
involving the average reflectance \cite{Rytov}. This follows directly from
Eq. (\ref{LL-0}) by virtue of the current conservation relationship, $%
|R_N|^2 + |T_N|^2 = 1$. The length $b$ in this equation is termed the
ballistic length.

Accordingly, in studies of the transport of the classical waves in
one-dimensional random systems, the following spatial scales arise in a
natural way:

\begin{itemize}
\item $l_T$ --- the \emph{transmission length} of a finite sample (\ref{LL-0}%
),

\item $l$ --- the \emph{localization length} (\ref{LL-3}) related to
transmission properties, and

\item $b$ --- \emph{ballistic length} (\ref{LL-21}).
\end{itemize}

The exponential decrease of transmission coefficient with the stack size is
only manifestation of Anderson localization. The phenomenon of localization
itself is the localized character of eigenstates in infinite disordered
system with sufficiently fast decaying correlations. The quantitative
characteristic of such a localization is the Lyapunov exponent which is
increment of the exponential growth of the currentless state with a given
value at certain point far from this point. The amplitude (\ref{amplitudes})
of the currentless state in the inhomogeneous medium in the basis of running
waves can be parameterized as

\begin{equation}  \label{currentless-1}
\vec e=e^{\xi}\left(
\begin{array}{c}
e^{i\theta} \\
\\
e^{-i\theta}%
\end{array}
\right)=R\left(
\begin{array}{c}
e^{i\theta} \\
\\
e^{-i\theta}%
\end{array}
\right),
\end{equation}
where $R(z)$ and $\theta(z)$ are the modulus and the phase of the considered
currentless solution correspondingly.

It is known\cite{LGP,PF} that at given initial values $\xi(0)$ ($R(0)$), and
$\theta(0)$, the function $\xi(z)$ at a sufficiently far point is
approximately proportional to its distance from the initial point. In
discrete terms, with the probability $1$ the positive limit exists

\begin{equation}  \label{Lyapunov-1}
\gamma=\lim_{N\to\infty}\frac{\xi(N)}{N}=\lim_{N\to\infty} \frac{1}{N}\ln%
\frac{R(N)}{R(0)},
\end{equation}
which is called Lyapunov exponent. Its reciprocal value we also call
localization length

\begin{equation}  \label{locLyap}
l_{\xi}=\frac{1}{\gamma},
\end{equation}
however index $\xi$ reminds that this localization length is defined through
Lyapunov exponent.

To compare the two localization lengths $l$ and $l_{\xi}$, we consider first
the continuous case were corresponding dynamical variable $\xi(z)$ depends
on continuous coordinate $z$. In this case, transmittance of the system with
length $L$ is exactly expressed as\cite{LGP,GMP}

\begin{equation}  \label{transmittance-1}
\mathcal{T}_{L}\equiv|T_{L}|^{2}=\frac{4} {e^{2\xi_{c}(L)}+e^{2\xi_{s}(L)}+2}%
,
\end{equation}
where $\xi_{c}(z)$ and $\xi_{s}(z)$ are two independent solutions satisfying
so called cosine and sine initial conditions $\theta_{c}(0)=0$ and $%
\theta(0)=\pi/2$ and having the same limiting behavior

\begin{equation}  \label{limits}
\gamma=\frac{1}{l_{\xi}}=\lim_{z\to\infty}\frac{\xi_{c}(z)}{z}%
=\lim_{z\to\infty}\frac{\xi_{s}(z)}{z}.
\end{equation}
Equations (\ref{transmittance-1}) and (\ref{limits}) evidently show that in
continuous case $l$ and $l_{\xi}$ exactly coincide.

In the discrete case (multilayered stack), corresponding expression for
transmittance reads

\begin{eqnarray}  \label{transmittance-2}
\mathcal{T}_{N}\equiv|T_{N}|^{2}={4}\left(
e^{2\xi_{c}(N)}+e^{2\xi_{s}(N)}+\right.  \notag \\
\left.2e^{\xi_{c}(N)+\xi_{s}(N)}
\sin\left(\theta_{c}(N)-\theta_{s}(N)\right)\right)^{-1}.
\end{eqnarray}
Here the last term in denominator differs from that in Eq. (\ref%
{transmittance-1}). Moreover, it can change its sign and generally speaking
can essentially reduce the denominator itself thus enlarging transmittance
and as a result enlarging localization length $l_{\xi}$ in compare to $l$.
Thus, Eqs. (\ref{transmittance-2}) and (\ref{Lyapunov-1}) enable us to state
only that $l\geq l_{\xi}$ in contrast to the continuous case where these two
localization lengths always coincide. In spite of that, studying of
localization in normal disordered multilayered stacks did not show any
difference in the two lengths. We will see below that such a difference
really manifests itself in the alternated metamaterial stacks.

In this review we are mainly interested in the transmission length $l_T$.
This quantity can be found directly by standard transmission experiments. At
the same time, it is sensitive to the size of the system and therefore is
best suited to the description of the transmission properties in both the
localized and ballistic regimes. More precisely, the transmission length
coincides either with the localization length $l$ or with the ballistic
length $b$, respectively in the cases of comparatively long stacks
(localized regime) or comparatively short stacks (ballistic regime). That is,

\begin{equation*}
l_T\approx \left\{
\begin{array}{ccc}
l &  & N\gg l \\
&  &  \\
b &  & N\ll b.%
\end{array}%
\right. .  \notag
\end{equation*}

\subsection{Transfer Matrices and Weak Scattering Approximation}
\label{subsec:transfer}
In this Subsection we describe some methods used for calculation of
transmission length and other transmission or/and localization
characteristics in various regimes. All of them are based on various
versions of transfer matrix approach.

\begin{figure}[h]
\centerline{\includegraphics[width=8cm]{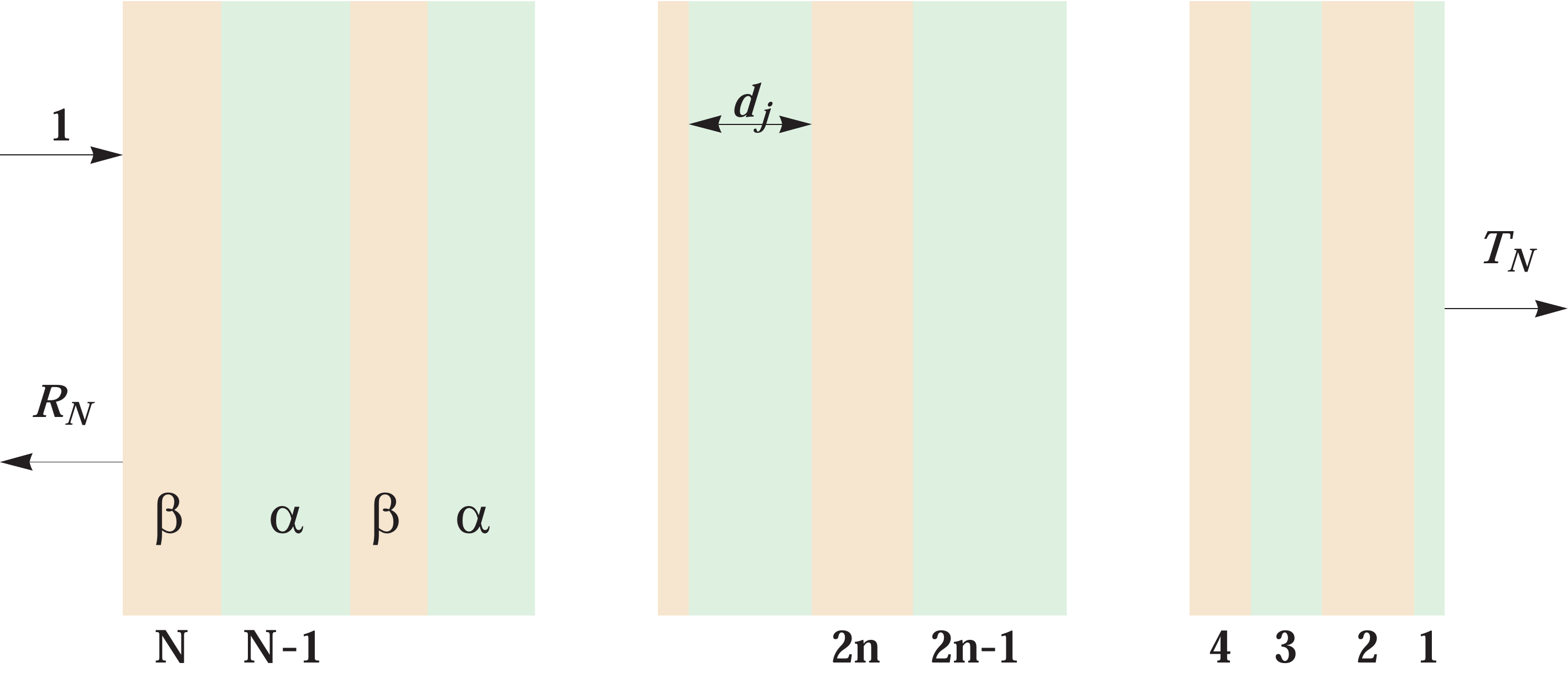}}
\caption{(Ref.~[\onlinecite{Asatryan07}]) Two component multilayered
alternative stack.}
\label{Meta_Fig1}
\end{figure}

Consider the M-stack alternatively comprised of even number $N$ of uniform
layers labeled by index $j=1,...,N$ from right to left, so that all odd
layers $j=2n-1,$ are of type ``$\alpha$'' and all even layers $j=2n$ are of
type ``$\beta$'', $n=1,2,...,N/2$ (see Fig.~\ref{Meta_Fig1}). In general
case the $j-$th layer is characterized by its dimensionless thickness $d_{j}$%
, dielectric permittivity $\varepsilon_{j}$ and magnetic permeability $%
\mu_{j}.$

The total transfer matrix (\ref{T-stack-2}) is factorized to the product

\begin{eqnarray}  \label{product}
\hat{T}(N)={\hat t}_{N}{\hat t}_{N-1}...{\hat t}_{2}{\hat t}_{1}
\end{eqnarray}
of the layer transfer matrices ${\hat t}_{j}.$

Note that for considered alternated stack, it is natural to join each pair
of subsequent layers with numbers $j=2n-1$ and $j=2n$ into one effective
cell number $n$. Then the total transfer matrix factorizes to the product of
$N/2$ transfer matrices of separate cells~\cite%
{Asatryan07,Izra09,IMT-10,TIM-FNT,Mak-12}.

Parameterizing the transfer matrix of the $j-$th layer by its transmission $%
t_{j}$ and reflection $r_{j}$ coefficients of a corresponding layer we
obtain the recurrence relations

\begin{eqnarray}
T_{j}=\frac{T_{j-1}t_{j}}{1-R_{j-1}r_{j}},\ \ \ T_{0}=1,
\label{rec-t}\\
R_{j}=r_{j}+\frac{R_{j-1}t_{j}^{2}}{1-R_{j-1}r_{j}}, \ \ \ R_{0}=0,
\label{rec-r}
\end{eqnarray}
where $T_{j}$ and $R_{j}$ are transmission and reflection coefficients of
the reduced stack comprised of only $j$ first layers. These relations
provide an \textit{exact} description of the system and will be used later
for direct numerical simulations of its transmission properties. Another
possible but less effective way is related to direct numeric calculation of
the total transfer matrix (\ref{product}).

Relations (\ref{rec-t}) and (\ref{rec-r}) serve as a starting point for the
weak scattering approximation (WSA) elaborated in~\cite{Asatryan07} and
based on assumption that the reflection from a single layer is small \emph{%
i.e.}, $\left\vert r_{j}\right\vert \ll 1$. This demand is definitely
satisfied in the case of weak disorder. Within WSA, instead of exact
relations (\ref{rec-t}), (\ref{rec-r}) we use for the transmission length
the following first order approximations

\begin{eqnarray}
\ln T_{j} &=&\ln T_{1,j-1}+\ln t_{j}+R_{j-1}r_{j},
\label{rec3}\\
R_{j} &=&r_{j}+R_{j-1}t_{j}^{2}, \ \ \ j=2,3,...,N.
\label{rec4}
\end{eqnarray}%
Note that in deriving Eq. (\ref{rec4}), we omit the first-order term $%
R_{j-1}^{2}t_{j}^{2}r_{j}.$ This is uncontrolled action. The omitted term
contributes only to the second order of $\ln T_{j}$ already after the first
iteration for not very large number of layers $j$. For sufficiently large $j,
$ it should be taken into account. Nevertheless as we will see below, this
approximation is excellent in all wavelength region.

Neglecting the last term in the right hand side of Eq. (\ref{rec3}) we come
to the so-called single-scattering approximation (SSA), which implies that
multi-pass reflections are neglected so that the total transmission
coefficient is approximated by the product of the single layer transmission
coefficients as well as total transmittance is approximated by the product
of the single layer transmittances that results in

\begin{equation*}  \label{SSA}
\ln|T_{N}|=\sum_{j=1}^{N}\ln|t_{j}|.
\end{equation*}%
In the case of very long stacks (\emph{i.e.}, as the length $N\rightarrow
\infty $), we can replace the arithmetic mean, $N^{-1}\sum_{j=1}^{N}\ln
|t_{j}|$, by its ensemble average $\langle \ln |t|\rangle .$ On the other
hand, in this limit the reciprocal of the transmission length coincides with
the localization length. Using the energy conservation law, $%
|r_{j}|^{2}+|t_{j}|^{2}=1$, which applies in the absence of absorption, the
reciprocal localization length in single-scattering approximation may be
written as

\begin{equation*}
\left( \frac{1}{l}\right) _{ssa}=\frac{1}{2}\langle |r|^{2}\rangle
\label{SSA-1}
\end{equation*}%
and is proportional to the mean reflectance of a single random layer \cite%
{LGP,Liansky-1995}.

The version of transfer matrix approach described above is based on
consideration of a single layer embedded into vacuum. This version and
related WSA were used in~\cite{Asatryan07,Asatryan10a,Asatryan10b,Asatryan12}
for analytical and numerical study of metamaterial M-stacks (see Section~\ref%
{sec:meta}).

Another version used in~\cite{Bliokh-mag} (Section~\ref{subsec:nonreciprocal}%
) is based on a separation of wave propagation inside a layer and through
the interface between layers (see \textit{e.g.} Ref.~\cite{BerryKlein}).
Here wave propagation inside the $j$-th layer is described by diagonal
transfer matrix

\begin{equation}  \label{space}
{\hat S}_j=\mathrm{diag}( {\text{e}}^{-i\varphi_j},{\text{e}}^{i\varphi_j}),
\end{equation}
where $\varphi_j=k_{j}d_{j}$ is the phase accumulated upon the wave
propagating from left to right through the $j$-th layer, and $k_{j}=\frac{%
\omega}{c}\sqrt{\varepsilon_{j}\mu_{j}}.$ The interfaces are described by
unimodular transfer matrices ${\hat F}^{0\alpha},{\hat F}^{\alpha\beta},{%
\hat F}^{\beta\alpha},{\hat F}^{\beta 0}$ corresponding, respectively, to
transitions (all from left to right) from vacuum to the medium `$\alpha$',
from the medium `$\alpha$' to the medium `$\beta$', from the medium `$\beta$%
' to the medium `$\alpha$', and from the medium `$\alpha$' to vacuum. Thus,
the total transfer matrix (\ref{T-stack-2}) of the structure in Fig.~\ref%
{Meta_Fig1} is

\begin{eqnarray}  \label{total transf matr}
\hat{T}(N)&=&{\hat F}^{0\alpha}{\hat F}_{N}{\hat S}_N{\hat F}_{N-1}{\hat S}%
_{N-1}~...~ {\hat F}_{2}{\hat S}_{2}{\hat F}_{1}{\hat S}_{1}{\hat F}^{\alpha
0},  \notag \\
\!\!\!\!\!{\hat F}_{2n-1}&\equiv& {\hat F}^{\beta\alpha}, ~{\hat F}%
_{2n}\equiv {\hat F}^{\alpha\beta}, \ n=1,2,...,N/2.
\end{eqnarray}

Using the group property of the interface transfer matrices: ${\hat F}%
^{\beta\alpha}={\hat F}^{\beta 0}{\hat F}^{0\alpha},$ and ${\hat F}%
^{\alpha\beta}={\hat F}^{\alpha 0}{\hat F}^{0\beta},$ the total transfer
matrix is factorized to the product (\ref{product}) where the layer transfer
matrices are

\begin{eqnarray*}  \label{layert}
\hat{t}_{2n}={\hat F}^{0\beta}{\hat S}_{2n}{\hat F}^{\beta 0}, \ \ \ \ {\hat
t}_{2n-1}={\hat F}^{0\alpha}{\hat S}_{2n-1}{\hat F}^{\alpha 0}.
\end{eqnarray*}
Such a representation is especially efficient in the shortwave limit where
the total transmission coefficient reduces to the product of the
transmission coefficients of only interfaces (see Ref.~\cite{BerryKlein} and
Section~\ref{subsec:nonreciprocal}).

Come now to application of the transfer matrix approach to calculation of
the Lyapunov exponent $\gamma$. Define for each layer the curentless vector $%
\vec{e}_{j}$ by Eq. (\ref{currentless-1}) with the corresponding values $%
\xi_{j}$ and $\theta_{j}$. In this terms Lyapunov exponent is written as

\begin{equation}  \label{Lyapunov-2}
\gamma =\lim_{j\to\infty}\frac{\xi_{j}}{j}=
\lim_{j\to\infty}(\xi_{j}-\xi_{j-1}).
\end{equation}
(we used Shtolz theorem). The vectors $\vec{e}_{j}$ and $\vec{e}_{j-1}$
satisfy the equation

\begin{equation}
\label{layer-transf}
\vec{e}_{j}=\hat{t}_{j-1}\vec{e}_{j-1}.
\end{equation}
Therefore the difference in the r.h.s. of Eq. (\ref{Lyapunov-2}) is some
function of $\theta_{j-1}$

\begin{equation}  \label{difference}
\xi_{j}-\xi_{j-1}=\Phi(\theta_{j-1}),
\end{equation}
which explicit form is determined by Eq.~\ref{layer-transf}.
Using the self averaging of the ratio $\xi_{j}/j$ and the fact that the
phase $\theta_{j}$ stabilizes\cite{LGP}, we finally obtain for Lyapunov
exponent

\begin{equation}  \label{Lyapunov-30}
\gamma=\langle \Phi(\theta)\rangle_{\text{st}},
\end{equation}
where average in the r.h.s. is taken over stationary distribution of the
phase $\theta$.

Continuous version of this result was obtained in~\cite{LGP} (see Eq.
(10.2)). Its discrete version in slightly different terms (see Section %
\ref{subsec:enlight}) was obtained in \cite{IKT}. Note that due to existence
of the closed formula (\ref{Lyapunov-30}) for Lyapunov exponent, the task of
analytical calculation of the localization length $l_{\xi}=\gamma^{-1}$ is a
simpler problem than that of transmission length $l_{T}$.

The next steps are standard (see \textit{e.g.} Refs.~[\onlinecite{GP,LGP}]):
using (\ref{layer-transf}) to get the dynamic equation for the phase $\theta$%
, write down corresponding Fokker-Planck equation for its distribution,
solve it and calculate the average (\ref{Lyapunov-30}). Moreover, in weakly
disordered systems, only the first and the second order terms should be
accounted for in the dynamic equations~\cite{APS,LGP}. For minimally disordered M-stacks defined in Section~\ref{sec:intro}, this program was successfully realized in~\cite{TIM-FNT,Mak-12} (see Section \ref{subsec:enlight} below).

\section{Suppression of Localization in Metamaterials}

\label{sec:meta} \numberwithin{equation}{section}
Over the past decade, the physical properties of metamaterials and their
possible applications in modern optics and microelectronics, have received
considerable attention (see e.g. Refs \cite%
{Sok,Bliokh,classification,Shalaev}). The reasons for such an interest are
unique physical properties of metamaterials including their ability to
overcome the diffraction limit~\cite{Veselago,Pendry}, potential role in
cloaking~\cite{Schurig}, suppression of spontaneous emission rate \cite{emis}%
, the enhancement of quantum interference \cite{inter}, etc. One of the
first study of the effect of randomness~\cite{Gorkunov} revealed that weak
microscopic disorder may lead to a substantial suppression of the wave
propagation through magnetic metamaterials over a wide frequency range.
Therefore the next problem was to study localization properties of
disordered metamaterial systems.

It was known that, in normal multilayered systems comprising right-handed
media, the localization length is proportional to the square of wavelength $%
\lambda$ in the long-wavelength limit, tends to a constant in a
short-wavelength regime, and oscillates irregularly in the intermediate
region~\cite{Ping07,John,Azbel,Baluni,deSterke}. Natural question arises:
how inclusion of metamaterial layers influences the localization and
transmission effects.

The study of localization in metamaterials was started in Ref. [%
\onlinecite{Dong}] where wave transmission through an alternating sequence
of air layers and metamaterial layers of random thicknesses was studied.
Localized modes within the gap were observed and delocalized modes were
revealed despite the one-dimensional nature of the model. Then comprehensive
study of transmission properties of M-stacks was done in\cite%
{Asatryan07,Asatryan10a,Asatryan10b,Asatryan12}. Here anomalous enhancement
of the transmission through minimally disordered (see Section~\ref{sec:intro}) M-stacks was revealed\cite%
{Asatryan07}, non-coincidence of the two localization lengths $l$ and $%
l_{\xi}$ was established\cite{Asatryan10a}, polarization\cite{Asatryan10b}
and dispersion\cite{Asatryan12} effects in transmission were studied.

Scaling laws of the transmission through a similar mixed multilayered
structure were investigated in Ref. [\onlinecite{random}]. It was shown that
the spectrally averaged transmission in a frequency range around the fully
transparent resonant mode decayed with the number of layers much more
rapidly than in a homogeneous random slab. Localization in a disordered
multilayered structure comprising alternating random layers of two different
left-handed materials was considered in Ref. [\onlinecite{Chan}]. Within the
propagation gap, the localization length was shorter than the decay length
in the underlying periodic structure (opposite of that observed in the
random structure of right-handed layers).

Detailed investigation of Lyapunov exponent (and therefore localization
length $l_{\xi}$) in various multilayered metamaterials was presented in\cite%
{Izra09,IMT-10,TIM-FNT,Mak-12}. In the weak disorder limit, explicit
expressions for Lyapunov exponent valid in all region of wavelengths for
various kinds of correlated disorder were obtained\cite{Izra09,IMT-10} and
analytical explanation of anomalous suppression of localization was done\cite%
{TIM-FNT,Mak-12}.

Dispersion effects in M-stacks comprised by metamaterial layers separated by
air layers with only positional disorder were considered in\cite%
{Mog-10,Reyes,Mog-11}. Here essential suppression of localization in the
vicinity of the Brewster angle and at the very edge of the band gap was
revealed\cite{Mog-10}, influence of both quasi-periodicity and structural
disorder was studied\cite{Reyes} and effects of some types of disorder
correlation on light propagation and Anderson localization were investigated%
\cite{Mog-11}.

In this Section we consider suppression of localization in sufficiently
disordered M-stacks. In the first four Subsections we consider the
model with non-correlated fluctuating thicknesses and dielectric
permittivities. This model possesses the main features cause by the presence
of metamaterials and at the same time remains comparatively simple. The
results concerning disorder correlations can be found in papers mention in
the previous paragraph and detailed recent survey\cite{IKM}. The
presentation is mostly based on works\cite%
{Asatryan07,Asatryan10a,Asatryan10b,Asatryan12,Mak-12}.


\subsection{Model}

\label{subsec:model}

We start with the model described at the beginning of Suection~\ref%
{subsec:transfer} and displayed in Fig.\ref{Meta_Fig1}. Electromagnetic
properties of the $j$-th layer with given dielectric permittivity $%
\varepsilon_{j}$ and magnetic permeability $\mu_{j},$ are characterized by
its impedance $Z_{j}$ and refractive index $\nu_{j}$

\begin{equation}  \label{ImpRefInd}
Z_{j}=\sqrt{\mu_{j}/\varepsilon_{j}}, \ \ \ \ \nu_{j}=\sqrt{%
\mu_{j}\varepsilon_{j}}.
\end{equation}

Being embedded into vacuum, each layer can be described by its reflection
and transmission coefficients with respect to wave with dimensionless length
$\lambda$ incident from the left

\begin{eqnarray}  \label{t}
r_{j} =\frac{\rho _{j}(1-e^{2i\beta _{j}})}{1-\rho _{j}^{2}e^{2i\beta _{j}} }%
, \ \ \ \ \ t_{j} =\frac{(1-\rho _{j}^{2})e^{i\beta _{j}}}{1-\rho
_{j}^{2}e^{2i\beta_{j}}}.
\end{eqnarray}%
Here $\rho_{j}=(Z_{j}-1)/(Z_{j}+1)$ is Fresnel coefficient, $\beta
_{j}=kd_{j}\nu _{j}$, and $k=2\pi /\lambda $ is dimensionless wavenumber.%
\newline

Within our model, dielectric permittivity, magnetic permeability and
thickness of the $j$-th layer have the forms

\begin{eqnarray}
\varepsilon _{j}=(-1)^{j}(1+\delta _{j}^{(\nu )})^{2},  \notag \\
\mu_{j}=(-1)^{j}, \ \ \ \ \ d_{j}=1+\delta_{j}^{d},  \label{eps}
\end{eqnarray}%
so that corresponding impedance and refractive index are

\begin{eqnarray}
Z_j&=&\sqrt{\mu_j/\varepsilon_j}= (1+\delta^{(\nu)}_j)^{-1}  \label{z-1-imp}
\\
\nu _{j}&=&(-1)^{j} (1+\delta _{j}^{(\nu )}).  \label{nu-1-imp}
\end{eqnarray}%
The thickness fluctuations $\delta _{j}^{(d)}$ are independent identically
distributed zero-mean random variables, as well as all refractive index
fluctuations $\delta _{j}^{(\nu)}$. To justify the weak scattering
approximation, we assume that all these quantities $\delta _{j}^{(d,\nu)}$
are small.

The considered model possesses some symmetry: statistical properties of the
fluctuations and absorption coefficient are the same for $L$ and $R$ layers.
As a consequence of this symmetry, the scattering coefficients of $R$ and $L$
layers are complex conjugate $t_{r}=t_{l}^{\ast }$ and $r_{r}=r_{l} ^{\ast }$%
, that results in the relations

\begin{equation}  \label{symmetry}
\langle g(t_{r})\rangle =\langle g(t_{l})\rangle ^{\ast },\ \ \ \langle
g(r_{r})\rangle =\langle g(r_{l})\rangle ^{\ast }.
\end{equation}
valid for any real-valued function $g$ in either the lossless or absorbing
cases. In more general models this symmetry can be broken.

The model with two parameters (here - thickness and refractive index) is in
a sense the simplest sufficiently disordered model. Further simplification
where only one of these quantities is random qualitatively changes the
picture. Indeed, the case of M-stack with only thickness disorder in the
absence of absorption is rather trivial: such stack is completely
transparent (a consequence of $Z_{j}\equiv 1$). On the other hand, M-stack
with only refractive-index disorder as it was revealed in\cite{Asatryan07},
manifests a dramatic suppression of Anderson localization - essential
enlightenment in the long wave region. This intriguing case is considered
below in Section~\ref{subsec:enlight}. So here we focus on the case where
both two types of disorder are simultaneously present.

Specific features of transmission and localization in the M-stacks look more
pronounced in comparison with those of homogeneous stack (H-stack) comprised
of solely either right-handed or left-handed layers. Therefore albeit
localization in disordered H-stacks with right-handed layers has been
studied by many authors \cite{Baluni,deSterke,Asatryan98,Ping07,Luna}, we
also consider this problem here in its most general formulation. This
consideration enables us to compare localization properties of M- and
H-stacks. To describe a H-stack composed of only $R$ ($L$) layers, all
multipliers $(-1)^{j}$ in Eqs. (\ref{eps}) and (\ref{nu-1-imp}) should be
replaced by 1 (-1). 

\subsection{Mixed Stack}

\label{subsec:mixed} 

Within the version (\ref{rec3}), (\ref{rec4}) of weak scattering
approximation, contributions from the even and odd layers are separated. As
a result the transmission length of a finite length M-stack may be cast in
the form\cite{Asatryan10a}

\begin{equation}  \label{FinalG3}
\frac{1}{l_T}=\frac{1}{l}+\left(\frac{1}{b}-\frac{1}{l}\right) f(N/\bar{l}),
\end{equation}
where

\begin{equation}  \label{f}
f(x)=\frac{ 1-\text{e}^{-x}}{x}.
\end{equation}

\noindent Localization length $l$, ballistic length $b$, and \emph{crossover
length} $\bar{l}$ are completely described by the three averages $<\ln|t|>$,
$<r>$, and $<t^{2}>$ composed of transmission $t$ and reflection $r$
coefficients of a single right-handed layer:

\begin{equation}
\frac{1}{l}=-\langle \ln |t|\rangle -\frac{|\langle r\rangle |^{2}+\mathrm{Re%
}\left( \langle r\rangle ^{2}\langle t^{2}\rangle ^{\ast }\right) }{%
1-|\langle t^{2}\rangle |^{2}},  \label{loc}
\end{equation}

\begin{eqnarray*}  \label{bal}
\frac{1}{b} =\frac{1}{l}-\frac{2/\bar{l}}{1-\exp (-2/\bar{l})}\times  \notag
\\
\left( \frac{|\langle r\rangle |^{2}+\mathrm{Re}\left( \langle r\rangle
^{2}\langle t^{2}\rangle ^{\ast }\right) }{1-|\langle t^{2}\rangle |^{2}}-%
\frac{|\langle r\rangle |^{2}}{2}\right)
\end{eqnarray*}

\begin{equation}  \label{cross}
\bar{l}=-\frac{1}{\ln |\langle t^{ 2}\rangle |}.
\end{equation}%
These Eqs. (\ref{FinalG3}) - (\ref{cross}) are valid in the presence of
absorption. However below to make our treatment more transparent, we
consider the lossless case.

The characteristic lengths $l$, $b$, and $\bar{l}$ are functions of
wavelength $\lambda$. The first two always satisfy the inequality $l(\lambda
)>b(\lambda )$, while in the long wavelength region the crossover length is
the shortest of the three, $b(\lambda )>\bar{l}(\lambda )$. In the case of a
fixed wavelength $\lambda ,$ for comparatively short stacks with $N\ll \bar{l%
}(\lambda )$ the function $f(N,\bar{l})\approx 1$, while for sufficiently
long stacks $N\gg \bar{l}(\lambda )$, it tends to zero $f(N,\bar{l})\approx 0
$. Correspondingly, transmission length coincides with ballistic length $%
l_T(\lambda )\approx b(\lambda )$ for short stacks $N\ll \bar{l}(\lambda )$
and with localization length $l_T(\lambda )\approx l(\lambda )$ for long
stacks $N\gg \bar{l}(\lambda )$ with the transition between the two ranges
of $N$ being determined by the crossover length $\bar{l}(\lambda )$. Thus
ballistic regime occurs when the stack is much shorter than the crossover
length $N\ll \bar{l}(\lambda )$. The localized regime is realized for the
stacks longer than localization length $N\gg l(\lambda ).$ For the stacks
with intermediate sizes $\bar{l}(\lambda )\lesssim N \lesssim l(\lambda),$
transmission length coincides with localization length, however they
correspond to the transition region between ballistic regime and localized
one.

Alternatively we can consider the stack with a given size $N$ and use the
wavelength as the parameter governing the localized and ballistic regimes.
To do this, we introduce two characteristic wavelengths, $\lambda _{1}(N)$
and $\lambda _{2}(N),$ defined by the relations

\begin{equation}  \label{wavelengths-1}
N=l(\lambda _{1}(N)),\ \ \ N=\bar{l}(\lambda _{2}(N)).
\end{equation}
In these terms, the localized regime occurs if $\lambda \ll \lambda _{1}(N) $%
, while in the long wavelength region, $\lambda \gg \lambda _{2}(N)$, the
propagation is ballistic. Intermediate range of wavelengths, $\lambda
_{1}(N)<\lambda <\lambda _{2}(N)$, corresponds to transition region between
the two regimes.

Consider now example of rectangular distribution, where the fluctuations $%
\delta _{j}^{(\nu )}$ and $\delta _{j}^{(d)}$ are uniformly distributed over
the intervals $[-Q_{\nu },Q_{\nu }]$ and $[-Q_{d},Q_{d}]$ respectively and
have the same order of magnitude $Q_{\nu }\sim Q_{d}$ so that the
dimensionless parameter

\begin{equation*}  \label{zeta}
\zeta =2\frac{Q_{d}^{2}}{Q_{\nu }^{2}}
\end{equation*}%
is of order of unity.

At the next step, we calculate the averages $<\ln|t|>$, $<|r|>$, and $%
<|t^{2}|>$ with the help of Eqs. (\ref{t}) - (\ref{nu-1-imp}), substitute
them into Eqs.,~(\ref{loc}) and (\ref{cross})and neglect the contribution of
terms of order higher than $Q_{d}^{2}$. The resulting general expressions
for localization, ballistic and crossover lengths are rather cumbersome so
we present here only their asymptotical forms.

In the short wavelength region, the main contribution to localization length
is related to the first term in the r.h.s. of Eq. (\ref{loc}) corresponding
to the single scattering approximation and the localization length is

\begin{equation}  \label{shortWave}
l(\lambda )=\frac{12}{Q_{\nu }^{2}}, \ \ \ \ \ \ \lambda\ll 1.
\end{equation}%
This means that the size $N$ of the short stack $NQ_{\nu}^{2}\ll 1$ is
always smaller than localization length and the short wave transmission
through short stack is always ballistic.

Opposite limiting case $NQ_{\nu}^{2}\gg 1$ corresponds to the long stacks.
Here both two regimes are realized and transition from localized propagation
to the ballistic one occurs at the long wavelength $\lambda\sim Q_{\nu}\sqrt{%
N}\gg 1$. Indeed, asymptotical expressions for all three characteristic
lengths read

\begin{equation}  \label{lloc}
l(\lambda)\approx \frac{3\lambda^2}{2\pi^2Q_{\nu}^2} \ \frac{3+\zeta}{1+\zeta%
},
\end{equation}

\begin{equation}  \label{crossover-3}
\bar{l}(\lambda )\approx \frac{3\lambda^2}{2\pi^2Q_{\nu}^2} \frac{1}{%
4(3+\zeta )},
\end{equation}
and

\begin{equation}  \label{bbal}
b(\lambda )\approx \frac{3\lambda ^{2}}{2\pi ^{2}Q_{\nu }^{2}}.
\end{equation}

Note that the single scattering approximation for localization length fails
in the long wave limit because both two terms in the r.h.s. of Eq. (\ref{loc}%
) contribute to the asymptotic (\ref{lloc}).

Thus in the symmetric weak scattering case, ballistic, localization, and
crossover length in the long wave region differ only by numerical
multipliers, satisfy the inequality $\bar{l}(\lambda )<b(\lambda )<l(\lambda
)$ mentioned above, and are proportional to $\lambda^{2}.$ Two
characteristic wavelengths (\ref{wavelengths-1}) corresponding to
localization length (\ref{lloc}) and crossover length (\ref{crossover-3}),
are proportional to $Q_{\nu}\sqrt{N},$  differ only by a numerical
multiplier and satisfy the inequality $\lambda _{1}(N)<\lambda _{2}(N)$. For
the sufficiently long stacks $NQ_{\nu}^{2}\gg 1,$ they are lying in long
wave region $\lambda_{1,2}\gg 1.$

Localization properties of infinite stack are described by Lyapunov exponent
(\ref{Lyapunov-1}) or by localization length (\ref{locLyap}). Within the
considered model (\ref{t}) - (\ref{nu-1-imp}), its long wave asymptotic
calculated with the help of well known transfer matrix approach reads

\begin{eqnarray}  \label{Lyapunov-3}
\!\!\!\!\gamma \approx\frac{\pi ^{2}\overline{\left(1+\delta^{(d)}\right)^{2}%
}}{2\lambda ^{2}}\frac{\overline{\epsilon ^{2}}-\overline{\epsilon }^{2}}{%
\overline{\epsilon }}, \ \ \ \epsilon =(1+\delta^{(\nu)})^2.
\end{eqnarray}

In the case of rectangular distributions of the fluctuations of the
dielectric constants and thicknesses described above, reciprocal Lyapunov
exponent reduces to

\begin{equation}  \label{Lyapunov-4}
l_{\xi}(\lambda)=\gamma^{-1} (\lambda)\approx\frac{3\lambda ^{2}}{2\pi
^{2}Q_{\nu }^{2}}
\end{equation}
coinciding with ballistic length $b(\lambda)$. Thus the disordered M-stack
in the long wavelength region presents a unique example of a one-dimensional
disordered system in which the localization length defined as transmission
decrement of sufficiently long stack, differs from the reciprocal of the
Lyapunov exponent.

The qualitative picture of transmission and localization properties of the
symmetric mixed stack described above, remains correct in much more general
case where statistical properties of the $r$ and $l$ layers are different
and distributions of the fluctuations and thicknesses are not rectangular.
The only distinction we expect, is that localization and crossover lengths
will have different wavelength dependence that will result in more
complicated structure of ballistic region like that considered below for
H-stack (see Section \ref{subsec:hom} below). \newline

\begin{figure}[h]
\rotatebox{0}{\scalebox{0.40}{\includegraphics{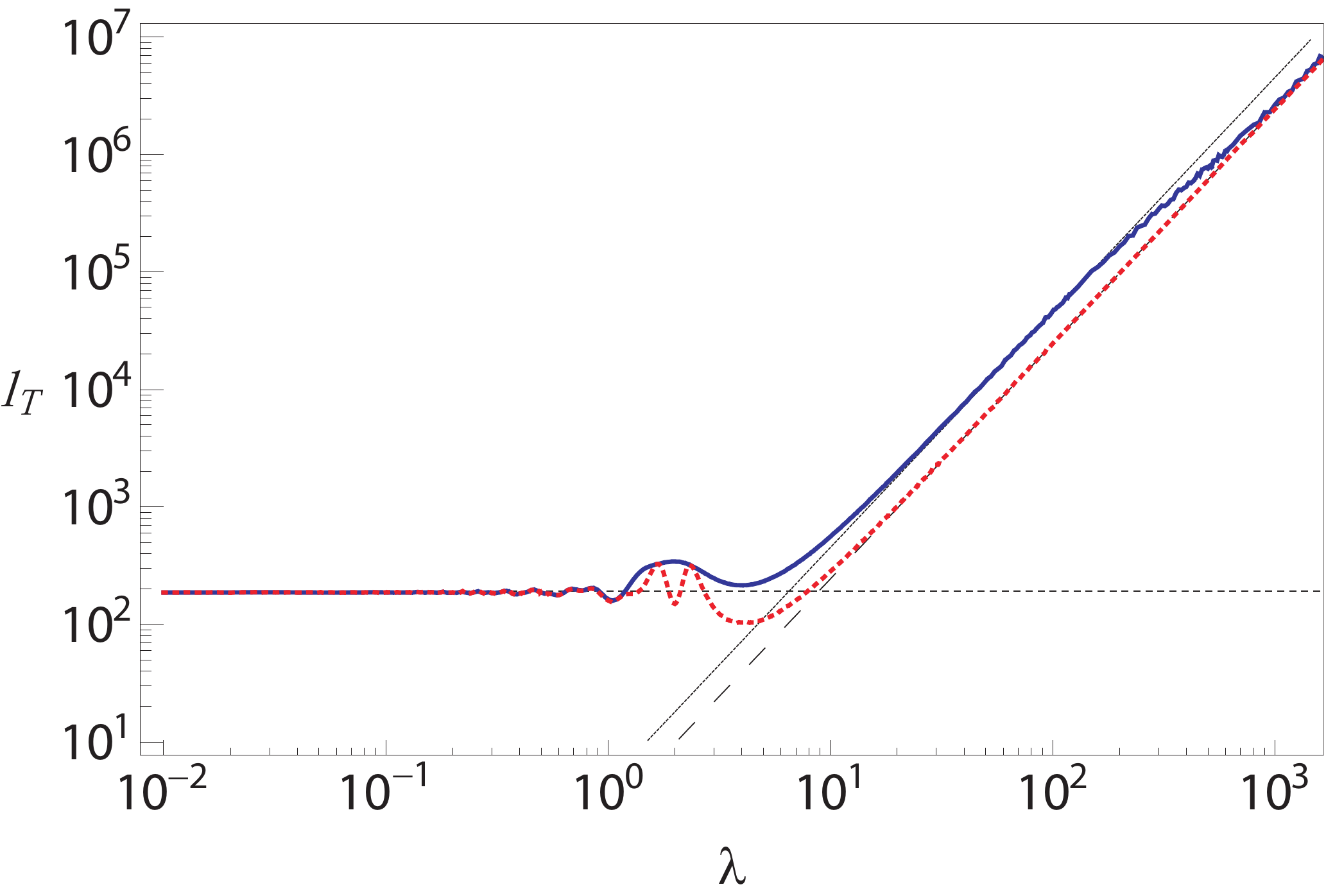}}}
\caption{(Ref.~[\onlinecite{Asatryan10a}], color online) Transmission length
$l_T$ vs $\protect\lambda$ for M-stack (thick solid line, direct simulation
and calculations based on WSA recurrence relations) and H-stack (thick
dashed line, direct simulation). Asymptotics of the localization length $l:$
the short wavelength asymptotic (thin dotted line), and the long wavelength
asymptotics---thin solid line for the M-stack and a thin dashed line for the
H-stack. }
\label{Meta-Fig2}
\end{figure}

To check the WSA theoretical predictions formulated above we provided a
series of numerical calculations. They were made for the lossless stack with
uniform distributions of the fluctuations $\delta ^{(d)},$ $\delta ^{(\nu )}$%
, with widths of $Q_{\nu }=0.25$ and $Q_{d}=0.2$, respectively and included
(a) direct simulations based on the exact recurrence relations (\ref{rec-t}%
), (\ref{rec-r}); (b) the weak scattering analysis for the transmission
length. In all cases, unless otherwise is mentioned, the ensemble averaging
is taken over $N_{r}=10^{4}$ realizations.\newline

Throughout this Subsection we considered only M-stacks. Nevertheless, to
emphasize the main features of the transmission in metamaterials, compare
transmission spectra for a M-stack of $N=10^{5}$ layers and a H-stack of
length $N=10^{3}$ plotted in the same Fig. \ref{Meta-Fig2}. Both stacks are
sufficiently long: for the shortest of them parameter $NQ_{\nu}^{2}$ is $%
62.5\gg 1$. There are two major differences between the results for these
two types of samples: first, in the localized regime ($N\gg l_T$), the
transmission length of the M-stack exceeds or coincides with that of the
H-stack; second, in the long wavelength region, the plot of the transmission
length of the M-stack exhibits a pronounced bend, or kink, in the interval $%
\lambda \in [ 10^{2},10^{3}]$, while there is no such feature in the H-stack
results.

Fig.~\ref{Meta-Fig2} demonstrates an excellent agreement of analytical and
numerical results: the curves obtained by direct numerical simulations and
by calculations based on the weak scattering approximation (WSA) are
indistinguishable (solid line). The short and long wavelength behavior of
the transmission length is also in excellent agreement with the calculated
asymptotics in both regimes. The characteristic wavelengths of this mixed
stack are $\lambda _{1}\approx 148$ and $\lambda _{2}\approx 839$.
Therefore, the region $\lambda \lesssim 148$ corresponds to localized
regime, whereas longer wavelengths, $\lambda \gtrsim 839$, correspond to the
ballistic regime. Thus the kink observed within the region $\lambda
_{1}\lesssim \lambda \lesssim \lambda _{2}$ describes crossover from the
localized to the ballistic regime. The long wave asymptotic of the ballistic
length, as we saw below, coincides with that of reciprocal Lyapunov
exponent. Therefore the difference between localization and ballistic
lengths of the M-stack simultaneously confirms the difference between
localization length and reciprocal Lyapunov exponent in localized regime.

More detail numerical calculations of transmission length, average
reflectance, and characteristic wavelengths of the M-stacks with various
sizes also demonstrate an excellent agreement between direct simulations and
WSA based calculations thus completely confirming the theory presented above%
\cite{Asatryan10a}.

Until now, we have dealt only with the transmission length $l_T(\lambda)$,
which was defined through an average value. However, additional information
can be obtained from the transmission length $l_N(\lambda)$ for a single
realization,

\begin{equation*}  \label{single-real}
\frac{1}{l_N}=-\frac{\ln \ \left\vert T_{N}\right\vert }{N}.
\end{equation*}

\begin{figure}[h]
\rotatebox{0}{\scalebox{0.40}{\includegraphics{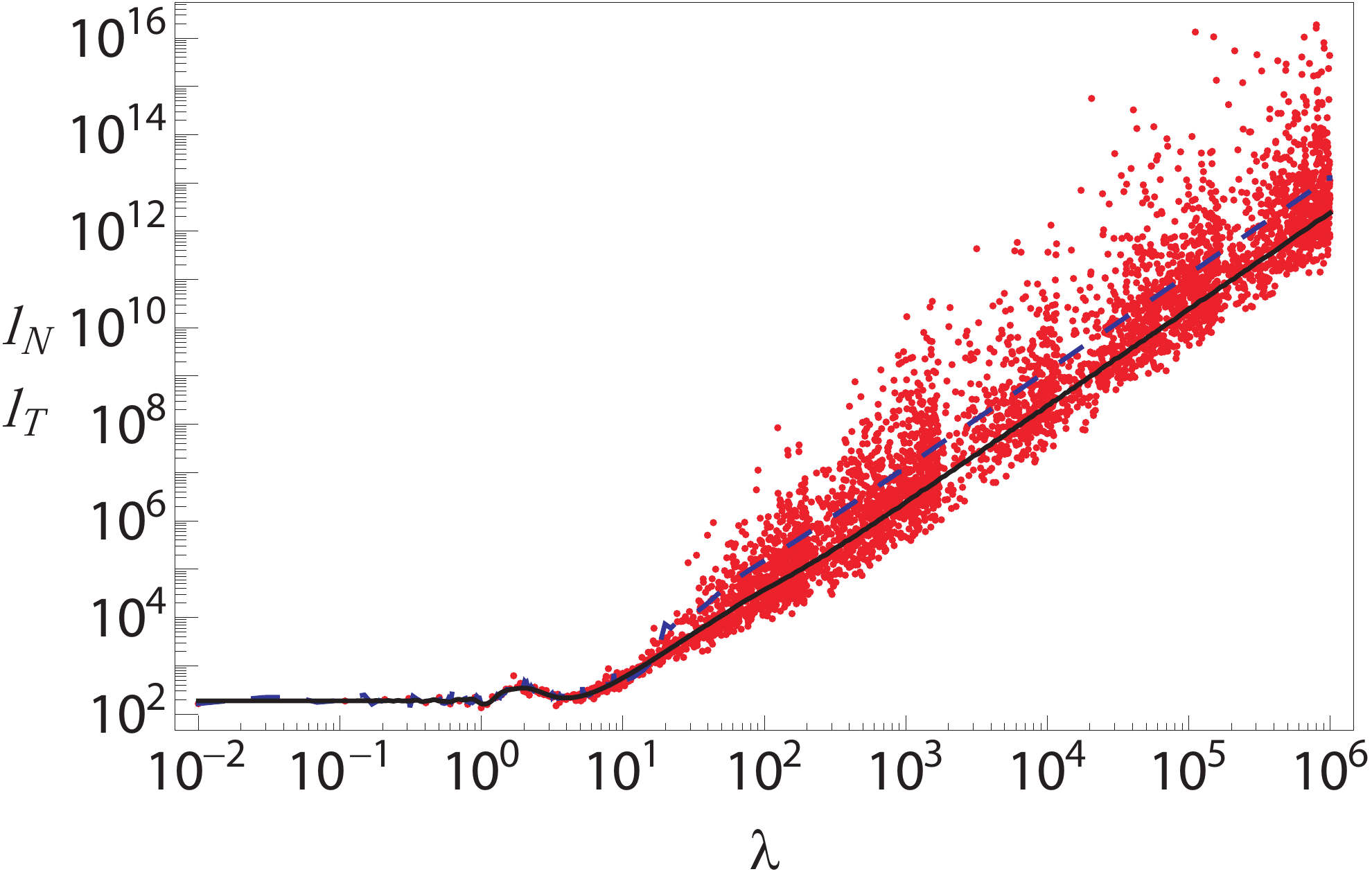}}}
\caption{(Ref.~[\onlinecite{Asatryan10a}], color online) Transmission
lengths $l_T$ (solid black line) and the transmission length for a single
realization $l_N$ (dashed blue line) vs $\protect\lambda$ for a M-stack with
$Q_{\protect\nu}=0.25$, $Q_d=0.2$ and $N=10^4$ layers. Each separate point
corresponds to a particular wavelength with its own realization of a random
stack.}
\label{Meta-Fig5New}
\end{figure}

In the localized regime, \emph{i.e.} for a sufficiently long M-stack with $%
N\gg l$, the transmission length for a single realization $l_N(\lambda)$ is
practically non-random and coincides with $l_T(\lambda)$ and $l$, while in
the ballistic region it fluctuates. The data displayed in Fig. \ref%
{Meta-Fig5New} enables one to estimate the difference between the
transmission length $l_T(\lambda)$ (solid line) and the transmission length $%
l_N(\lambda)$ for a single randomly chosen realization (dashed line), and
the scale of the corresponding fluctuations. Both curves are smooth,
coincide in the localized region, and differ noticeably in the ballistic
regime. The separate discrete points in Fig. \ref{Meta-Fig5New} present the
values of the transmission length $l_N(\lambda)$ calculated for different
randomly chosen realizations. It is evident that fluctuations in the
ballistic region become more pronounced with increasing wavelength.


\subsection{Homogeneous Stack}

\label{subsec:hom} 
For an H-stack composed entirely of either normal material or metamaterial
layers, the transmission length obtained within the WSA is

\begin{eqnarray}  \label{transhom}
\frac{1}{l_T} = \frac{1}{l}+\frac{1}{N} \times \ \ \ \ \ \ \ \ \ \ \ \ \ \ \
\ \   \notag \\
\!\!\!\! \mathrm{Re } \left\{ \frac{\langle r\rangle^2}{(1-\langle
t^2\rangle)^2} \left[ 1-\mathrm{exp} \left( -\frac{N}{\bar{l}}-i\frac{N}{%
\bar{l}_{b}} \right) \right] \right\},
\end{eqnarray}
where $\bar{l}$ is crossover length (\ref{cross}) and $\bar{l}_{b}$ is
ballistic crossover length defined by equation

\begin{equation*}  \label{addcross}
\frac{1}{\bar{l}_{b}}=-\mathrm{Im}\ln\langle t^{2}\rangle.
\end{equation*}
H-stack localization length $l$ is

\begin{equation}  \label{lochom}
\frac{1}{l}= -\langle \ln|t| \rangle - \mathrm{Re }\frac{\langle r \rangle^2%
}{1-\langle t^2\rangle}
\end{equation}
where $r,t$ are transmission and reflection coefficients of a $R$ layer (for
$L$ layers, they should be replaced by $r^{*}$ and $t^{*}$ however this does
not change the final result due to real part operation $\mathrm{Re }$).

Here we consider the simplest lossless model ($\sigma=0$) with only
refractive index disorder (i.e., $Q_{d}=0$). In contrast to M-stack case(see
Section~\ref{subsec:mixed} below), where a minimal model manifesting all
common features of the M-stack transmission properties necessarily includes
additional random parameter (in previous Subsection it is layer thickness),
for H-stack it is sufficient to include only one such parameter. As earlier,
we assume uniform distribution of refractive index fluctuations with the
width $2Q_{\nu}$. In this case, the short wave asymptotic of the
localization length coincides with that of M-stack (\ref{shortWave}) and
similarly to the M-stack, transmission through short H-stacks with $%
N\lesssim Q_{\nu}^{-2}$ is always ballistic. So below we consider long
stacks $NQ_{\nu}^{2}\gg 1$.

In the long wave region $\lambda\gg 1,$ the three characteristic lengths
entering Eq. (\ref{transhom}) asymptotically are

\begin{eqnarray}  \label{hlengths}
l=\displaystyle{\frac{3\lambda ^{2}}{2\pi ^{2}Q_{\nu }^{2}}}, \ \ \ \ \ \ \
\bar{l}=\frac{\lambda^2}{8\pi^2Q_{\nu}^2}, \ \ \ \ \ \ \ \ \bar{l}_{b}=\frac{%
\lambda}{4\pi}.
\end{eqnarray}
The main contribution to the long wave and short wave asymptotic of the
localization length is related to the first term in Eq. (\ref{lochom}).
Thus, the localization length of the H-stack in these two limits is well
described by the single scattering approximation. The long wave asymptotic
of the H-stack localization length differs from that of M-stack and
coincides with that of its reciprocal Lyapunov exponent (\ref{Lyapunov-4})
and ballistic length (\ref{bbal}).

We calculated also H-stack Lyapunov exponent. It is described by the same
equation (\ref{Lyapunov-3}) as that for M-stack, thus the reciprocal
Lyapunov exponents for both types of stacks have the same asymptotic form (%
\ref{Lyapunov-4}). This coincidence was established analytically in a wider
spectral region in \cite{IMT-10}.

Long H-stacks with $N\gg Q_{\nu}^{{-2}}$ in the long wave region $\lambda\gg
1$ manifest both ballistic and localized behavior. Transition between these
regimes is governed by two characteristic wavelengths defined by Eq. (\ref%
{wavelengths-1}). Similarly to the M-stack case, they are proportional to $%
Q_{\nu}\sqrt{N},$  differ only by a numerical multiplier and satisfy the
inequality $\lambda _{1}(N)<\lambda _{2}(N)$.

At starting part of long wave region $1\ll \lambda\ll \lambda_{1}(N)$
transmission length $l_T$ coincides with the localization length $l$ and has
asymptotic described by Eq. (\ref{hlengths}). Then after passing transition
region $\lambda_{1}(N)\ll \lambda\ll \lambda_{2}(N)$ the ballistic regime  $%
\lambda\gg \lambda_{2}(N)$ starts. In this regime, transmission length
coincides with ballistic length $b(\lambda)$ described by equation

\begin{eqnarray}  \label{bal-H-1}
\frac{1}{b(\lambda )}&=& \frac {2\pi ^{2}Q_{\nu }^{2}} {3\lambda ^{2}} \left[%
1+\frac{NQ_{\nu }^{2}}{12} \left( \frac{\sin \displaystyle{\frac{%
\lambda_{3}(N)}{2\lambda}}} {\displaystyle{\frac{\lambda_{3}(N)}{2\lambda}}}
\right)^{2}\right],  \notag \\
&&\lambda_{3}(N)=4\pi N,
\end{eqnarray}%
obtained by expansion of the exponent $\exp\left(-N/\bar{l}\right)$ in Eq. (%
\ref{transhom}).

Due to appearance of additional characteristic wavelength $\lambda_{3}(N)$
determined by equation $N=\bar{l}_{b}(\lambda_{3}(N))$ where $\bar{l}_{b}$
is the ballistic crossover length (\ref{hlengths}), ballistic region is
naturally divided onto two subregions. The first of them defined by
inequalities $\lambda_{2}(N)\ll \lambda\ll \lambda_{3}(N)$ is near ballistic
region where ballistic length coincides with localization length

\begin{equation}  \label{bal-H-2}
b_{n}(\lambda )=\displaystyle{\frac{3\lambda ^{2}}{2\pi ^{2}Q_{\nu }^{2}}} .
\end{equation}%
Thus crossover from localized regime to ballistic one is not accompanied by
any change of transmission length. In the ballistic transition region $%
\lambda\sim\lambda_{3}(N)$, the second term in Eq. (\ref{bal-H-1}) becomes
essential leading to oscillations of ballistic length. Finally in the far
ballistic region $\lambda\gg\lambda_{3}(N),$ expansion of the sine in Eq. (%
\ref{bal-H-1}) shows that for long stacks the second term in this equation
dominates and far ballistic length is

\begin{equation}  \label{bal-H-3}
\frac{1}{b_{f}(\lambda )}=\frac{2\pi ^{2}Q_{\nu }^{2}}{3\lambda ^{2}}
\left(1+\frac{NQ_{\nu }^{2}}{12}\right)\approx \frac{N\pi ^{2}Q_{\nu }^{4}}{%
18\lambda ^{2}}.
\end{equation}

The region $\lambda\geq \lambda_{3}(N)$ possesses a simple physical
interpretation. Indeed, in this subregion, the wavelength essentially
exceeds the stack size and so we may consider the stack as a single weakly
scattering uniform layer with an effective dielectric permittivity \cite%
{Asatryan10a}

\begin{equation*}  \label{effective-5b}
\varepsilon_{\mathrm{eff}}= \left (1+\frac{Q_{\nu }^{2}}{3}\right).
\end{equation*}%
Substitution this value to the text-book formula for reflectivity of the
uniform sample leads immediately to the far long wavelength ballistic length
(\ref{bal-H-3}). We note that because of the effective uniformity of the
H-stack in the far ballistic region, the transmission length on a single
realization is a less fluctuating quantity than that in the near ballistic
region, where it fluctuate strongly as over entire ballistic region for
M-stacks.
\begin{figure}[h]
\rotatebox{0}{\scalebox{0.40}
{\includegraphics{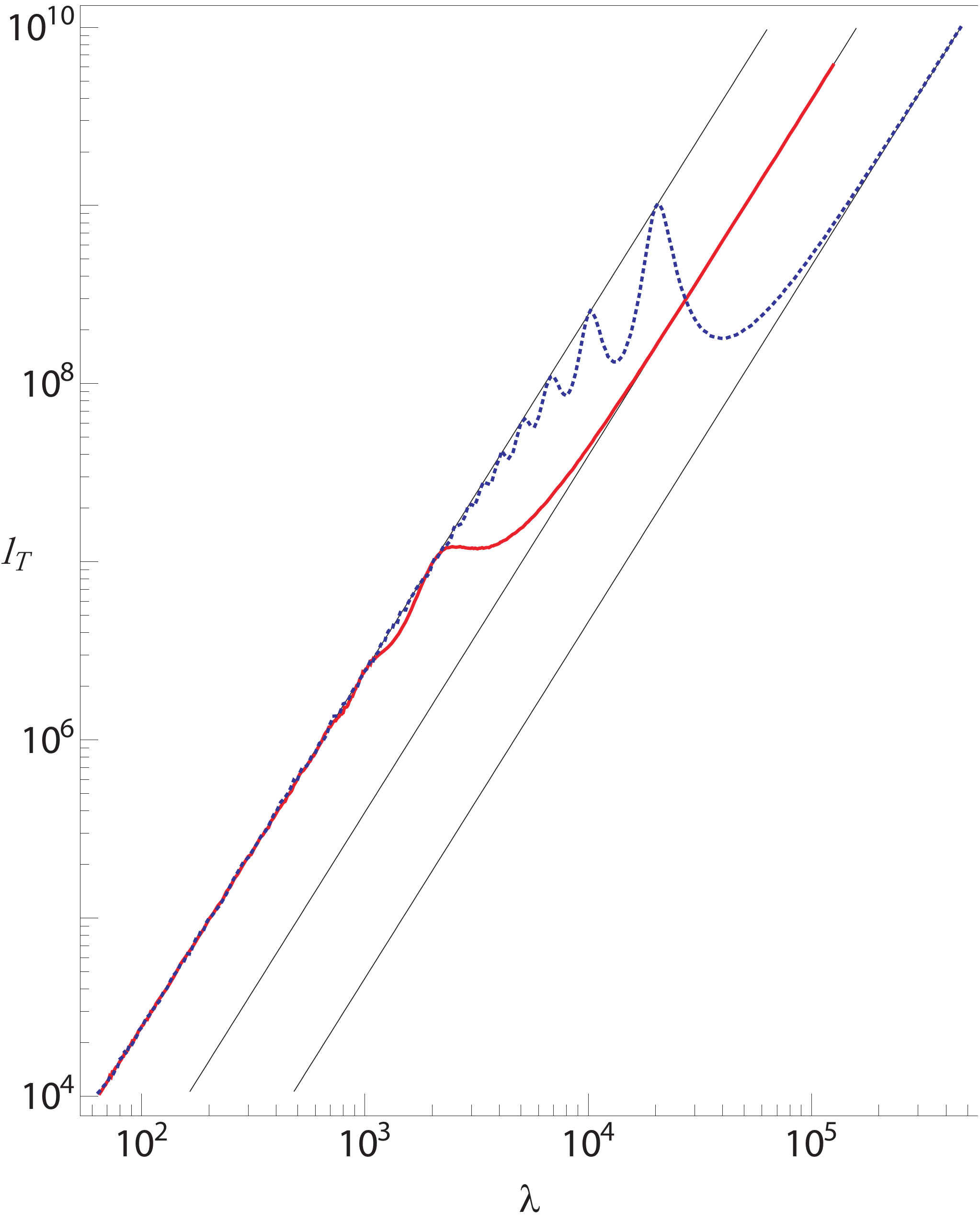}}}
\caption{(Ref. [\onlinecite{Asatryan10a}], color online) 
Transmission length $l_T$ vs $\lambda$ for H-stacks of $N=10^{3}$ 
(solid line), and $10^{4}$ (dotted line) layers (numerical simulation and WSA). Long wave asymptotics for the ballistic length in the near and far ballistic regions are plotted in thin solid lines.}
\label{Meta-Fig5}
\end{figure}

Numerical calculations for H-stack, show an excellent agreement between
direct simulations and calculations based on WSA: corresponding curves can
not be distinguished. Figure \ref{Meta-Fig2} explicitly demonstrates that
transmission length preserves the same analytical form in localized long
wave region and near ballistic region. For the considered stacks with $%
N=10^{3}$, the transmission spectrum features corresponding to transition
between two ballistic subregions can not be manifested. Indeed, the
transition occurs at the wavelength $\lambda\sim 10^{4}$ that is out of
range in this figure.

To study the crossover from near to far ballistic behavior, consider the
transmission lengths of H-stacks with $N=10^{3}$ and $10^{4}$ over the
wavelength range extended up to $\lambda \sim 10^{6}$ plotted in Fig.~\ref%
{Meta-Fig5}. The transition from the localized to the near ballistic regime
occurs without any change in the analytical dependence of transmission
length, however the crossover from the near to the far ballistic regime is
accompanied by a change in the analytical dependence that occurs at $\lambda
=\lambda _{2}(N)$, which for these stacks is of the order of $10^{4}$ and $%
10^{5}$ respectively. The crossover is accompanied by prominent oscillations
described by Eq. (\ref{bal-H-1}). Finally, we note that the vertical
displacement between the moderately long and extremely long wavelength
ballistic asymptotes does not depend on wavelength, but grows with the size
of the stack, according to the law

\begin{equation*}
\ln \frac{b_{n}}{b_{f}}=\ln \frac{NQ_{\nu }^{2}}{12},  \label{shift}
\end{equation*}%
which stems from Eqs. (\ref{bal-H-2}), (\ref{bal-H-3}).

Detailed study of the average reflectivity of the H-stacks with various
lengths at all long wave region\cite{Asatryan10a} also completely confirm
theoretical predictions formulated above.

\begin{figure}[h]
\rotatebox{0}{\scalebox{0.40}{\includegraphics{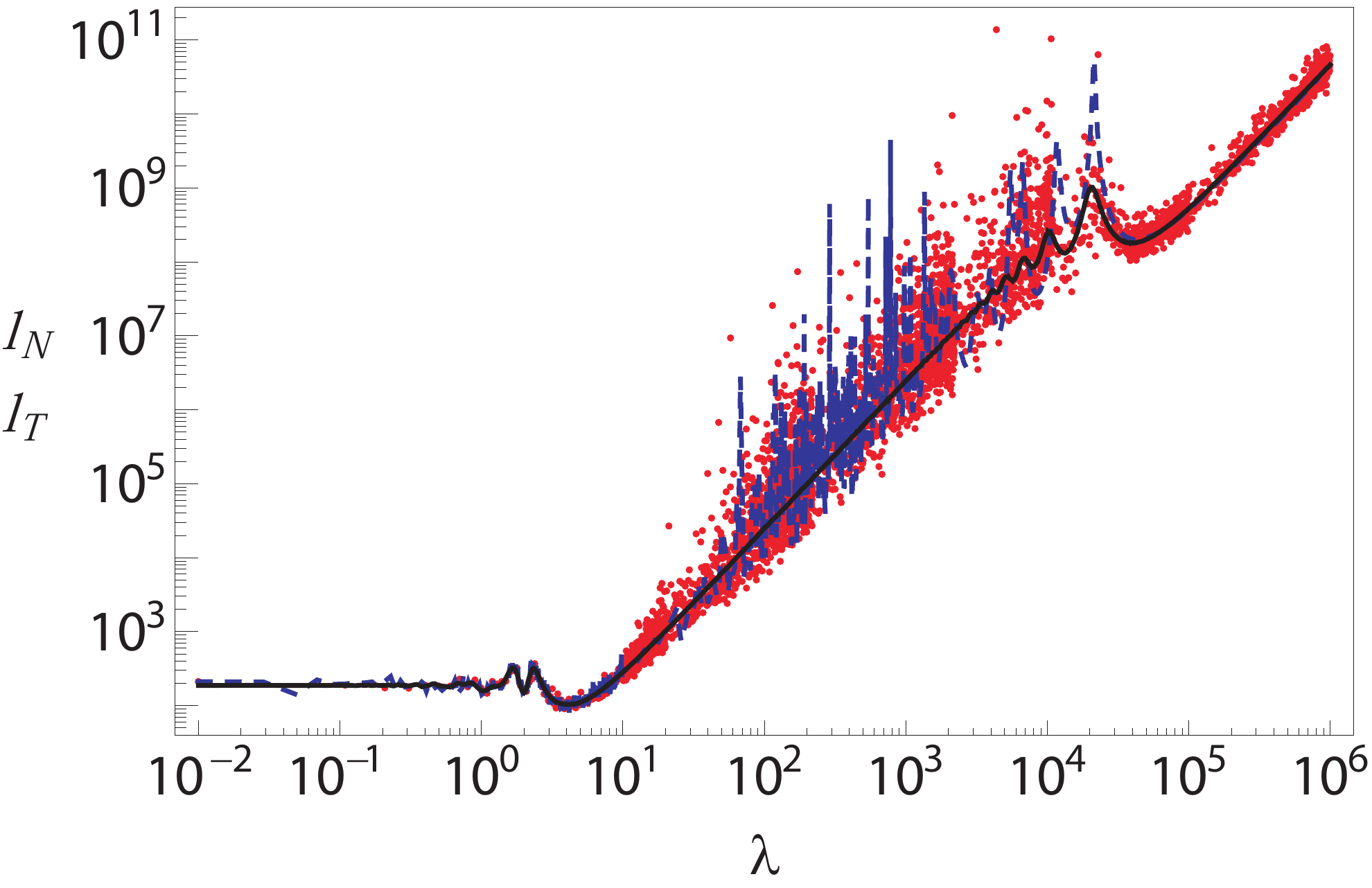}}}
\caption{(Ref.~[\onlinecite{Asatryan10a}], color online) Transmission
lengths $l_T$ (solid black line) and the transmission length for a single
realization $l_N$ (dashed blue line) vs $\protect\lambda$ for a H-stack with
$Q_{\protect\nu}=0.25$, $Q_d=0.2$ and $N=10^4$ layers. Each separate point
corresponds to a particular wavelength with its own realization of a random
stack.}
\label{Meta-Fig8New}
\end{figure}

Consider now statistical properties of the H-stack transmission length on a
given realization $l_N(\lambda)$. For very long stacks $N\to\infty$ this
length becomes practically non-random in both localized region due to
self-averaging of Lyapunov exponent, and far ballistic region because due to
self-averaging nature of the effective dielectric permittivity. For less
long stacks, transmission length $l_T$ fluctuates also even in the far
ballistic region, however for sufficiently long stacks these fluctuations
are essentially suppressed since they must vanish in the limit as $N\to\infty
$. This is demonstrated by Fig. \ref{Meta-Fig8New} where the transmission
length $l_T$ (solid line) and the transmission length $l_N(\lambda)$ for a
single randomly chosen realization (dashed line) are plotted. Like the
M-stack case, the H-stack single realization transmission length in the near
ballistic region is a complicated and irregular function, similar to the
well known ``magneto fingerprints'' of magneto-conductance of a disordered
sample in the weak localization regime \cite{Altshu}. This statement is
supported by displayed in Fig. \ref{Meta-Fig8New} the set of separate
discrete points, each of them presenting transmission length calculated for
a different randomly chosen realization.


\subsection{Transmission Resonances}

\label{subsec:res} 

An important signature of the localization regime is the presence of
transmission resonances (see, for example, Refs. \cite%
{Lifshits,Soven,Bliokh-2}), which appears in sufficiently long, open systems
and which are the ``fingerprints'' of a given realization of disorder. These
resonances manifest themselves as narrow peaks of transmittivity $|T_N|^{2}$
on a given realization as a function of wavelength $\lambda$. Figure~\ref%
{PRL_Fig5}\cite{Asatryan07} presents a single realization of the
transmittance $|T_N|^{2}$ as a function of $\lambda$ for a M-stack (dashed
line) and for the corresponding H-stack of $N=10^{3}$ layers (solid line).
It is evident that the resonance properties exhibited by homogeneous and
mixed media, serve as another (in addition to the behavior of the
localization length) discriminating characteristic of these two media.
Indeed there are no resonances for the M-stack for $\lambda \gtrsim 4$,
while the disordered homogeneous stack exhibits well pronounced resonances
over the entire spectrum.

\begin{figure}[h]
{\scalebox{0.3}{\includegraphics{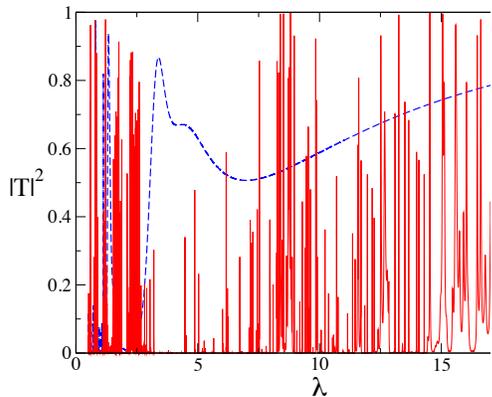}}}
\caption{(Ref.~[\onlinecite{Asatryan07}], color online) Transmittance $|T|^2$ vs $\protect%
\lambda$ for a single realization ($Q=0.25$, $N=10^3$). Solid: normal
H-stack, dotted: M-stack. }
\label{PRL_Fig5}
\end{figure}

Note that the dotted curve in Fig.~\ref{PRL_Fig5} describes resonance
properties of periodic $Q_{d}=0$ comparatively short M-stack with only
refractive index disorder (RID). Important feature of such a stack is the
lack of phase accumulation over its total length: in the particular
realization of Fig.~\ref{PRL_Fig5}, the accumulated phase of the wave in the
mixed stack never exceeds $\pi/2$. Therefore to subdue such a suppression of
the phase accumulation, one need or to enlarge essentially the stack size,
or to switch on an additional (thickness $Q_{d}$ or magnetic permittivity $%
\mu$) disorder.

\begin{figure}[h]
\rotatebox{0}{\scalebox{0.40}{\includegraphics{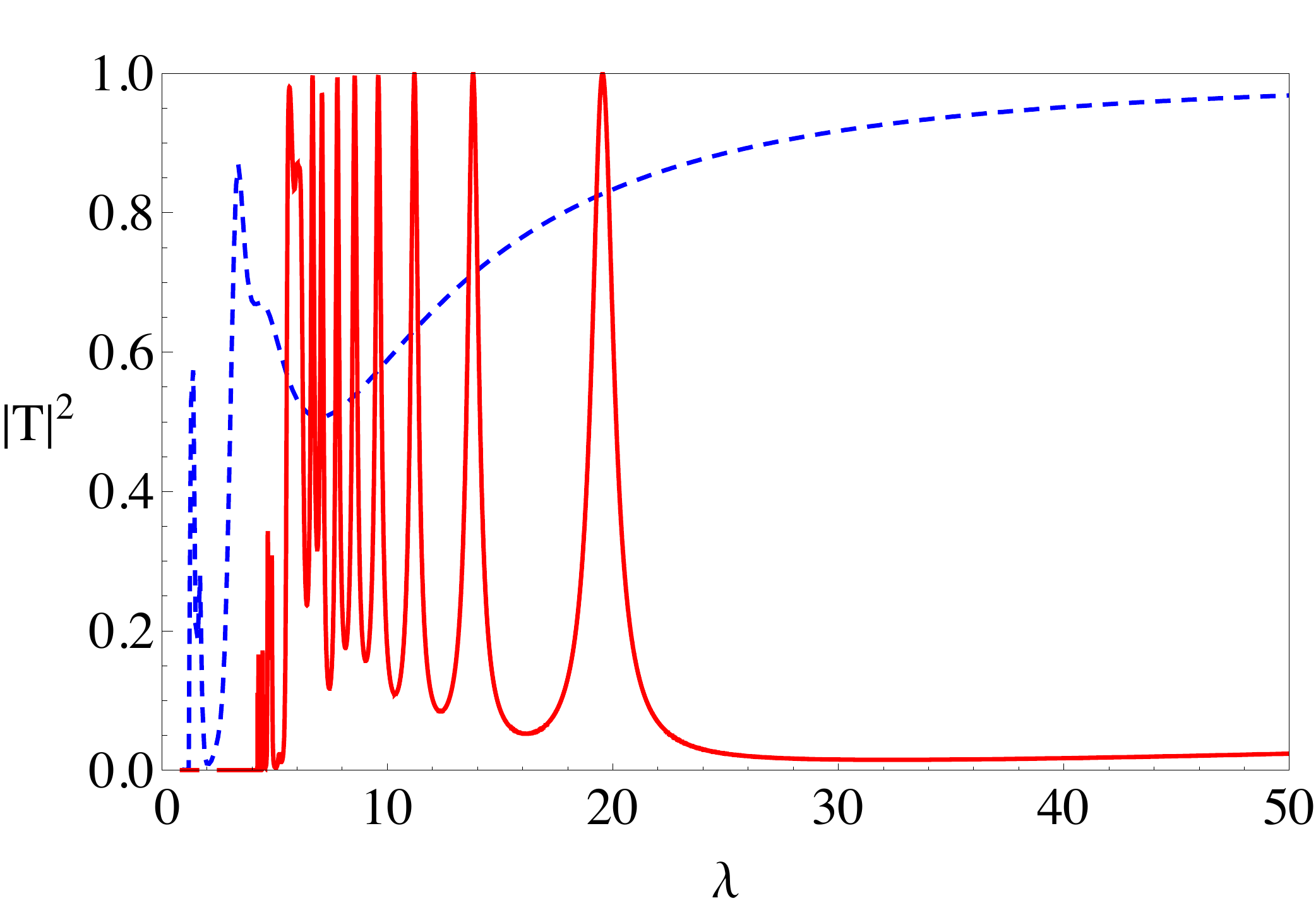}}}
\caption{ (Ref.~[\onlinecite{Asatryan10a}], color online) Single realization
transmittance $|T|^{2}$ vs wavelength $\protect\lambda$ for RID M-stacks
with $Q_\protect\nu =0.25$ and $Q_{d}=0$ for $N=10^{5}$ layers (solid line)
and $N=10^{3}$ layers (dotted line).}
\label{Meta-Fig12}
\end{figure}

The first possibility is demonstrated in Fig~\ref{Meta-Fig12} where
transmittance spectra $|T|^{2}(\lambda)$ for a realization, of two different
M-stack with two lengths $N=10^{3}$ and $N=10^{5}$ and only refractive index
disorder is displayed. It is readily seen that while the resonances in the
shorter stack (dashed line) at $\lambda \geq 5$ do not exist at all, they do
appear in the same region for the longer sample (solid line).

\begin{figure}[h]
\rotatebox{0}{\scalebox{0.40}{\includegraphics{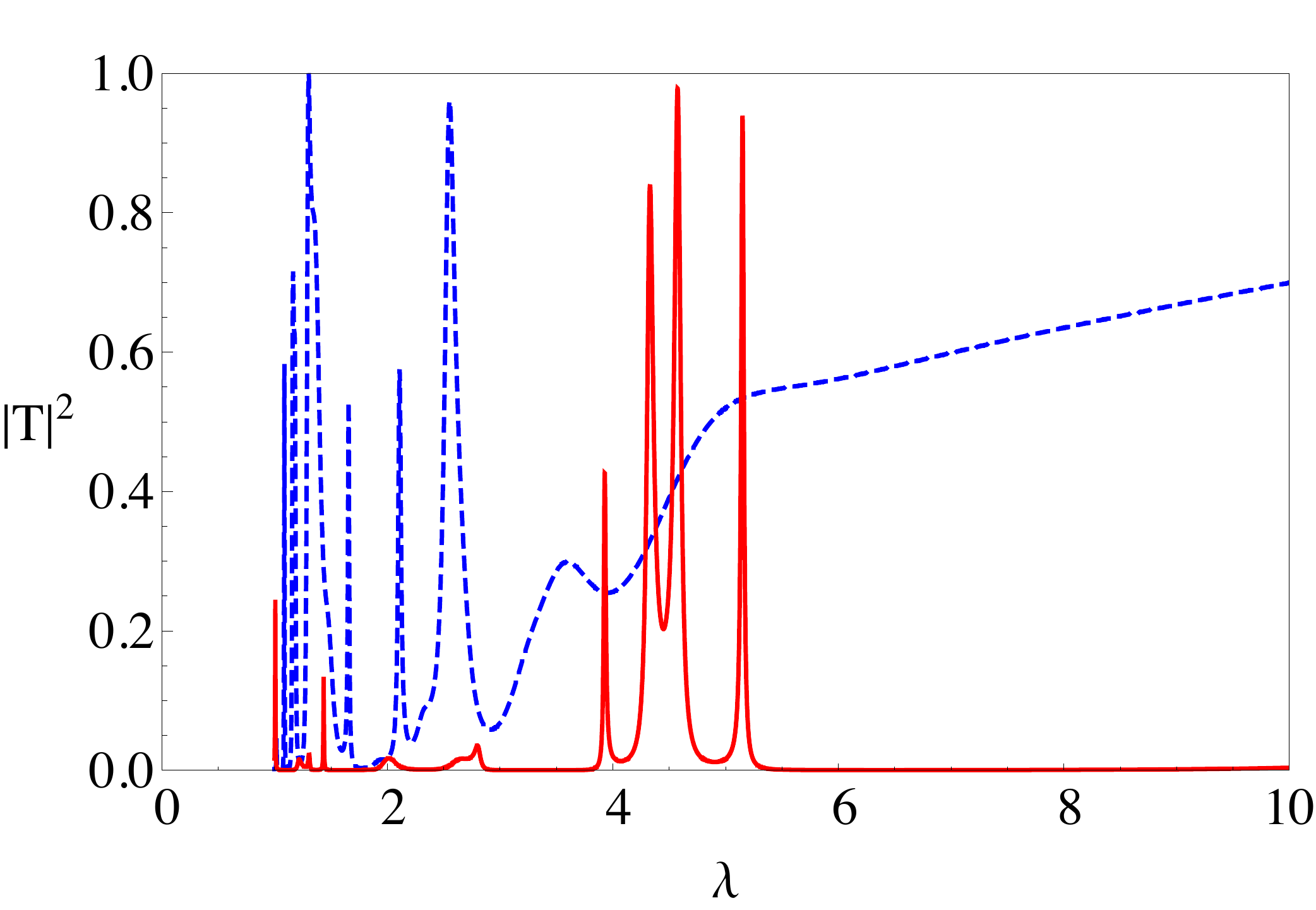}}}
\caption{ (Ref.~[\onlinecite{Asatryan10a}], color online) Single realization
transmittance $|T|^2$ vs $\protect\lambda$ for M-stack of $N=10^3$ layers
with $Q_\protect\nu=0.25$. Solid line corresponds to an M-stack with $Q_d=0.2
$, and the dashed line to M-stack with no thickness disorder, \emph{i.e.}, $%
Q_d=0.0$.}
\label{Meta-Fig13}
\end{figure}

The second way to generate transmission resonances is to introduce
additional disorder. This is confirmed by the transmittance spectra for a
realization, of two M-stacks of the same size $N=10^{3}$ with only
refractive index disorder (dashed line), and both (thickness and refractive
index) types of disorder (solid line), plotted in Fig.~\ref{Meta-Fig13}. It
is clear that while the RID M-stack with this length, is too short to
exhibit transmission resonances at $\lambda >3$, resonances do emerge at
longer wavelengths for the M-stack with thickness disorder.

Transmission resonances are responsible for the difference between two
quantities that characterize the transmission, namely transmittance
logarithm $\langle \ln |T|^{2}\rangle $ and logarithm of average
transmittance $\ln \langle |T|^{2}\rangle $. The former reflects the
properties of a typical realization, while the latter value is often very
sensitive to the existence of almost transparent realizations associated
with the transmission resonances. Moreover, in some cases namely small
number of such realizations contribute mainly to the average transmittance.

Thus the ratio of the two quantities mentioned above

\begin{equation*}
s=\frac{\langle \ln |T|^{2}\rangle }{\ln \langle |T|^{2}\rangle }.  \notag
\label{ratio}
\end{equation*}
is a natural characteristic of the transmission resonances. In the absence
of resonances, this value is close to unity, while in the localization
regime $s>1$. In particular, this ratio takes the value $4$ in the
high-energy part of the spectrum of a disordered system with Gaussian
white-noise potential\cite{LGP}.

\begin{figure}[h]
\rotatebox{0}{\scalebox{0.40}{\includegraphics{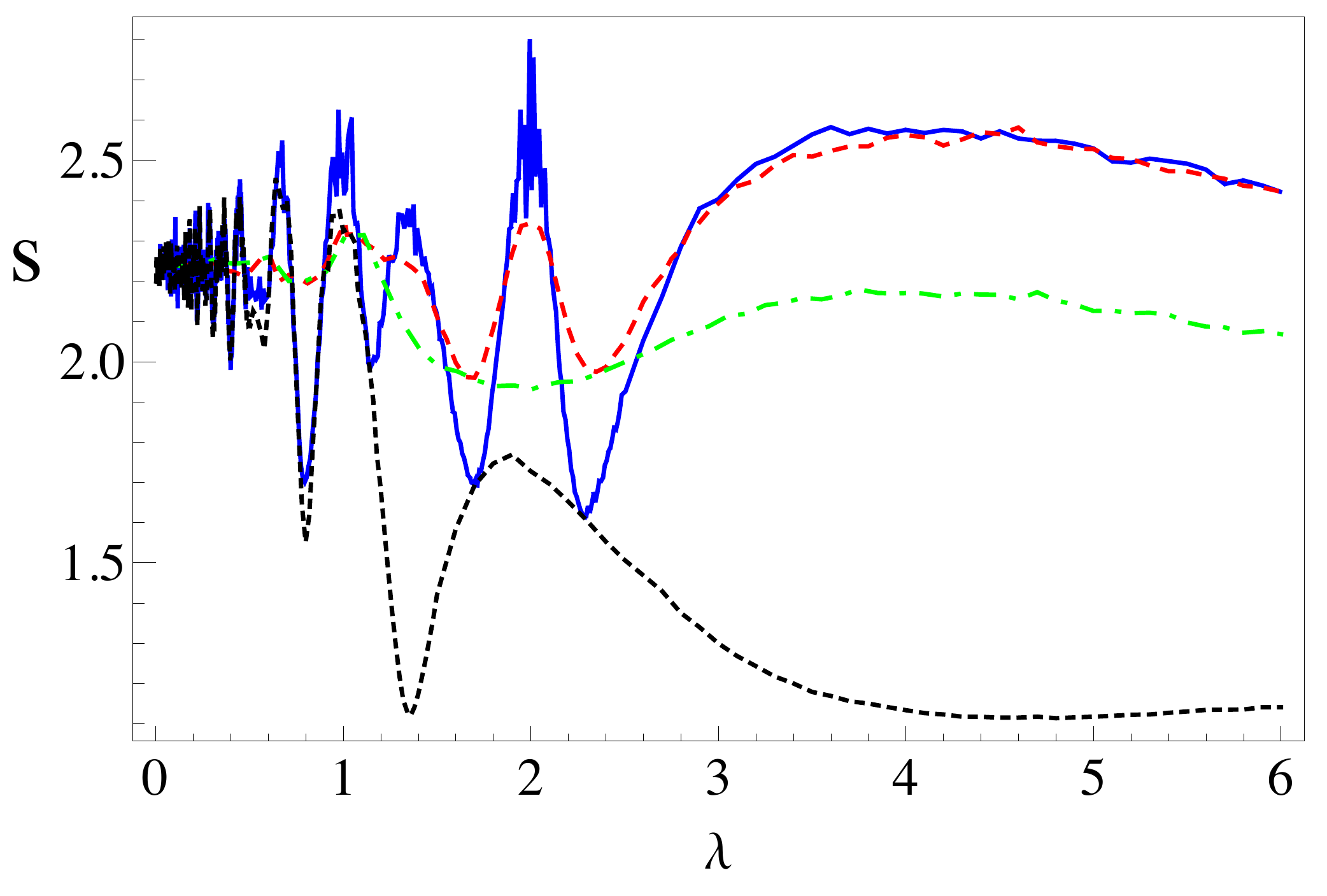}}}
\caption{(Ref.~[\onlinecite{Asatryan10a}], color online) Ratio $s(\protect%
\lambda)$ vs. wavelength $\protect\lambda$ for $Q_{\protect\nu}=0.25$ and
the stack length $N=10^3$. Solid and dashed curves are for the RID H-stack
and H-stack with $Q_d =0.2$, respectively. The middle dashed-dotted curve is
for an M-stack with $Q_d =0.25$, and the bottom dotted line is for a RID
M-stack.}
\label{Meta-Fig11}
\end{figure}

Consider the ratio $s(\lambda )$ as a function of the wavelength for RID M-
and H-stacks and for the corresponding stacks with thickness disorder
plotted in Fig.~\ref{Meta-Fig11}. In all cases, the stack length is $%
N=10^{3}.$ It is evident that for the RID M-stack $s(\lambda )\approx 1$,
\emph{i.e.}. the length of this M-stack is too short for the localization
regime to be realized. In other three cases, however, $s(\lambda )\gtrsim 2$%
, which means that the localization takes place even in such comparatively
short stack.

\subsection{Polarization Effects}

\label{subsec:POL}

The results obtained above for normal incidence can be easily generalized to
the case of oblique incidence. Here all characteristic lengths and
wavelengths depend on angle of incidence $\theta$ and \textit{s}- and
\textit{p}-polarizations should be considered separately. Qualitatively new
features appear: essential enlightening in the vicinity of Brewster angle
and appearance of supercritical regime induced by total internal reflection.
We describe these new properties within the frameworks of the model defined
in the previous Subsection \ref{subsec:model}.

General expressions for transmission length for both M-stacks (Eqs. (\ref%
{FinalG3}) - (\ref{cross})) and H-stacks (Eqs. (\ref{transhom}) - (\ref%
{lochom})) as well as general expressions (\ref{t}) for transmission and
reflection coefficients of a single layer remain the same as in the case of
normal incidence. However explicit expressions for the parameters entering
these coefficients are changed. Fresnel interface reflection coefficient is
now given by

\begin{eqnarray}  \label{ro}
\rho = \frac{\mathcal{Z}\cos\theta_{\nu}-\cos\theta} {\mathcal{Z}%
\cos\theta_{\nu}+\cos\theta}, \\
\mathcal{Z}=\left\{
\begin{array}{ccc}
Z^{-1}, &  & s-\text{polarization} \\
&  &  \\
Z, &  & p-\text{polarization}.\nonumber%
\end{array}
\right.
\end{eqnarray}
Here characteristic angle $\theta_{\nu}$ and the layer impedance $Z$
relative to the background (free space) according Eq. \ref{z-1-imp} are

\begin{eqnarray*}  \label{zlab}
\cos\theta_{\nu}=\sqrt{1-\frac{\sin^{2}\theta}{\nu^{2}}}, \ \ \ \ Z= \sqrt{%
\frac{\mu}{\varepsilon}}= \frac{1}{1+\delta_{\nu}}.
\end{eqnarray*}
Then the phase shift $\beta$ is now

\begin{equation}  \label{phase}
\beta=k d \nu \cos\theta_{\nu}, \ \ \ \ k=2\pi/\lambda.
\end{equation}
Characteristic angle conserves its direct geometrical meaning for incidence
angle $\theta\leq\theta_{c}$ ( subcritical incidence angle) where critical
angle is

\begin{equation*}  \label{theta-c}
\theta_{c}= \sin^{-1}(1-Q_{\nu}).
\end{equation*}
For the supercritical incidence angle $\theta\geq\theta_{c}$, the values of $%
\theta_{\nu}$ are complex.

Below we mention only final asymptotical expressions for some characteristic
lengths of the problem in the typical cases. We take into account both two
types of disorder however in all final results we keep only the leading
terms and omit the higher order corrections with respect to the refractive
index and thickness fluctuations $Q_{\nu,d}$.\newline

In the short wave limit, localization length is the same for M- and
H-stacks. In the subcritical region of incidence angles it is

\begin{equation*}  \label{shLasEa}
\frac{1}{l}\approx \frac{Q_{\nu}^2}{12\cos^4\theta}\left\{
\begin{array}{ccc}
1 &  & s-\text{polarization}, \\
&  &  \\
\cos^{2} 2\theta &  & p-\text{polarization}.%
\end{array}%
\right.
\end{equation*}
Note that for \emph{p}-polarization, this expression acquires angle
dependent multiplier that vanishes at the Brewster angle $\theta=\pi/4$.
Accounting for the next term we obtain the localization length at the
Brewster angle

\begin{equation*}  \label{c}
{l}=\frac{45}{4Q^4_{\nu}},
\end{equation*}
which is $Q_{\nu}^{-2}$ times larger than that far from the Brewster angle
and than that for $s-$polarization in the same shortwave limit.

At the incidence angle $\theta>\theta_{c}$, total internal reflection occurs
and the WSA fails. If the supercriticality $\theta-\theta_{c}$ is not
extremely small, then the exponent $2i\beta$ in Eq. (\ref{t}) is real and
negative and thus the magnitude of the single layer transmission coefficient
is exponentially small. This results in the attenuation length for both
polarizations

\begin{eqnarray*}  \label{ShAs2}
\frac{1}{l_{att}} = \mathrm{Im }\langle \beta \rangle =k \ \mathrm{Im}
\langle d \sqrt{\sin^{2}\theta-\nu^{2}}\rangle=  \notag \\
\!\!\!\!\frac{k\sin^{2}\theta}{8Q_{\nu}} (\pi-2\theta_{0}- \sin
2\theta_{0}), \ \ \ \sin\theta_{0}=\frac{\sin\theta_{c}}{\sin\theta}.
\end{eqnarray*}
Due to $\propto k$ dependence, in the short wave limit $l_{att}\to \infty$
and transmission length in supercritical region of the angles of incidence
coincides with the attenuation length. However for the same reason at long
waves attenuation contribution can be neglected and the main contribution to
the transmission length is due to Anderson localization.

In the long wave region, H- and M-stacks demonstrate different behavior and
we describe them separately.\newline

\noindent\underline {a) Homogeneous stacks.}\newline
For \textit{s}-polarization, long wave asymptotic of the transmission length
is similar to that for normal incidence (\ref{bal-H-1})

\begin{eqnarray*}  \label{ltrE}
\frac{1}{l_T}=\frac {2\pi^2Q_\nu^2} {3\lambda^{2}\cos^2\theta}\times\ \ \ \
\ \ \ \ \   \notag \\
\left [1+ \frac {NQ_\nu^2} {12}\left( \frac {\sin \displaystyle{\frac {2\pi
N\cos\theta}{\lambda} }} {\displaystyle{\frac{2\pi N\cos\theta}{\lambda}}}
\right)^2 \right].
\end{eqnarray*}
This expression describes localized region as well as both ballistic
subregions.

In the case of $p-$polarized wave, the localization length is given by

\begin{eqnarray*}  \label{LHN2}
\frac{1}{l} = \frac {2\pi^2Q_\nu^2\cos^{2}2\theta} {3\lambda^{2}\cos^2\theta}%
+ \frac{\pi^2Q_{\nu}^4}{6\cos^4\theta}\times  \notag \\
\left( 1-\frac{19}{6}\cos2\theta+\frac{7}{15}\cos4\theta + \frac{19}{30}%
\cos6\theta \right).
\end{eqnarray*}
At Brewster angle $\theta=\displaystyle{\frac{\pi}{4}}$ the first term
vanishes and transmission length is

\begin{equation}  \label{tpm4}
\frac{1}{l}=\frac{16\pi^2Q^4_{\nu}}{45\lambda^{2}}.
\end{equation}

\noindent\underline {b) Mixed stacks.}\newline
Reciprocal transmission length for $s$-polarized wave is

\begin{eqnarray}  \label{mixE}
\frac{1}{l_T} & = & \frac{k^2Q^2_{\nu}}{3\cos^2\theta} \left (\frac{1}{2}-%
\frac{1-f(N\alpha_{s})}{3+\zeta\cos^4\theta} \right),  \notag \\
\alpha_{s}&=&\frac {k^2Q^2_{\nu}} {3\cos^2\theta} (3+\zeta\cos^4\theta),
\end{eqnarray}
where the function $f$ and parameter $\zeta$ are defined in Eqs. (\ref{f})
and (\ref{zeta}) correspondingly. Equation (\ref{mixE}) describes the
transition from localization to ballistic propagation at long wavelengths.
In the limit $N\to\infty$ transmission length tends to localization length

\begin{eqnarray*}  \label{mixEL}
l = \frac{3\lambda^{2}\cos^2\theta} {2\pi^2Q^2_{\nu}} \ \frac{%
3+\zeta\cos^4\theta}{1+\zeta\cos^4\theta},
\end{eqnarray*}
while the opposite extreme, i.e., as $N \to 0$, gives the ballistic length

\begin{eqnarray*}  \label{mixEB}
b = \frac{3\lambda^{2}\cos^2\theta}{2\pi^2Q^2_{\nu}},
\end{eqnarray*}
which coincides with that for a H-stack in \emph{s}-polarization.\newline

\begin{figure}[h]
\center{
\rotatebox{0}{\scalebox{0.4}{\includegraphics{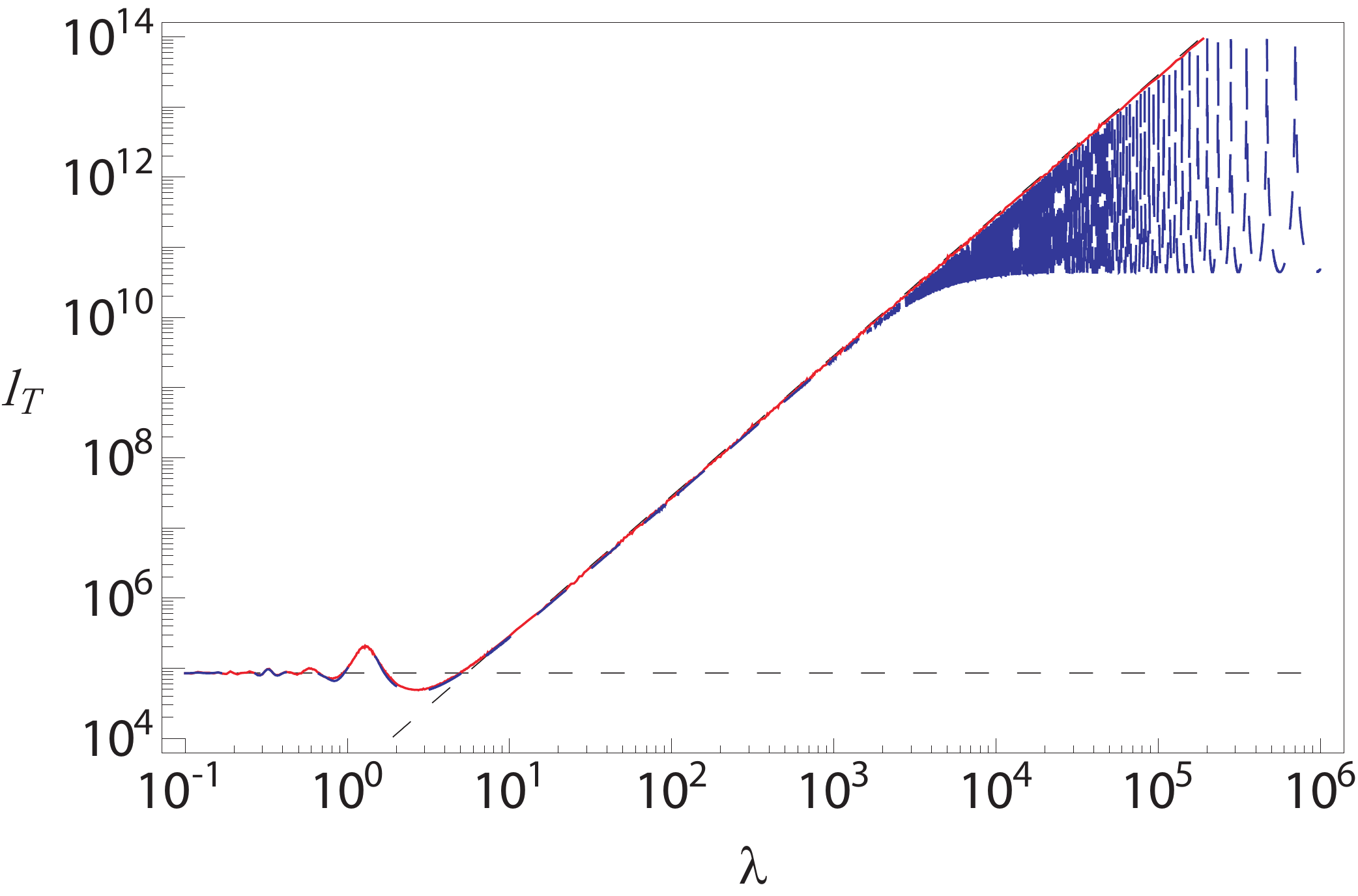}}}}
\caption{(Ref.~[\onlinecite{Asatryan10b}], color online) Transmission length
$l_T$ versus $\protect\lambda$ for a M- stack in $p$-polarized light with $%
Q_{\protect\nu}=0.1$, $Q_d=0.2$ and $N=10^{6}$, at the Brewster angle $%
\protect\theta=45^{0}$ red solid line. The blue dashed line shows results
for \emph{s}-polarization and a H-stack, re-plotted for comparison. }
\label{POL-Fig5}
\end{figure}

For \textit{p}-polarized waves incident at angles away from the Brewster
angle, the transmission length is given by:

\begin{eqnarray}  \label{lmp}
\frac{1}{l_T} = \frac {k^2Q^2_{\nu}\cos^{2} 2\theta} {3\cos^2\theta}\times \
\ \ \ \ \   \notag \\
\left (\frac{1}{2}-\frac{1-f(N\alpha_{p})} {2+\cos^{2}
2\theta+\zeta\cos^4\theta} \right), \\
\ \ \ \ \ \ \alpha_{p}=\frac {k^2Q^2_{\nu}} {3\cos^2\theta} (2+\cos^{2}
2\theta+\zeta\cos^4\theta).  \notag
\end{eqnarray}
The localization length is deduced from Eq. (\ref{lmp}) by taking the limit
as $N\rightarrow \infty$

\begin{eqnarray*}
l = \frac{3\lambda^{2}\cos^2\theta}{2 \pi^2Q^2_{\nu}\cos^2 2\theta} \frac{%
2+\cos^2 2\theta+\zeta\cos^4\theta} {\cos^2 2\theta+\zeta\cos^4\theta}.
\label{lmpL}
\end{eqnarray*}
Correspondingly, the ballistic length is obtained by calculating the limit
as $N\rightarrow 0$

\begin{eqnarray*}
{b} = \frac{3\lambda^{2}\cos^2\theta}{2 \pi^2Q^2_{\nu}\cos^2 2\theta}.
\label{lmpb}
\end{eqnarray*}

At the Brewster angle $\theta=\pi/4$, accounting for the higher order
corrections to r.h.s. of Eq. (\ref{lmp}) we obtain the transmission length
the same result (\ref{tpm4}) that for H-stack.

All analytical predictions are confirmed by numerical calculations. As in
the case of normal incidence theoretical curves based on WSA approximations
mostly can not be distinguished from those obtained by direct simulations.
The results obtained mostly similar to those of normal incidence. Therefore
here we mention only some of them which differ from presented above.\newline

\begin{figure}[h]
\center{
\rotatebox{0}{\scalebox{0.4}{\includegraphics{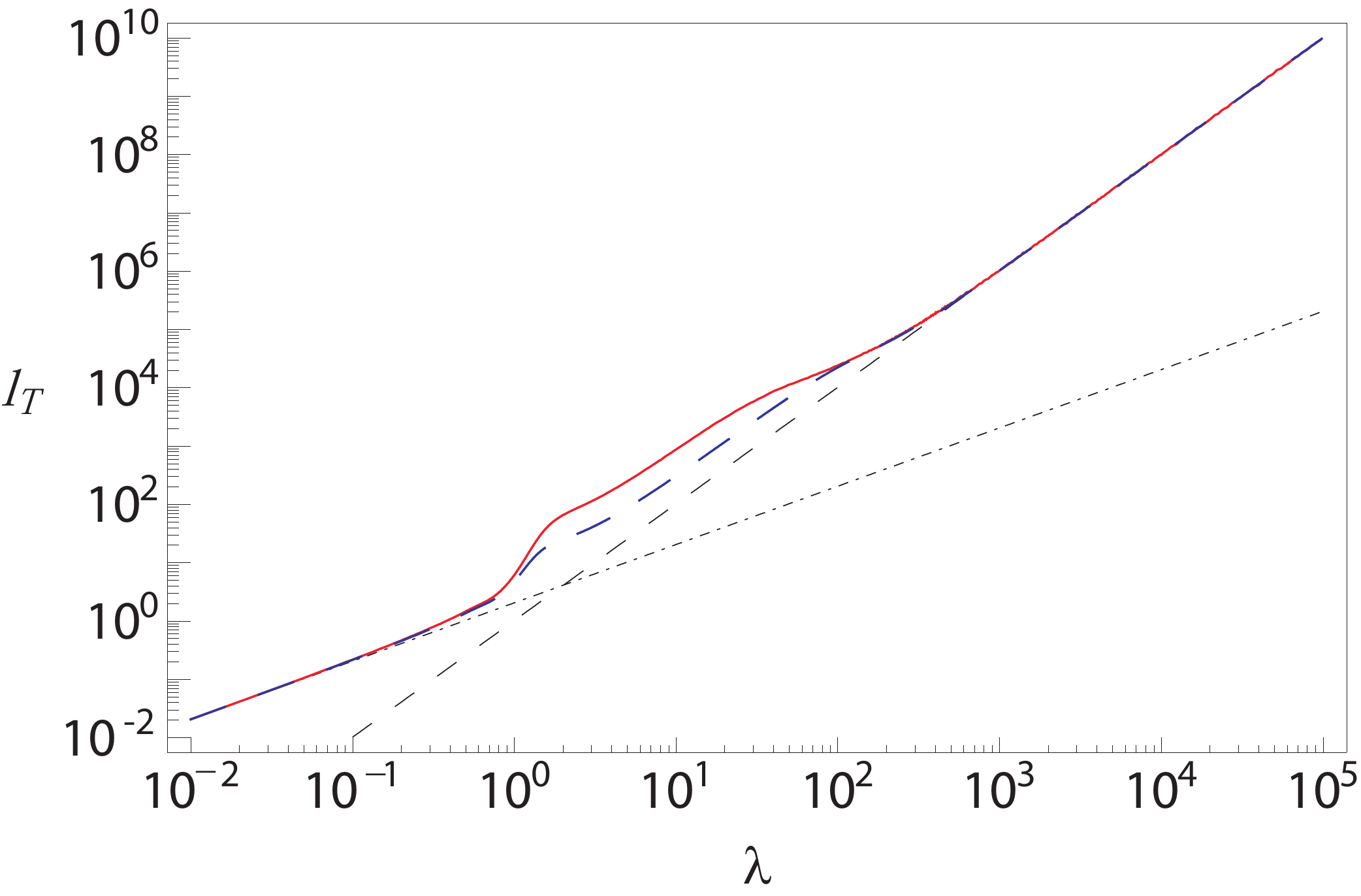}}}}
\caption{(Ref.~[\onlinecite{Asatryan10b}], color online) Transmission length
$l_T$ versus $\protect\lambda$ for a M-stack in $s$-polarized light with $Q_{%
\protect\nu}=0.1$, $Q_d=0.2$ and $N=10^{4}$, and for the supercritical
incidence angle $\protect\theta=75^{\circ}$. Red solid curve: numerical
simulations; Blue dash curve: analytic form. }
\label{POL-Fig8}
\end{figure}

In Fig. \ref{POL-Fig5} the transmission length spectrum of an M-stack of
length $N=10^6$ in \emph{p}-polarized light with other parameters $%
Q_{\nu}=0.1$, $Q_{d}=0.2$, $N_{r}=10^{4}$, and the incidence angle $%
\theta=\pi/4$ is displayed. The chosen angle of incidence is less than the
critical angle $\theta_{c}=\arcsin 0.9=64.16^{\circ}$ and coincides with the
Brewster angle for the single layer with mean refractive index $\nu=\pm 1$.
The results of the numerical simulation and the WSA analytical forms
coincide and are displayed by a single red solid line. Localization occurs
for $\lambda\leq\lambda_1\approx 19$, while the transition from localization
to ballistic propagation occurs at $\lambda\sim\lambda_1$. In contrast to
the case of \textit{s}-polarization, this transition is not accompanied by a
change of scale and is given by the same wavelength dependence. Transition
from near to far ballistic length is accompanied by oscillations of
transmission length which are much more pronounced in comparison to the case
of normal incidence.\newline

Consider now a supercritical case where the angle of incidence $%
\theta=75^{\circ}$ exceeds the critical angle. In Fig. \ref{POL-Fig8} we
present the transmission length spectrum for \emph{s}-polarized light is
presented. The results of both the exact numerical calculation (red solid
line) and the analytic form (long dashed blue curve) are displayed. The
short wave (dashed dotted line)  and the long wave (black dashed line)
asymptotic of the transmission length, respectively coincide with the
numerical results for $\lambda\leq 1$ and $200\leq\lambda$. In the
intermediate region $1\leq\lambda\leq 200$, however, the theoretical
description underestimates the actual transmission length since the WSA is
no longer valid for the chosen, supercritical angle of incidence. For \emph{p%
}-polarization, the results are qualitatively the same, but with the
discrepancy at the intermediate wavelengths even more pronounced.

\begin{figure}[h]
\center{
\rotatebox{0}{\scalebox{0.45}{\includegraphics{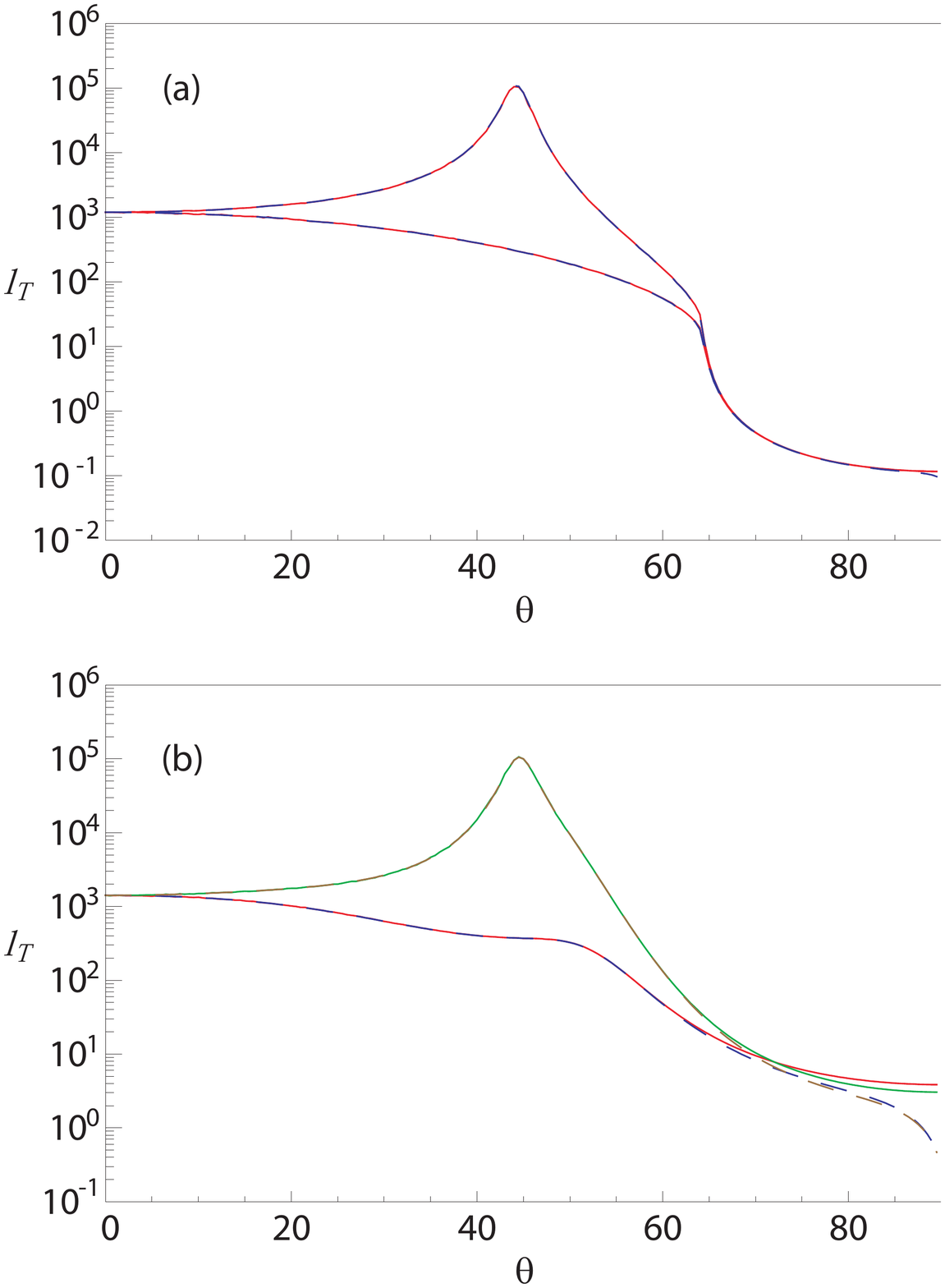}}}}
\caption{(Ref.~[\onlinecite{Asatryan10b}], color online) Transmission length
$l_T$ versus incidence angle $\protect\theta$ for a mixed stack with $Q_{%
\protect\nu}=0.1$, $Q_d=0.2$, for (a) $\protect\lambda=0.1$ (upper panel),
and (b) $\protect\lambda=1$ (lower panel). The top and bottom curves are
respectively for \emph{p-} and \emph{s}-polarizations. }
\label{POL-Fig22}
\end{figure}

We consider also the angular dependence of the transmission length for mixed
stacks. In Fig.~\ref{POL-Fig22} the transmission length $l_T$ as a function
of the angle $\theta$ for a stack of length $N=10^{6}$ at the two
wavelengths $\lambda=0.1$ and $\lambda=1$ is displayed. In either case, the
calculated transmission length does not exceed the stack length and so, for
subcritical angles, our calculations display the true localization length.
For the shorter wavelength $\lambda=0.1$, the form of the transmission
length for both polarizations is similar to that observed for homogeneous
stacks.

Fig. \ref{POL-Fig22}(b) displays results for an intermediate wavelength $%
\lambda=1$ with the lower solid red and blue dashed curves respectively
displaying the results of numerical simulations and analytical predictions
for \emph{s}-polarization, (bottom curves), while the upper solid green and
brown dashed curves display simulations and analytical predictions for \emph{%
p}-polarization. The agreement between simulations and the theoretical form
is again excellent for angles of incidence less then the critical angle, $%
\theta<\theta_{c}$, while for angles greater then the critical angle, the
discrepancies that are evident are again explicable by the breaking down of
the WSA at extreme angles of incidence.

\subsection{Dispersive Metamaterials}

\label{subsec:disp} 
Real metamaterials always are dispersive materials. Here we consider a
dispersive model of the stack composed of metalayers with the same thickness
$d$ and random dielectric permittivity and the magnetic permeability
described by Lorentz oscillator model

\begin{eqnarray}
\varepsilon (f) &=&1-\frac{f_{ep}^{2}-f_{e}^{2}}{f^{2}-f_{e}^{2}+i\gamma f},
\label{eps-m} \\
\mu (f) &=&1-\frac{f_{mp}^{2}-f_{m}^{2}}{f^{2}-f_{m}^{2}+i\gamma f}.
\label{mu-m}
\end{eqnarray}%
Here $f$ is circular frequency, $f_{m}$ and $f_{e}$ are the resonance
frequencies and $\gamma $ is the phenomenological absorption parameter. In
this model, disorder enters the problem through random resonance frequencies
so that

\begin{equation*}
f_{e}=\bar{f}_{e}(1+\delta _{e}),\ \ \ \ f_{m}=\bar{f}_{m}(1+\delta _{m}),
\label{resonances}
\end{equation*}%
where $\bar{f}_{e,m}=\langle f_{e,m}\rangle $ are the mean resonance
frequencies (with the angle brackets denoting ensemble averaging) and $%
\delta _{e,m}$ are independent random values distributed uniformly in the
ranges $[-Q_{e,m},Q_{e,m}]$. The characteristic frequencies $f_{mp}$ and $%
f_{ep}$ are non-random. Therefore, in lossless media ($\gamma =0$), both the
magnetic permeability and the dielectric permittivity vanish with their mean
values, $\bar{\varepsilon}(f)=\langle \varepsilon (f)\rangle $ and $\bar{\mu}%
(f)=\langle {\mu }(f)\rangle ,$ at frequencies $f=f_{ep}$ and $f=f_{mp}$
respectively, i.e.,

\begin{equation*}  \label{characteristic}
\mu (f_{mp})=\bar{\mu}(f_{mp})=0,\ \ \ \ \varepsilon (f_{ep})=\bar{%
\varepsilon}(f_{ep})=0.
\end{equation*}

Following Ref. \cite{Shelby,Smith2010}, in our numerical calculations we
choose the layer thickness $d=0.003$m and the values of characteristic
frequencies $f_{mp}=10.95\text{GHz}$, $f_{m0}=\bar{f_{m}}=10.05\text{GHz}$, $%
f_{ep}=12.8\text{GHz}$, $f_{e0}=\bar{f_{e}}=10.3\text{GHz,}$ and $\gamma =10%
\text{MHz,}$ which fit the experimental data given in Ref. \cite{Shelby}.
That is, we are using a model based on experimentally measured values for
the metamaterial parameters. Then we choose the maximal widths of the
distributions of the random parameters $\delta _{e,m}$ as $Q_{e,m}= 5\times
10^{-3}$ corresponding to weak disorder.

\begin{figure}[tbh]
\centerline{\rotatebox{0}{\includegraphics[width=8.6cm]{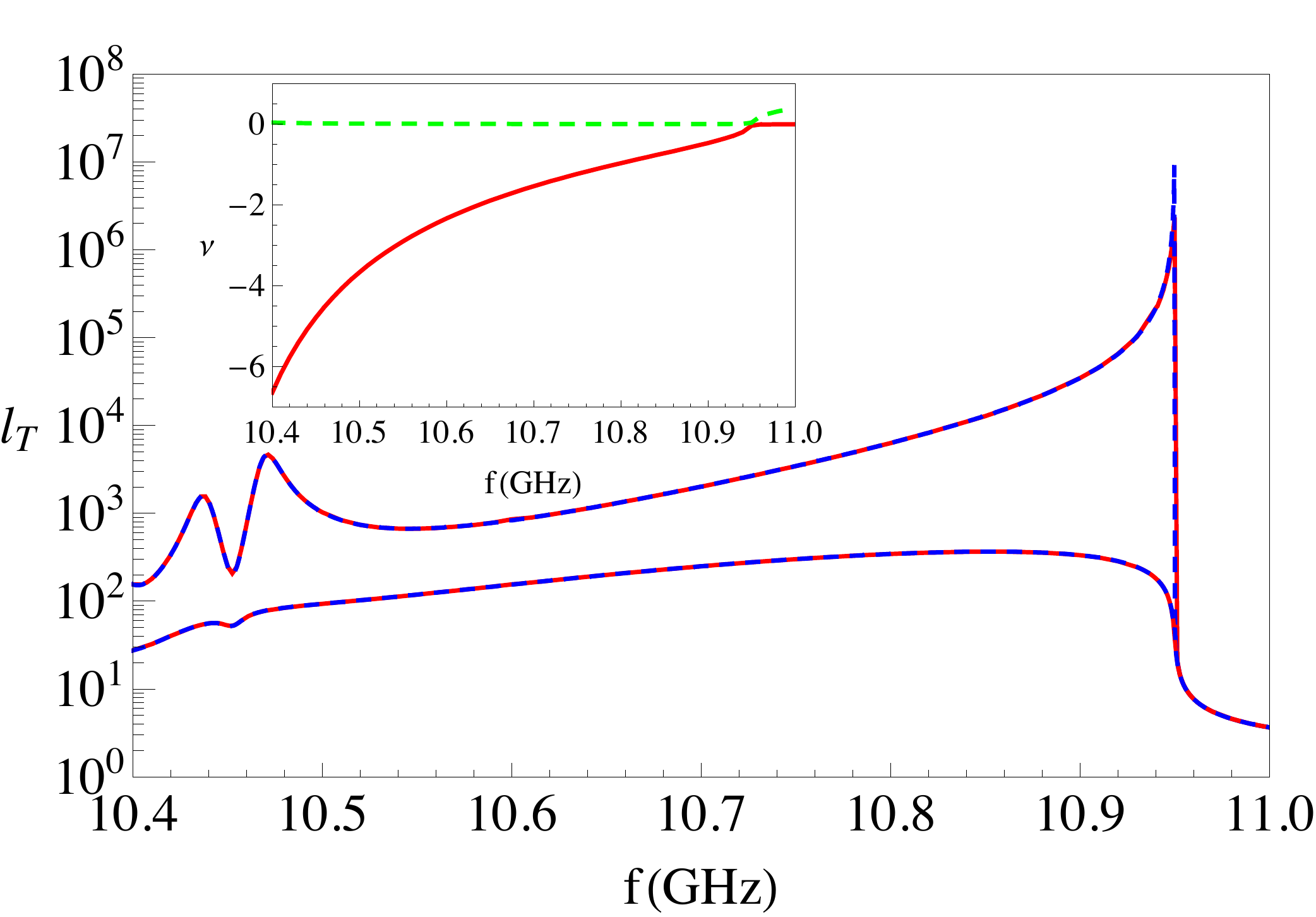}}}
\caption{(Ref.~[\onlinecite{Asatryan12}], color online) Transmission length $l_T$ vs
frequency $f$ at normal incidence ($\protect\theta _{a}=0$) for a
metamaterial stack without absorption (top curve) and in the presence of the
absorption (bottom curves). Red solid curves display numerical simulations
while blue dashed curves show the analytical predictions. Inset: the real
(red solid line) and imaginary (green dashed line) part of the metamaterial
layer refractive index.}
\label{Fig2_Zero}
\end{figure}

We focus our study on the frequency region $10.40\text{GHz}<f<11.00\text{GHz}
$. In the absence of absorption and disorder, for these frequencies the
dielectric permittivity and the magnetic permeability of the metamaterial
layers vary over the intervals $-26.9<\varepsilon <-2.9$ and $-1.64<\mu
<0.055$. The refractive index is negative in the frequency range $10.40\text{%
GHz}<f<f_{mp}=10.95\text{GHz}$, as shown in the inset of Fig.\ref{Fig2_Zero}%
. However, at $f_{mp}=10.95\text{GHz}$, the magnetic permeability changes
sign and the metamaterial changes from being double negative (DNM) to single
negative (SNM). As we show later, such changes have a profound effect on the
localization properties.

We study the transmission of a plane wave either \emph{s} - or \emph{p} -
polarized and incident on a random stack from free space with an angle of
incidence $\theta_{0}$.

In the previous Subsections, we have described and used an effective WSA
method developed and elaborated in Refs \cite%
{Asatryan07,Asatryan10a,Asatryan10b}, for studying the transport and
localization in random stacks composed of the weakly reflecting layers.

In the dispersive case, the reflection from a single layer located in free
space is not necessarily weak, in which instance the method seems
inapplicable. However, we can replace each layer by the same layer
surrounded by infinitesimally thin layers of a background medium with
permittivity and permeability given by the mean values of $\bar{\varepsilon}%
(f)\equiv \langle \varepsilon (f)\rangle $ and $\bar{\mu}(f)\equiv \langle
\mu (f)\rangle $ respectively. In the considered case of weakly disordered
stacks, we can use the WSA approximation for all layers beside two "leads"
connecting the stack with free space from the very left and the very right
its ends. The localization characteristics which are intrinsic properties of
the stack do not feel the leads. Their role is restricted by only change the
coupling conditions to the random stack through the angle of incidence
transforming it from its given value $\theta_{0}$ outside the lead to the
frequency dependent refracted value $\theta_{b}$ inside the lead. These
angles are related by Snell law $\sin \theta _{0}=\sin \theta_b \sqrt{ \bar{%
\varepsilon}(f)\bar{\mu}(f)}.$ It is important to note that, while in the
localized regime the input and output leads are of no significance, they do
play a crucial role when localization breaks down (see below).

The single layer scattering\ is described by Eqs. (\ref{t}) where according
Eqs.~(\ref{ImpRefInd}) and(\ref{phase})

\begin{equation}  \label{shift-disp}
\beta _{n}=kd\nu _{n}\cos \theta _{n}, \ \ \ \nu _{n}=\sqrt{\varepsilon
_{n}\mu _{n}},
\end{equation}%
and $k=\displaystyle{\frac{2\pi }{\lambda}=\frac{2\pi f}{c}}$ is the free
space wave number. The interface Fresnel reflection coefficient $\rho _{n}$
is given by

\begin{equation}  \label{Fresnel-disp}
\rho _{n}=\frac{Z_{b}\cos \theta _{b}-Z_{n}\cos \theta _{n}}{Z_{b}\cos
\theta _{b}+Z_{n}\cos \theta_{n}},
\end{equation}%
The impedances $Z_{b}$ and $Z_{n}$ are

\begin{eqnarray*}  \label{impedances}
Z_{b}&=&\left\{
\begin{array}{ccc}
\sqrt{\bar{\mu}/\bar{\varepsilon}} &  & p\text{-polarization}, \\
&  &  \\
\sqrt{\bar{\varepsilon}/\bar{\mu}} &  & s\text{-polarization}, \\
&  &
\end{array}%
\right. ,  \notag \\
Z_{n}&=&\left\{
\begin{array}{ccc}
\sqrt{\mu _{n}/\varepsilon _{n}} &  & p\text{-polarization}, \\
&  &  \\
\sqrt{\varepsilon _{n}/\mu _{n}} &  & s\text{-polarization}. \\
&  &
\end{array}%
\right. ,
\end{eqnarray*}%
and the angles $\theta _{b}$ and $\theta _{n}$ satisfy Snell's law

\begin{eqnarray}  \label{cosa}
\nu _{n}\sin \theta _{n}=\bar{\nu}\sin \theta _{b}&=&\sin\theta_0,\ \ \ \
\bar{\nu}=\sqrt{\bar{\varepsilon}\bar{\mu}}.
\end{eqnarray}

General WSA expressions (\ref{lochom}) and (\ref{loc}) for localization
length of mono-type and mixed stacks remain valid for the stacks composed of
dispersive stacks. To study localization properties of such stacks we should
insert there the same single layer scattering coefficients (\ref{t}) with
dispersive phase shift (\ref{shift-disp}) and Fresnel coefficient (\ref%
{Fresnel-disp}).

Dispersion affects essentially the transport properties of the disordered
medium. In particular, it can lead to suppression of the localization either
at some angle of incidence, or at a selected frequency, or even in a finite
frequency range. Below we consider the two first cases for the H-stack
composed of $L$-layers. The third case will be considered further in Section %
\ref{subsec:enlight}.\newline

In the presence of dispersion, the long-wave asymptotic of the localization
length is

\begin{equation}  \label{l-disp}
\frac{1}{l}\!=\!\frac{\pi ^{2}d^{2}}{2\lambda^{2}(f)}\!\left( \frac{\langle
\mu ^{2}\rangle -\langle \mu \rangle ^{2}}{\langle \mu \rangle ^{2}}\! +\!%
\frac{\langle \varepsilon ^{2}\rangle -\langle \varepsilon \rangle ^{2}}{%
\langle \varepsilon \rangle ^{2}}\right) ,
\end{equation}%
where $\mu$ and $\varepsilon$ are given by Eqs. (\ref{mu-m}), (\ref{eps-m}),
and frequency-dependent wavelength in the medium

\begin{equation*}  \label{wavelength}
\lambda (f)=\frac{c}{f\sqrt{\varepsilon (f)\mu (f)}}.\ \ \ \
\end{equation*}
and can be large even when the wavelength of the incident signal, $\lambda=%
\displaystyle{\frac{c}{f}},$ is small.

Accordingly, the inverse localization length

\begin{equation*}  \label{llarge}
l^{-1}\propto f^{2}\varepsilon (f)\mu (f)
\end{equation*}%
becomes small not only at low frequencies $f\rightarrow 0$ but also in the
vicinity of $\mu $- or $\varepsilon $- zero points. For example, as the
frequency approaches the $\mu $-zero point from below, i.e., $f\rightarrow
f_{mp}^{-}$, in a H-stack of metamaterial layers, $\mu (f)$, for any
realization, is proportional to the difference $(f_{mp}-f)$ and the
expression for localization length diverges as $(f_{mp}-f)^{-1}$. Formally,
this divergence can be treated as delocalization, however the limiting value
$1/l=0$ means nothing but the absence of exponential localization. Moreover,
when the localization length becomes larger than the size of the stack,
ballistic transport occurs and the transmission coefficient is determined by
transmission length, rather than by the localization length.

To calculate the transmission coefficient for this case we consider, for the
sake of simplicity, a stack with only $\varepsilon $-disorder. Here the
transfer matrix of the $n$-th layer at $f=f_{mp}$ has the form

\begin{equation*}  \label{transmittance-3}
\mathcal{T}_{n}\equiv \mathcal{T}(\epsilon _{n})=\left\Vert
\begin{array}{ccc}
1+\epsilon _{n} &  & \epsilon _{n} \\
-\epsilon _{n} &  & 1-\epsilon _{n}%
\end{array}%
\right\Vert ,
\end{equation*}
where $\epsilon _{n}=ikd\varepsilon _{n}/2.$

As a consequence of the easily verified group property

\begin{equation*}  \label{transmittance-4}
\mathcal{T}(\epsilon _{1})\mathcal{T}(\epsilon _{2})=\mathcal{T}(\epsilon
_{1}+\epsilon _{2}),
\end{equation*}%
it follows that the stack transfer matrix $\mathcal{T}$ is

\begin{equation*}
c \mathcal{T}=\prod_{n=1}^{N}\mathcal{T}(\epsilon _{n})=\left\Vert
\begin{array}{ccc}
1+\mathcal{E} &  & \mathcal{E} \\
&  &  \\
-\mathcal{E} &  & 1-\mathcal{E}%
\end{array}%
\right\Vert ,
\end{equation*}%
where

\begin{equation*}  \label{[LargE]}
\mathcal{E}=\frac{ikL}{2}\frac{1}{N}\sum_{n=1}^{N}\varepsilon _{n},\ \ \ \ \
L=Nd.
\end{equation*}%
In a sufficiently long stack, $\mathcal{E}\approx \frac{1}{2}ikL\bar{\epsilon%
}$ and the transmittance T$=\left\vert \mathcal{T}_{11}\right\vert ^{-2}$ is
given by

\begin{equation*}  \label{transmittance-6}
\text{T}=\frac{1}{1+\left( \displaystyle{\frac{kL\bar{\varepsilon}(f)}{2}}%
\right) ^{2}}.
\end{equation*}%
Thus, at the frequency $f_{mp}$, the transmittance of the sample is not an
exponentially decreasing function of the length $L$ (as is typical for 1D
Anderson localization). It decreases much more slowly, namely, according to
the power law $\text{T} \propto L^{-2}$. The explanation of such a decrease
is that at a $\mu $-zero point ($f=f_{mp}$), the refractive index $\nu _{n}$
vanishes together with the phase shift $\beta _{n}=kd\nu _{n}\cos \theta_{n}
$ across the layer, thereby destroying the interference, which is the main
cause of localization. Another form of the explanation is that the effective
wavelength inside the stack tends to infinity when $\mu \rightarrow 0$ and
exceeds the stack length. Obviously, such a wave is insensitive to disorder
and therefore cannot be localized.

In the limit as the frequency approaches the $\mu $-zero frequency, from
above, i.e., $f\rightarrow f_{mp}^+$, the medium is single-negative and $%
\varepsilon \mu <0$. For frequencies $f$ not too close to $f_{mp},$ the
radiation decays exponentially inside the sample due to tunneling, and in
the absence of dissipation the decay rate is:

\begin{equation}  \label{l-disp-1}
l_{att}=\frac{1}{kd\sqrt{-\langle \mu \rangle \langle \varepsilon \rangle }}.
\end{equation}
Thus, as we approach the $\mu $-zero frequency from the right, the
formally-calculated localization length diverges as $l\propto
(f-f_{mp})^{-1/2}$ i.e. much more slowly than for the left-hand limit for
which $l\propto (f_{mp}-f)^{-1}.$ The transport properties in the vicinity
of the $\varepsilon $-zero frequency $f_{ep}$ can be considered in a similar
manner. Waves are also delocalized in the more exotic case when both
dielectric permittivity and magnetic permeability vanish simultaneously. The
vanishing of both $\mu$ and $\varepsilon$ simultaneously can happen at Dirac
points in photonic crystals \cite{Hu}.

The use of off-axis incidence from free space for frequencies for which $\mu$
or $\varepsilon$ are zero is not an appropriate mechanism for probing the
suppression of localization. In such circumstances, tunneling occurs and the
localization properties of the stack are not ``accessible'' from free space.
Nevertheless, suppression of localization can be revealed using an internal
probe, e.g., by placing a plane wave source inside the stack, or by studying
the corresponding Lyapunov exponent. Both approaches show total suppression
of localization at the frequencies at which dielectric permittivity or
magnetic permeability vanish.

In such circumstances, each layer which is embedded in a homogeneous medium
with material constants given by the average values of the dielectric
permittivity and magnetic permeability, is completely transparent, with this
manifesting the complete suppression of localization. However the
``delocalized'' states at the zero-$\mu$ or zero-$\varepsilon$ frequencies
are in a sense trivial, corresponding to fields which do not change along
the direction normal to the layers.\newline

Another example of the suppression of localization is related to the
Brewster anomaly. As we saw above, in a non-dispersive mixed stack with only
thickness disorder, delocalization of $p$-polarized radiation occurs at the
Brewster angle of incidence. At this angle, the Fresnel coefficient $\rho$ (%
\ref{ro}) and, therefore, the reflection coefficient (\ref{t}) as well,
vanish for any frequency, thus making each layer completely transparent.

In the presence of dispersion, the same condition $\rho =0$ leads to more
intriguing results. In this instance, frequency-dependent angles, at which a
layer becomes transparent, exist not only for $p$-polarization, but also for
an $s$-polarized wave. This means that the Brewster anomaly occurs for both
polarizations, with the corresponding angles, $\theta _{p}$ and $\theta _{s},
$ being determined by the conditions

\begin{eqnarray}
\tan ^{2}\theta _{p} &=&\frac{\varepsilon (\varepsilon \overline{\mu }-%
\overline{\varepsilon }\mu )}{\overline{\varepsilon }(\varepsilon \mu -%
\overline{\varepsilon }\,\overline{\mu })},  \label{brewS} \\
\tan ^{2}\theta _{s} &=&\frac{\mu (\overline{\varepsilon }\mu -\varepsilon
\overline{\mu })}{\overline{\mu }(\varepsilon \mu -\overline{\mu }\,%
\overline{\varepsilon })}.  \label{brewP}
\end{eqnarray}

The right hand sides of these equations always have opposite signs.
Therefore from Brewster conditions (\ref{brewS}) and (\ref{brewP}) one can
find either the Brewster angle and corresponding polarization for a given
frequency, or the Brewster frequency and corresponding polarization for a
given angle of incidence.

While, for a stack with only thickness disorder, the condition $\rho =0$ can
be satisfied for all layers simultaneously, when $\varepsilon $ and/or $\mu $
fluctuate, the conditions (\ref{brewS}) or (\ref{brewP}) define the
frequency-dependent Brewster angles which are slightly different for
different layers. These angles occupy an interval within which the stack is
not completely transparent, but has anomalously large transmission lengths
\cite{Sipe,Asatryan10b}.

When only the dielectric permittivity is disordered and $\mu =\overline{\mu }%
, $ the Brewster conditions (\ref{brewS}), (\ref{brewP}) simplify to

\begin{eqnarray}  \label{brewPe}
\tan ^{2}\theta _{s} &=&-1, \\
\tan ^{2}\theta _{p} &=&\frac{\varepsilon }{\overline{\varepsilon }}\approx
1.
\end{eqnarray}%
In this case, the Brewster condition is satisfied only for \textit{p}%
-polarization. For weak disorder,  the Brewster angle of incidence from the
effective medium is $\theta _{p}\approx \pi /4$. For a given frequency $f$,
angle of incidence from free space, $\theta_0$, should be found from Snell's
law (\ref{cosa}), and for a given $\theta_0 $, the Brewster frequency $f_{p}$
follows from

\begin{equation}  \label{BFp}
\sqrt{\overline{\varepsilon }(f_{p})\overline{\mu }(f_{p})}=\frac{\sin
\theta_0 }{\sin \theta _{p}}=\sqrt{2}\sin \theta_0.
\end{equation}%
Note that this equation may be satisfied at multiple frequencies depending
on the form of the dispersion.

The case of only magnetic permeability disorder, $\varepsilon =\overline{%
\varepsilon }$, is described by similar equations which are obtained by
replacement $s\leftrightarrow p$ in Eqs. (\ref{brewSe}) - (\ref{BFp}).

For disorder in both the permeability and the permittivity, the existence of
a Brewster anomaly angle depends, in accordance with Eqs. (\ref{brewS}) and (%
\ref{brewP}), on the sign of the quantity $\xi =(\overline{\varepsilon }\mu
-\varepsilon \overline{\mu })/(\varepsilon \mu -\overline{\varepsilon }%
\overline{\mu })$. If $\xi >0$, the Brewster angle exists for $s$%
-polarization, while if $\xi <0$, it exists for $p$-polarization. In the
case $\xi =0$, the layer and the medium in which it is embedded are
impedance matched, and thus the layer is completely transparent.

The features of transmission length mentioned above are completely confirmed
by numerical calculations. Consider first the case of normal incidence on a
stack of $N=10^{7}$ layers, in which we randomize only the dielectric
permittivity ($Q_{m}=0$) with $Q_{e}=0.5\times 10^{-2}$. In Fig.~\ref%
{Fig2_Zero} the transmission length $l_T$ as a function of frequency $f$ is
displayed. The upper curves present the lossless case, while the lower
curves show the effects of absorption (see~\cite{Asatryan12} for details).

The red, solid curves and the blue, dashed curves display results from
numerical simulations and the WSA theoretical prediction respectively. The
top curves represent the genuine localization length for all frequencies
except those in the vicinity of $f\approx f_{mp}=10.95\text{GHz}$ where the
transmission length dramatically increases.

In the absence of absorption, for frequencies $f>10.95\text{GHz}$, the
metamaterial transforms from being double negative to single negative (see
inset in Fig. \ref{Fig2_Zero}). The refractive index of the metamaterial
layer changes from being real to being pure imaginary, the random stack
becomes opaque, and the transmission length substantially decreases. Such a
drastic change in the transmission length (by a factor of $10^{5}$) might be
able to exploited in a frequency controlled optical switch. Across the
frequency interval $10.4GHz<f<11.0$GHz, theoretical results are in an
excellent agreement with those of direct simulation. Moreover, for all
frequencies except in the region $10.4Ghz<f<10.5$GHz, the single scattering
approximation excellently describes the $l_T$ behavior. Quite surprisingly,
the asymptotic equations (\ref{l-disp}) and (\ref{l-disp-1}) are in the
excellent agreement with the numerical results even over the frequency range
$10.9\text{Ghz}<f<11.0\text{Ghz}$, including in the near vicinity of the
frequency $f_{mp}=10.95\text{GHz}$ at which $\mu $ vanishes.

Absorption substantially influences the transmission length (the lower curve
in Fig.\ref{Fig2_Zero})\cite{Asatryan12} and smoothes the non-monotonic
behavior of the transmission length for $f<10.5\text{GHz}$. The small dip at
$f\approx 10.45 \text{GHz}$ correlates with the corresponding dip in the
transmission length in the absence of absorption. The most prominent effect
of absorption occurs for frequencies just below the $\mu $-zero frequency $%
f_{mp}=10.95\text{GHz}$. While in the absence of absorption, the stack is
nearly transparent in this region, turning on the absorption reduces the
transmission length by a factor of $10^{2}$--$10^{3}$ for $f>10.7\text{GHz}$%
. In contrast, for frequencies $f>10.95\text{GHz}$, the transmission lengths
in the presence and absence of absorption are nearly identical because here
the stack is already opaque and its transmittance is not much affected by an
additional small amount of absorption.

The case where both disorders of the dielectric permittivity and magnetic
permeability are present, is qualitatively similar to that of the single
disorder case considered above.

\begin{figure}[tbh]
\centerline{\rotatebox{-0}{\includegraphics[width=8.6cm]{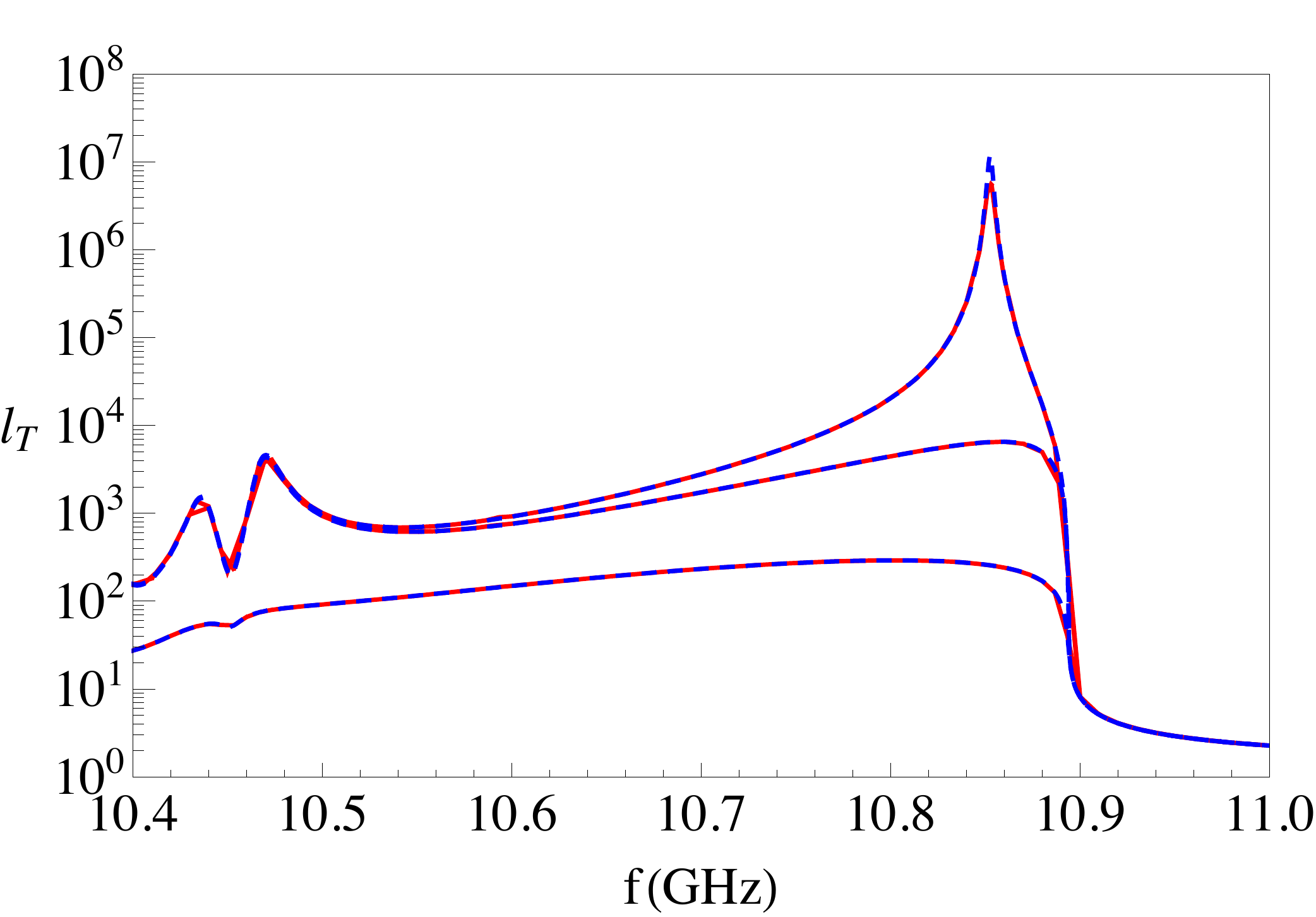}}}
\caption{(Ref.~[\onlinecite{Asatryan12}], color online) Transmission length $l_T$ vs
frequency $f$ for $\protect\theta _{a}=30^{\circ }$ for a metamaterial
stack: without absorption, \textit{p}-polarization (top curves), \textit{s}
polarization (middle curves); in the presence of absorption (bottom curves).}
\label{Fig3_Zero}
\end{figure}

In the case of oblique incidence, polarization effects become important. In
Fig.\ref{Fig3_Zero}, the transmission length frequency spectrum is displayed
for the same metamaterial H-stack with only dielectric permittivity disorder
for the angle of incidence $\theta_0 =30^{\circ }$. Here for frequencies $%
f<10.55\text{GHz}$, the transmission length is largely independent of the
polarization. Moreover it does not differ from that for normal incidence
(compare with the top curve in Fig.\ref{Fig2_Zero} ). This is due to the
high values of the refractive indices at these frequencies ($|\nu_{n}|>4)$,
resulting in almost zero refraction angles (\ref{cosa}) for angles of
incidence that are not too large.

The transmission length manifests a sharp maximum at an angle close to the
Brewster angle, as commented upon in Refs \cite{Sipe,Asatryan10b}. This is
indeed apparent in Fig. \ref{Fig3_Zero} for the frequency $f\approx 10.85%
\text{GHz}$. Because only $\varepsilon $ fluctuates, the Brewster condition
is satisfied only for \textit{p}-polarization (\ref{brewPe}) at a single
frequency $f_{p}\approx 10.852\text{GHz}$. The introduction of additional
permeability disorder (not shown) reduces the maximum value of the
localization length by two orders of magnitude.

Comparison of Figs \ref{Fig2_Zero} and \ref{Fig3_Zero} shows that the
frequency of the maximal suppression of localization decreases as the angle
of incidence increases. At normal incidence it coincides with the $\mu $%
-zero frequency $f_{mp}$ while for oblique incidence at $\theta_0 =30^{\circ
}$ it coincides with the Brewster frequency $f_{p}$ for \textit{p}%
-polarization.

Absorption strongly diminishes the transmission providing the main
contribution to the transmission length while the permittivity disorder has
little influence on the transmission length. In this case, the results for
both two polarizations are therefore practically indistinguishable.

The transmission properties of a stack with only magnetic permeability
disorder at oblique incidence, are similar to those for the case of only
dielectric permittivity disorder. The key difference is that there is a
Brewster anomaly for \textit{s}-polarization while for \textit{p}%
-polarization it is absent.

\begin{figure}[htb]
\centerline{\rotatebox{-0}{\includegraphics[width=8.6cm]{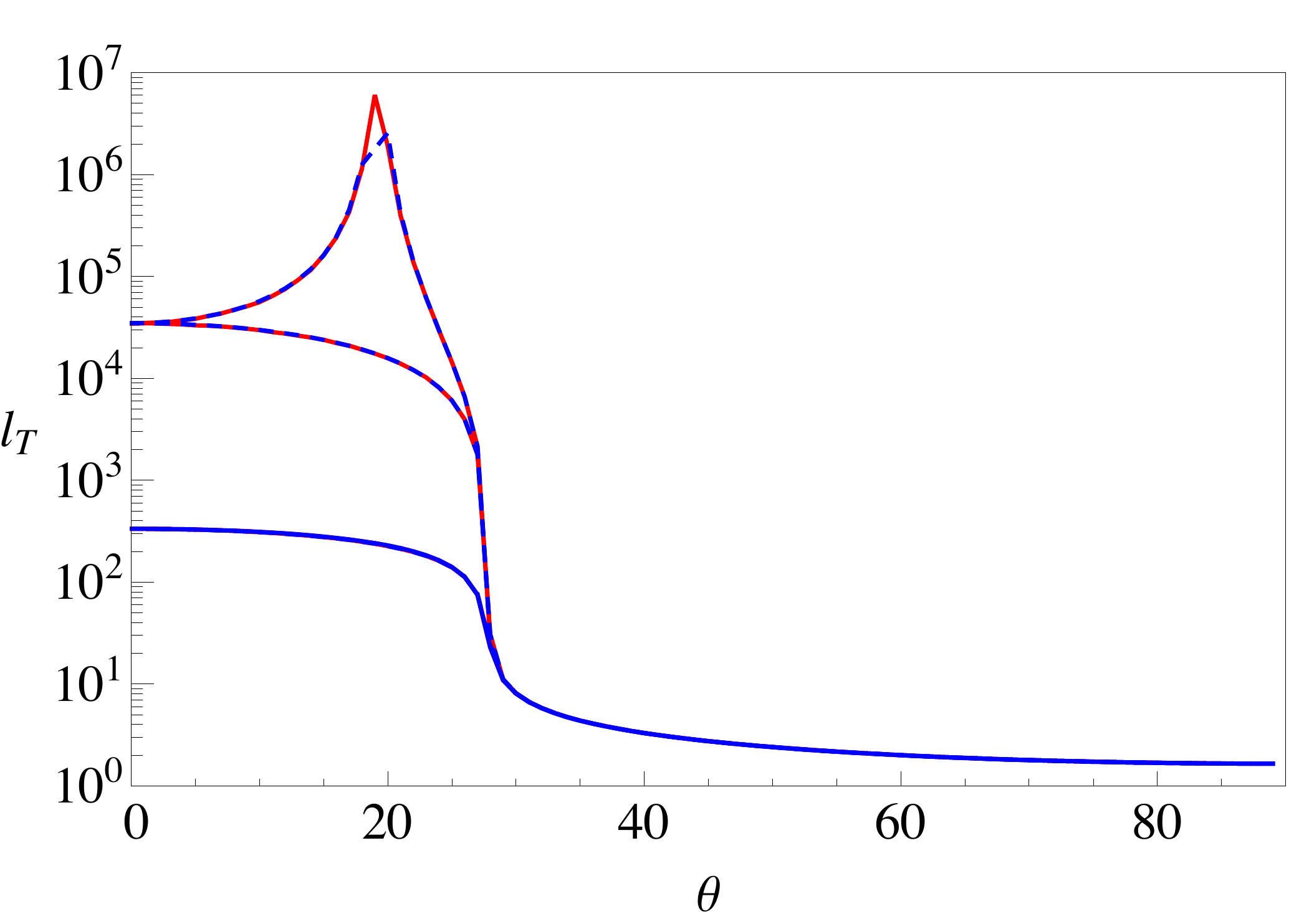}}}
\caption{(Ref.~[\onlinecite{Asatryan12}], color online) Transmission length $l_T$ vs angle
of incidence for a homogenous metamaterial stack at $f=10.7\text{GHz}$ with
permittivity disorder: in the absence of absorption---upper curve and for $p$
polarization; middle curve is for $s$ polarization and in the presence of
absorption and for both polarizations, lower curves.}
\label{Fig5_Zero}
\end{figure}

We consider also the dependence of the transmission length on the angle of
incidence at a fixed frequency. The results for both polarizations are
displayed in Fig. \ref{Fig5_Zero}. Here we have plotted the transmission
length of the stack with only dielectric permittivity disorder with $%
Q_{e}=0.5\times 10^{-2}$ at the frequency $f=10.90\text{GHz}$. The upper and
middle curves in this figure correspond to the results for $p$- and $s$%
-polarized waves respectively in the lossless case. For $s$-polarized light,
the transmission length decreases monotonically with increasing angle of
incidence, while for $p$- polarized wave it increases with increasing angle
of incidence. Such behavior reflects the existence of a Brewster angle for $p
$-polarization at the Brewster angle $\theta_0 =20^{\circ }$. The red solid
curve shows the results of simulations, while the blue dashed line is the
analytic prediction.

As in the previous cases, in the presence of absorption, the results for
both polarizations are almost identical (the lower curves in Fig. \ref%
{Fig5_Zero}). For angles $\theta_0 <30^{\circ }$, the transmission length is
dominated by absorption, while for angles $\theta_0 >30^{\circ }$ tunneling
is the dominant mechanism. The results for permeability disorder are very
similar to those for permittivity disorder.

For normal H-stacks, the transmission length manifests exactly the same
behavior as for H-stacks comprised of metamaterial layers.\newline


\subsection{Anomalous Suppression of Localization}

\label{subsec:enlight}

In this Section, we consider the stacks with only refractive index disorder
(RID) i.e. the stacks with $\delta_{d}=\delta_{\mu}=0$. In this limit, there
is nothing special for H-stacks. Their transmission length demonstrates
qualitatively and quantitatively the same behavior as was observed in the
presence of both refractive index and thickness disorder. Corresponding
formulae for the transmission, localization, and ballistic lengths can be
obtained from the general case by taking the limit as $Q_{d}\rightarrow 0$.

In the case of M-stacks, however, the situation changes markedly. Here
suppression of localization in the long wave region becomes anomalously
large enhancing transmission length on some orders of magnitude and even
changing its functional dependence on the wavelength\cite{Asatryan07}.
Instead of the universal $\propto\lambda^2$ dependence, long wave
asymptotic of both localization length $l$ and reciprocal Lyapunov exponent $l_{\xi}$
follows a power law $\propto\lambda^m$ with much larger exponent $m.$

\begin{figure}[h]
\rotatebox{0}{\scalebox{0.3} {\includegraphics{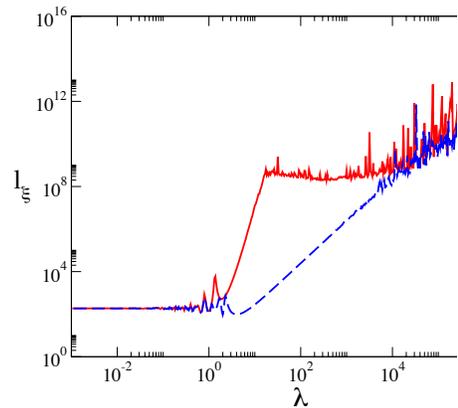}}}
\caption{(Ref.~[\onlinecite{Asatryan07}], color online) Localization length $l_{\protect\xi%
}$ vs. wavelength $\protect\lambda$ for $Q=0.25$ and $N=10^9$ layers; solid
line is for the M-stack, while the dashed line is for the corresponding
(normal) H-stack. }
\label{SMR}
\end{figure}

Let us start with some numerical results demonstrating such an anomalous
growth of the long wave localization lengths $l, \ l_{\xi}$ of the minimally disordered M-stack with only RID. In Fig.\ref{SMR} localization length $l_{\xi}$ for M-stack with $Q=0.25$ is
plotted. Solid line in Fig.~\ref{SMR} corresponds to $l_{\xi}$ for the
propagation in a M-stack and a single realization of $N=10^9$ layers, while
the dashed line is for the corresponding H-stack with the same parameters.
Within the localization region $l_{\xi}(\lambda)<10^{8}$, M-stack reciprocal
Lyapunov exponent grows in the long wave region essentially faster than that
of H-stack. While for H-stack is described by standard exponent $m=2$, its
value for M-stack was estimated as $m=6$ and the phenomenon itself was named
as $\lambda^{6}$ anomaly. The observed anomalous suppression of localization
was attributed to a lack of phase accumulation over the sample, due to the
cancelation of the phase that occurs in alternating $L$- and $R$-layers\cite%
{Asatryan07}.

Anomalous suppression of localization is manifested also in the case of
oblique incidence. The next Figure \ref{Fig10A} displays transmission length
spectra for a M-stack with only refractive index disorder for an angle of
incidence of $\theta=30^\circ$. There is a striking difference between the
two polarizations: in the case of \emph{p}-polarized light, there is strong
localization at long wavelengths ($\lambda\leq10^2$), with the localization
length showing $\propto\lambda^2$ dependence. In contrast, the localization
length for \emph{s}-polarized light is much larger and is estimated as $%
\approx\lambda^6$ dependence as occurs for normal incidence. Note that for
\emph{s}-polarization, anomalous enlightening manifests itself only in
localization regions in Fig. \ref{Fig10A} which are bounded from above by
the wavelength limits $\lambda\leq5,9$, and $12$ for stacks of length $%
N=10^5,10^{7}$, and $8 \times 10^{8}$ respectively.

This asymmetry between the polarizations suggests that the suppression of
localization is due not only to the suppression of the phase accumulation
but also to the vector nature of the electromagnetic wave. Because of the
symmetry of Maxwell's equations between the electric and magnetic fields, it
is to be expected that for a model in which there is disorder in the
magnetic permeability (with $\varepsilon=\pm1$) the situation will be
inverted with anomalous enlightening for \emph{p}-polarized waves and with
\emph{s}-polarization showing strong localization.\newline

\begin{figure}[h]
\center{
\rotatebox{0}{\scalebox{0.4}{\includegraphics{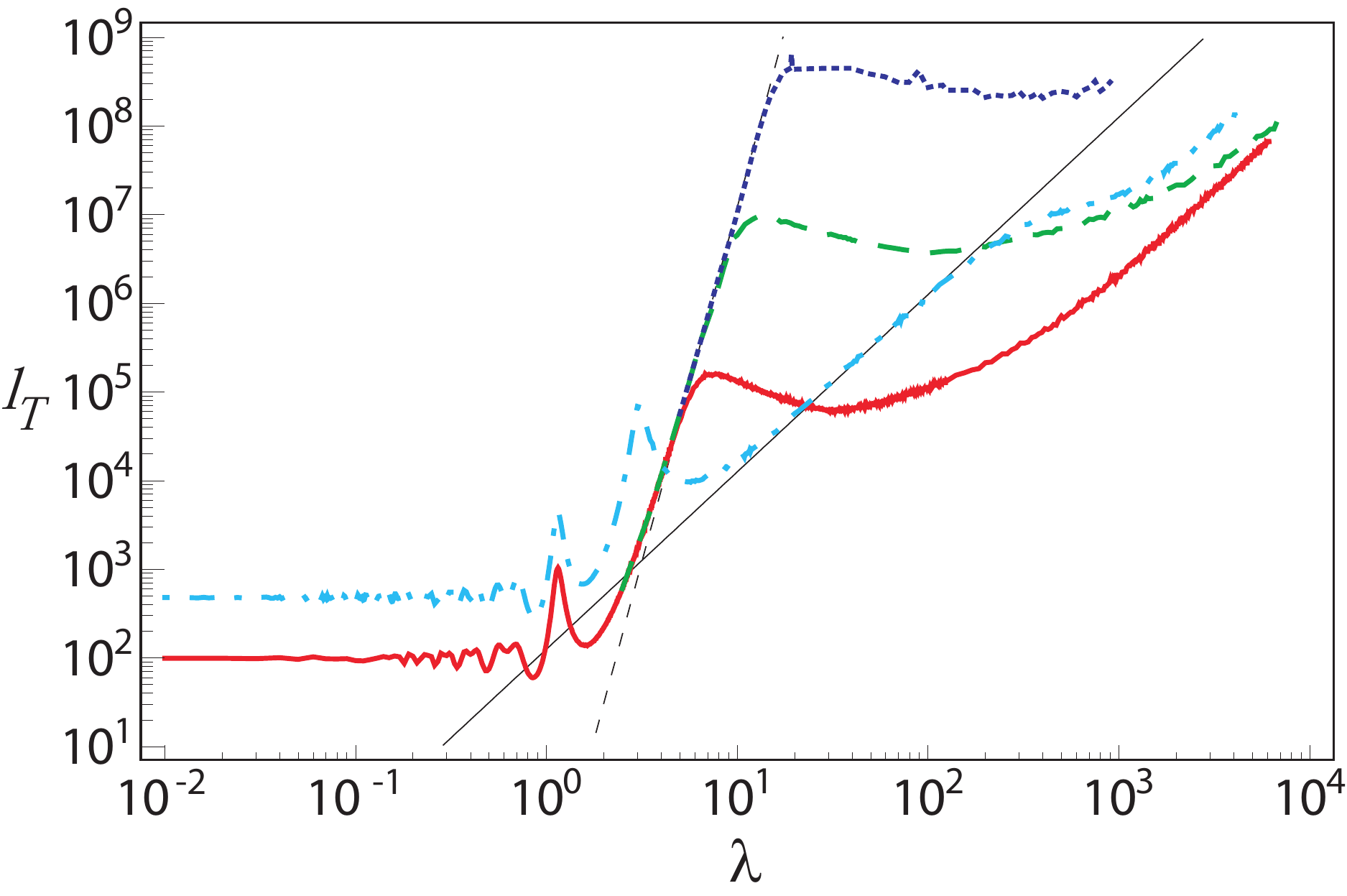}}}}
\caption{(Ref.~[\onlinecite{Asatryan10b}], color online) Transmission length
$l_T$ versus $\protect\lambda$ for a M-stack with $Q_{\protect\nu}=0.25$, $%
Q_d=0$ and $\protect\theta=30^0$ for \emph{p}-polarized light (cyan dashed
dotted curve, $N=10^6$) and \emph{s}-polarized light (red solid curve, $%
N=10^5$; green dashed curve, $N=10^7$; blue dotted curve, $N=8\times10^8$). }
\label{Fig10A}
\end{figure}

The results of calculations\cite{Asatryan10a} provided for much longer
stacks up to $N=10^{12}$ qualitatively completely coincided with the
previous ones. However more detailed studies quantitatively occurred
slightly different. Generation of a least squares fitting $l_T=A\lambda ^{m}$
to the transmission length data, led to a bit surprising conclusions. The
best fits were $m\approx 6.25$ for $N=10^{7}$, $m\approx 7.38 $ for $N=10^{9}
$, and even $m\approx 8.78$, for $N=10^{12}.$ This shows that the question
about a genuine value of exponent $m$ remains still open.

\begin{figure}[h]
\centerline{\rotatebox{0}{\includegraphics[width=7.4cm]{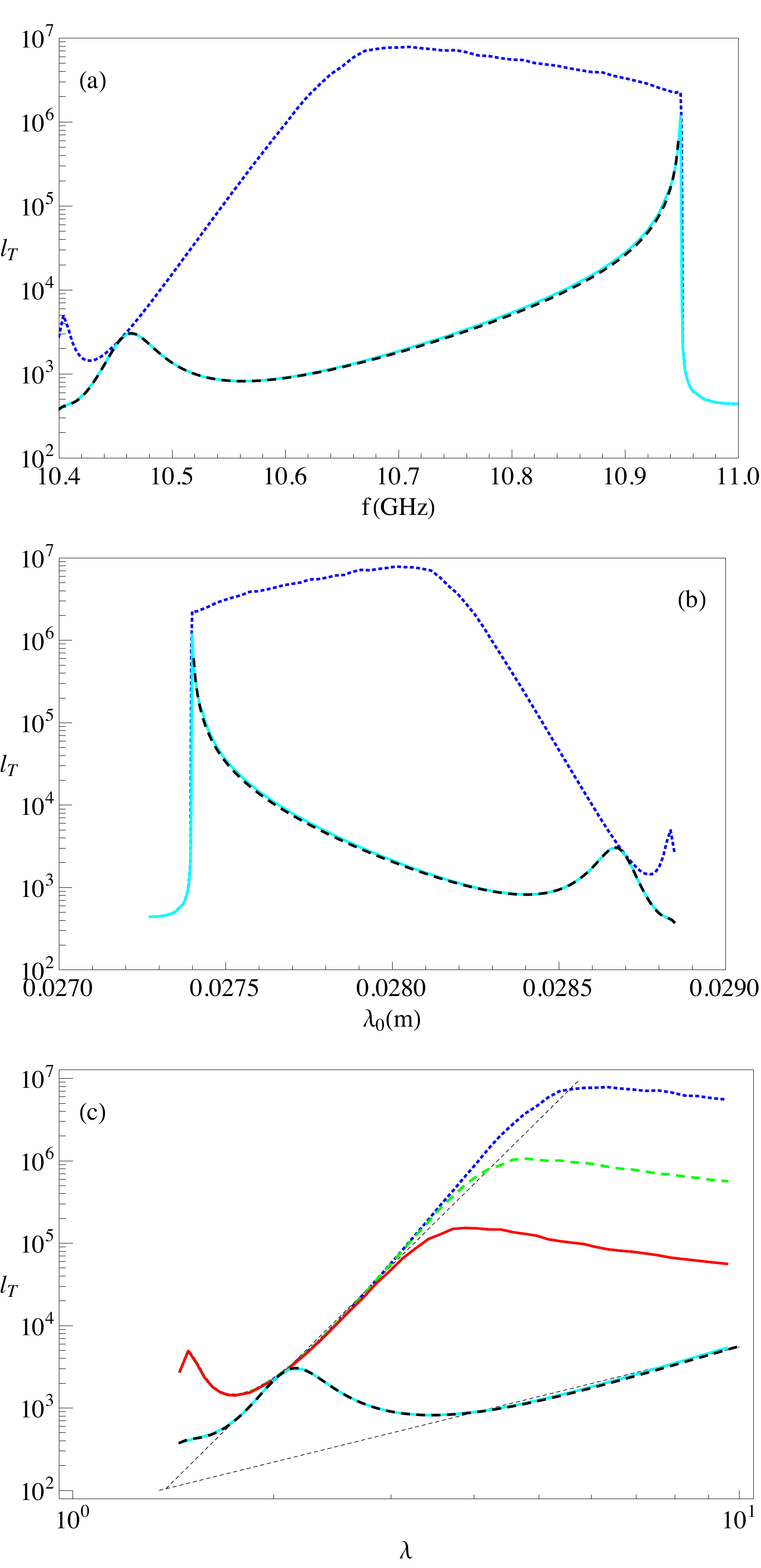}}}
\caption{(Ref.~[\onlinecite{Asatryan12}], color online) (a)Transmission length $l_T$ vs
frequency $f$ for a mixed stack with $N=10^7$ layers (top dotted blue
curve), and only dielectric permittivity disorder. The bottom curves on all
panels (a,b,c) are for a stack with $N=10^7$ layers with both permittivity
and permeability disorder (cyan, solid curve display simulation results
while the dashed, black curve is for the analytic prediction; (b) is the
same as in (a) but plotted as a function of the free space wavelength $%
\protect\lambda_0 $ while on panel (c) we plot the transmission length as a
function of the averaged wavelength inside of the stack normalized to the
thickness of the layer, for $N=10^7$ layers (blue dotted top curve), $N=10^6$
layers (dashed green curve) and for $N=10^5$ layers (red solid curve)
respectively. }
\label{Fig8}
\end{figure}

Consider now the long wave behavior of the localization length in the
presence of dispersion. In the panel a) of the Fig.\ref{Fig8}, the
transmission length spectrum is plotted in the case of normal incidence, for
a small permittivity disorder of $Q_{e}=0.5\times 10^{-2}$. One can
immediately observe significant (up to four orders of magnitude) suppression
of localization in the frequency region $10.50\text{GHz}<f<10.68\text{GHz}$.
However, this suppression seems to have nothing common with observed above
anomalous enlightening. Indeed, in this case the localization length grows
with increasing frequency, while in the previous studies \cite%
{Asatryan07,Asatryan10a,Asatryan10b}, similar growth has been observed with
increasing incident wavelength. This is demonstrated in Fig.\ref{Fig8}b
where the same transmission length spectrum is plotted as a function of free
space wavelength. Thus, the localization length decreases by four orders of
magnitude, manifesting as an enhancement, rather than the suppression, of
localization with increasing wavelength.

Although at the first sight these findings are in sharp contrast with the
previous ones, they are correct and physically meaningful. In the model
studied earlier\cite{Asatryan07,Asatryan10a,Asatryan10b}, the wavelength of
the incident radiation largely coincided with the wavelength inside each
layer. In dispersive medium considered here, these two wavelengths differ
substantially. Accordingly, in Fig.\ref{Fig8}c, we plot the transmission
length as a function of wavelength within the stack and obtain results which
are very similar to those in Refs. \cite{Asatryan07,Asatryan10a,Asatryan10b}%
. To emphasize this similarity, we have plotted the transmission length
spectrum for three different stack lengths: $N=10^{5},10^{6},10^{7}$. It is
easily seen that the suppression of localization in the dispersive media is
qualitatively and quantitatively similar to that predicted in Ref. \cite%
{Asatryan07}. Corresponding exponent $m$ of anomalous enlightening estimated
with the help of these results, is $m\approx 8.2$.\newline

Enhanced suppression of localization exists in the strictly periodic
alternative M-stacks with a constant layer thickness and only refractive
index disorder. By other words, in
mixed stacks having constant layer thickness, the dielectric permittivity disorder alone is not sufficiently strong to
localize low-frequency radiation by a standard way. There are many ways to
violate these conditions. It is possible to add thicknesses fluctuations\cite%
{Asatryan07,Asatryan10a}, or magnetic permeability fluctuations\cite%
{Asatryan12}, or to introduce a weak difference between two constant
thicknesses of $R$- and $L$-layers, or not to change any parameter but
rearrange randomly the same numbers $N/2$ of $R$- and $L$-layers\cite%
{Asatryan10a}. Each such a violation immediately destroys anomalous
suppression of localization  and restores standard long wave asymptotic $l\propto \lambda^{2}
$.

The analytical results obtained above in Section~\ref{sec:meta},
survive in the $\delta _{d}\rightarrow 0$ limit and predict $l\propto
\lambda ^{2}$ asymptotic. However more detailed investigation shows that WA in its form (\ref{rec3}), (\ref{rec4}) fails in this limit\cite{Asatryan10a}.

As was mentioned above, localization length $l_{\xi}$ manifests qualitatively the same behavior as transmission length $l_{T}.$ At the same time, its calculation is simpler than that of $l_{T}$. Lyapunov exponent in minimally disordered M-stacks was calculated in\cite{Mak-12} using some version of the method described in Refs. [\cite{GP,LGP,PF,GMP}] and at the end of Section~\ref{subsec:transfer}. The remaining part of this Subsection contains slightly modified details and results of this calculation [\cite{Mak-12}].

Consider the
electromagnetic wave of frequency $\omega =ck,$ in infinite array comprised
of two types of lossless alternative $\alpha $ and $\beta $ layers of equal
dimensionless thickness $\Delta _{j}=1$ with random only dielectric
permittivities. Enumerate the layers so that $j$-th layer occupy the
interval $j-1\leq z<j$ and choose all odd layers of $\alpha $ type and all
even of $\beta $ type. For alternative array, it is natural to choose an
elementary cell composed of two adjacent layers, as the main basic element
of the array\cite{Asatryan07,Mak-12}. The $n$-th cell occupies interval $%
2n-2\leq z<2n$ and consists of $(2n-1)$-th and $2n$ layers of type $\alpha $
and $\beta $ correspondingly. Each layer is characterized by its type $%
\alpha $ ($\beta $), magnetic permeability $\mu _{\alpha }=1$ ($\mu _{\beta
}=\pm 1$), refractive index $\nu _{\alpha }(n)$ ( $\nu _{\beta }(n)$),
impedance $Z_{\alpha }(n)=1/\nu _{\alpha }(n)$ ($Z_{\beta }(n)=\pm 1/\nu
_{\alpha }(n)$), and wave number $k_{\alpha ,\beta }=k\nu _{\alpha ,\beta }$
of the wave.

Within such a model,two systems are considered: the H-array when both $\alpha
$ and $\beta$ layers are made of right-handed materials, and M-array where $%
\alpha$ layers are right-handed material while $\beta$ layers are of
left-handed material. We emphasize that on the contrary of H-stack notion
where all the layers have the same statistical properties, H-array is
composed of two different materials with different statistical properties
for odd and even layers. Disorder is incorporated into the model via
dielectric permittivities $\varepsilon_{\alpha,\beta}$ only, so that
refractive index $\nu$ is a sole fluctuation parameter and the upper index
in its fluctuations $\delta^{(\nu)}_{\alpha,\beta}(n)$ can be omitted

\begin{equation}  \label{CD-nmu}
\nu_{\alpha}(n)=1+\delta_{\alpha}(n), \ \ \ \ \nu_{\beta}(n)=\pm[%
1+\delta_{\beta}(n)].
\end{equation}
Refractive index fluctuations $\delta_{\alpha,\beta}(n)$ are assumed to be
delta-correlated with zero mean value $\langle\delta_{\alpha,\beta}(n)%
\rangle=0$, and variance $\sigma^2$,

\begin{equation}  \label{CD-CorrDef}
\langle\delta_{\alpha}(n)\delta_{\beta}(n^{\prime})\rangle=\sigma^2\delta_{%
\alpha\beta}\delta_{nn^{\prime}}\,,
\end{equation}
where angular brackets mean the ensemble average.

To calculate Lyapunov exponent of the electromagnetic wave of the frequency $%
\omega$, consider two component vector

\begin{equation*}  \label{electric-S}
\vec{S}_{n}=\left(%
\begin{array}{c}
Q_{n} \\
\\
P_{n}%
\end{array}
\right)
\end{equation*}
with components

\begin{equation}  \label{QPs}
Q_{n}=E(2n-2), \ \ \ P_{n}=\frac{c}{\omega}E^{\prime}(2n-2)
\end{equation}
proportional to the field and its derivative at the left edge of the $n$-th
cell. These components are real. Therefore they automatically correspond to
the currentless field and can be parametrized as

\begin{equation}  \label{real-currentless}
\vec{S}_{n}=e^{\xi_{n}}\left(%
\begin{array}{c}
\cos\theta_{n} \\
\\
\sin\theta_{n}%
\end{array}
\right)
\end{equation}
(compare with Eq. (\ref{currentless-1})). Note that this is currentless state in the basis of standing waves while in
Section~\ref{subsec:transfer} the basis of running waves was used.

Using Maxwell equations and appropriate boundary conditions on the
interfaces of the layers, one obtains dynamic equation

\begin{eqnarray}  \label{CD-mapQP}
\vec {S}_{n+1}= \hat{T}\vec {S}_{n}.
\end{eqnarray}
Here $\hat{T}_{n}$ is the unimodular matrix with elements

\begin{equation}  \label{CD-ABCDn}
\begin{array}{ccc}
T_{11} & = & \cos\varphi_{\alpha}\cos\varphi_{\beta}-Z_{\alpha}^{-1}Z_{%
\beta}\sin\varphi_{\alpha}\sin\varphi_{\beta}, \\
T_{12} & = & Z_{\alpha}\sin\varphi_{\alpha}\cos\varphi_{\beta}+Z_{\beta}\cos%
\varphi_{\alpha}\sin\varphi_{\beta}, \\
T_{21} & = & -Z_{\alpha}^{-1}\sin\varphi_{\alpha}\cos\varphi_{\beta}-Z_{%
\beta}^{-1}\cos\varphi_{\alpha}\sin\varphi_{\beta}, \\
T_{22} & = & \cos\varphi_{\alpha}\cos\varphi_{\beta}-Z_{\alpha}Z_{%
\beta}^{-1}\sin\varphi_{\alpha}\sin\varphi_{\beta}%
\end{array}%
.
\end{equation}
They depend on the cell number $n$, due to randomized refractive indices (%
\ref{CD-nmu}) entering both the impedances $Z_{\alpha,\beta}(n)$ and phase
shifts $\varphi_{\alpha,\beta}(n)$,

\begin{subequations}
\label{CD-phi-ab}
\begin{eqnarray*}
\varphi_{\alpha}(n)&=&\frac{1}{2}k_{\alpha}(n)=\varphi[1+\delta^{\nu}_{%
\alpha}(n)], \\
\varphi_{\beta}(n)&=&\frac{1}{2}k_{\beta}(n)=\pm\varphi[1+\delta^{\nu}_{%
\beta}(n)],
\end{eqnarray*}
with $\varphi=k/2$.

In $\xi_{n},$ $\theta_n$ terms, dynamic equations read

\end{subequations}
\begin{eqnarray}  \label{CD-mapTheta}
\xi_{n+1}-\xi_{n}&=& \Phi(
\te_{n}),\\ \tan\theta_{n+1}&=&\frac{T_{21}+T_{22}\tan\theta_n} {T_{11}+T_{12}\tan%
\theta_n}, \ \ \ \ \ \ \ \ \ \ \ \ \ \
\end{eqnarray}
where now

\begin{equation}
\label{Phi-New}
\Phi(\te)=\frac{1}{2}
\ln\frac{(T_{11}+T_{12}\tan\te)^{2}+(T_{21}+T_{22}\tan\te)^{2}}
{1+\tan^{2}\te}
\end{equation}

Going to the limit $n\to\infty$ and using Eqs. (\ref{difference}) and (\ref%
{Lyapunov-30}) for localization length $l_{\xi}=\gamma^{-1}$ we obtain

\begin{equation}  \label{CD-Lyap}
\frac{1}{l_{\xi}}=\left\langle\Phi(\te)\right%
\rangle_{st},
\end{equation}
where averaging in the r.h.s. is taken over the stationary distribution of
the phase $\theta$.

In the case of weak disorder,

\begin{equation*}  \label{CD-WeakDis}
\sigma^2\ll1\quad\mathrm{and}\quad(\sigma\varphi)^2\ll1,
\end{equation*}
this distribution $\rho(\theta)$ can be explicitly found within the
framework of a proper perturbation theory. Expanding the exact $\theta$-map (%
\ref{CD-mapTheta}) up to the second order in perturbation~\cite{APS} and
taking into account the uncorrelated nature of the disorder (see Eq.~(\ref%
{CD-CorrDef}, one obtains,

\begin{eqnarray}  \label{CD-theta}
\theta_{n+1}-\theta_n=-\phi-\delta_{\alpha}(n)U(\theta_n)\mp  \notag \\
\delta_{\beta}(n)U(\theta_n-\phi/2) -\sigma^2W(\theta_n),
\end{eqnarray}
where


\begin{eqnarray}  \label{CD-UW}
U(\theta)=\varphi+\sin\varphi\cos(2\theta-\varphi), \ \ \ \ \ \ \ \ \ \ \ \
\ \ \ \   \notag \\
W(\theta)=\varphi[\cos(2\theta-2\varphi)\pm\cos(2\theta-2\phi)]+ \ \ \ \ \ \
\   \notag \\
\sin\varphi[\sin\theta\sin(\theta-\varphi)\pm\sin(\theta-\phi/2)\sin(\theta-%
\varphi-\phi/2)]+  \notag \\
\sin^2\varphi\sin(4\theta-2\varphi-\phi)\cos\phi, \ \ \ \ \ \ \ \ \ \ \ \ \
\ \ \ \ \
\end{eqnarray}
``plus" stands for the H-array, and ''minus" refers to the M-array, and

\begin{equation}  \label{CD-gamma}
\phi=\left\{%
\begin{array}{ccc}
k &  & \text{H-array} \\
&  &  \\
0 &  & \text{M-array}%
\end{array}%
\right.
\end{equation}
is the unperturbed Bloch phase shift $\phi$ over a unit $(\alpha,\beta)$
cell.

Now one should write down the Fokker-Plank equation related to the dynamic
equations (\ref{CD-theta})

\begin{eqnarray}  \label{CD-eqFP}
&&\frac{d^2}{d\theta^2}\left[U^2(\theta)+U^2(\theta-\phi/2)\right]%
\rho(\theta)  \notag \\
&&+2\frac{d}{d\theta}\left[\frac{\phi}{\sigma^2}+W(\theta)\right]%
\rho(\theta)=0,
\end{eqnarray}
find it normalized $\pi$-periodic solution and calculate average in the
r.h.s. of (\ref{CD-Lyap}).

For H-array, this program can be easily realized. Indeed in such a structure
the Bloch phase (\ref{CD-gamma}) is non zero, and for weak disorder the term
in Eq.~(\ref{CD-eqFP}) containing $\phi/\sigma^2$ prevails over the others.
Therefore, the phase distribution within the main order of perturbation
theory is uniform

\begin{equation}  \label{CD-RhoUniform}
\rho(\theta)=1/\pi.
\end{equation}
Substituting this probability density into definition (\ref{CD-Lyap}) and
using Eqs.~(\ref{CD-theta}), (\ref{CD-UW}) one gets

\begin{equation*}  \label{CD-LyapHomogen}
1/l_{\xi}\equiv\gamma=\sigma^2\sin^2\varphi.
\end{equation*}
In the long wave limit where the phase shift $\varphi$ is small, this result
yields the asymptotics

\begin{equation*}  \label{CD-LyapHomOmega}
l_{\xi}\approx\frac{\lambda^2}{\pi^{2}\sigma^{2}}, \ \ \ \lambda\gg 1.
\end{equation*}
This result gives rise to standard $\lambda-$dependence, $%
l_{\xi}\propto\lambda^2$ when $\lambda\to\infty$. In the case of uniform
distribution of $\delta$ over interval $[-Q_{\nu},Q_{\nu}]$ considered in
the Section~\ref{sec:meta}, it exactly coincides with the long wave
asymptotic (\ref{hlengths}) of the localization length $l$.

\begin{figure}[t!!!]
\begin{center}
\includegraphics[scale=0.5]{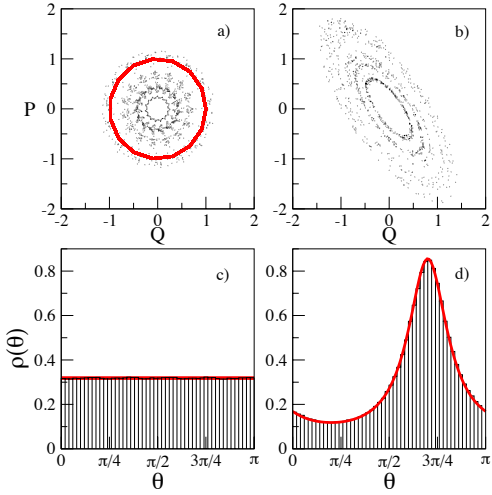}
\end{center}
\caption{(Ref.~[\onlinecite{Mak-12}], color online) a) phase space
trajectory generated by Eq.~(\protect\ref{CD-mapQP}) for H-array with $N=10^4
$, $\protect\varphi=\protect\pi/15$, for zero disorder (solid circle), and
for $\protect\sigma^2=0.003$ (scattered points); b) one trajectory for
M-array with $N=10^6$, $\protect\varphi=2\protect\pi/5$, $\protect\sigma%
^2=0.003$. c) $\protect\rho(\protect\theta)$ from Eq.~(\protect\ref{CD-mapQP}%
) for H-array (histogram), and Eq.~(\protect\ref{CD-RhoUniform}) (horizontal
line); d) $\protect\rho(\protect\theta)$ from Eq.~(\protect\ref{CD-mapQP})
for M-array (histogram), and Eq.~(\protect\ref{CD-RhoNonUniform}) (solid
curve).}
\label{EPL-Fig01}
\end{figure}

The principally different situation emerges for the M-array. In this case
the Bloch phase (\ref{CD-gamma}) is identically zero. As a result, $%
W(\theta)=-U(\theta)U^{\prime}(\theta)$ in Eq.~(\ref{CD-UW}), and Eq.~(\ref%
{CD-eqFP}) leads to a highly nonuniform phase distribution

\begin{equation}  \label{CD-RhoNonUniform}
\rho(\theta)=\frac{1}{\pi}\sqrt{\varphi^2-\sin^2\varphi}\Big/U(\theta).
\end{equation}
Fig.~\ref{EPL-Fig01} displays perfect agreement between analytical
expressions (\ref{CD-RhoUniform}), (\ref{CD-RhoNonUniform}) and data
obtained by the iteration of the exact map (\ref{CD-mapQP}).

To calculate the Lyapunov exponent via Eq.~(\ref{CD-Lyap}), one needs to
perform an average with the distribution $\rho(\theta)$ given by Eq.~(\ref%
{CD-RhoNonUniform}). Surprisingly, usage of Eqs.~ (\ref{CD-Lyap}), (\ref%
{CD-theta}) and (\ref{CD-RhoNonUniform}) results in zero Lyapunov exponent%
\cite{TIM-FNT} in the main (second order) approximation $\sim\sigma^{2}$.
Therefore, the Lyapunov exponent is determined by next orders of the
perturbation theory.

Unfortunately the direct evaluation of high order terms in $%
\rho(\theta)$ is rather cumbersome because of huge technical complexity\cite%
{TIM-FNT}. The crucial step which enables authors of   Ref.~[\onlinecite{Mak-12}] to resolve the problem is the following. It is known that  essential calculation difficulties are often related to the non-proper choice of dynamic variables. To understand how these variables should be chosen, let us analyze the numerical data displayed in
Fig.~\ref{EPL-Fig01}. The b-panel in this figure demonstrates that the
trajectory (\textit{i.e.} the sequence of points $(Q_{n},P_{n})$ has the
form of fluctuating ellipse specified by angle with respect to axes, and by
fixed aspect ratio. This results in strongly non-uniform phase distribution (d-panel in Fig~\ref{EPL-Fig01}). Therefore, one should introduce new variables $%
\widetilde{Q}_{n}$, $\widetilde{P}_{n}$ by rotating and rescaling the axes $%
Q,P$, so that the trajectory transforms into fluctuating circle. Then, one can expect that the distribution of a new phase $\Theta _{n}$ in the considered approximation will be uniform.

\begin{figure}[t!!!]
\begin{center}
\includegraphics[scale=0.55]{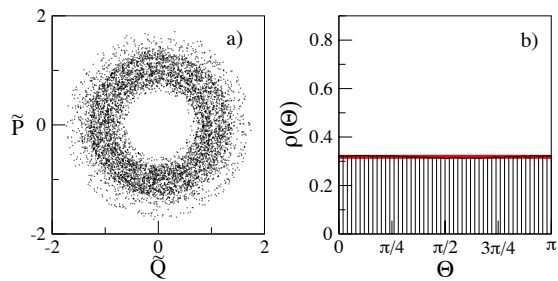}
\end{center}
\caption{(Ref.~[\onlinecite{Mak-12}], color online) (a) Phase space
trajectory in new variables $(\widetilde{Q},\widetilde{P})$; (b)
distribution $\protect\rho(\Theta)$ generated by the transformed map with
Eqs.~(\protect\ref{CD-ABCDnew}) and (\protect\ref{CD-TauAlpha}), for $%
\protect\gamma=0$, $\protect\varphi=2\protect\pi/5$, $\protect\sigma^2=0.02$
and $N=10^7$.}
\label{EPL-Fig02}
\end{figure}

To follow this recipe, let us rotate the vector $\vec{S}\to {\tilde {\vec S}}=%
\hat{R}\vec{S}$ with the help of unimodular matrix

\begin{equation*}  \label{CD-Rotation}
\hat{R}=\left\Vert%
\begin{array}{ccc}
\sqrt{\eta}\cos\tau &  & \sqrt{\eta}\sin\tau \\
&  &  \\
-\displaystyle{\frac{\sin\tau}{\sqrt{\eta}}} & & \displaystyle{\frac{\cos\tau}{\sqrt{\eta}}}
\end{array}%
\right\Vert,
\end{equation*}
where the angle $\tau$ describes rotation of the axes in $\vec{S}$-space,
with further rescaling the axes due to free parameter $\eta$. In new
coordinates the expressions (\ref{CD-mapQP}) and (\ref{CD-Lyap}) conserve
their forms, however, with the rotated transfer matrix

\begin{equation}  \label{CD-ABCDnew}
\tilde{\hat T}=\hat{R}\hat{T}\hat{R}^{-1}.
\end{equation}

\begin{equation*}  \label{CD-QP-RTheta-new}
\tilde{\vec {S}}_{n}=e^{\Xi_{n}} \left(
\begin{array}{c}
\cos\Theta_{n} \\
\\
\sin\Theta_{n}%
\end{array}
\right).
\end{equation*}
Now the distribution $\rho(\Theta)$ for new phase $\Theta$ can be found
starting from the quadratic expansion of Eq.~(\ref{CD-mapTheta}) with new
coefficients (\ref{CD-ABCDnew}) and $\phi=0$,

\begin{eqnarray}  \label{CD-ThetaNew}
\Theta_{n+1}-\Theta_{n}&=&[\eta_{\alpha}(n)-\eta_{\beta}(n)]V(\Theta_n)+
\notag \\
&&\sigma^2 V(\Theta_n)V^{\prime}(\Theta_n).
\end{eqnarray}
Here the function $V(\Theta)$ is

\begin{eqnarray}  \label{CD-V}
V(\Theta)=\sin\varphi\sin(2\tau-\varphi)\sin2\Theta  \notag \\
+\frac{\eta}{2}[\varphi -\sin\varphi\cos(2\tau-\varphi)][\cos2\Theta-1]
\notag \\
-\frac{1}{2\eta}[\varphi+\sin\varphi\cos(2\tau-\varphi)][\cos2\Theta+1].
\end{eqnarray}
The stationary Fokker-Plank equation corresponding to $\Theta$-map (\ref%
{CD-ThetaNew}) reads

\begin{equation*}  \label{CD-eqFPnew}
\frac{d}{d\Theta}\left[V^2(\Theta)\frac{d}{d\Theta}\rho(\Theta)+V(\Theta)V^{%
\prime}(\Theta)\rho(\Theta)\right]=0.
\end{equation*}
From this equation one gets that the phase distribution is uniform, $%
\rho(\Theta)=1/\pi$, and the trajectory is, indeed, a fluctuating circle
provided that

\begin{equation}  \label{CD-UniformCond}
\frac{d}{d\Theta}{V(\Theta)V^{\prime}(\Theta)}=0.
\end{equation}
With the use of Eqs.~(\ref{CD-V}) and (\ref{CD-UniformCond}), we now can
obtain the desired expressions for the angle $\tau$, parameter $\eta$ and
function $V(\Theta)$ (which is actually no more $\Theta$-dependent),

\begin{eqnarray}  \label{CD-TauAlpha}
\tau=\frac{\varphi}{2}, \ \ \ \eta^2=\frac{\varphi+\sin\varphi}{%
\varphi-\sin\varphi},  \notag \\
V(\Theta)=\sqrt{\varphi^2-\sin^2\varphi}.
\end{eqnarray}
The results presented in Fig.~\ref{EPL-Fig02} confirm success of the chosen
approach: in new variables the trajectory is a fluctuating circle and the
phase distribution is uniform.

The Lyapunov exponent $\gamma$ can be now obtained via Eq.~(\ref{CD-Lyap})
with the change $\theta_n\to\Theta_n$. Taking into account that $\gamma$
vanishes within quadratic approximation in disorder, we expand the $\Theta$%
-map of the form (\ref{CD-mapTheta}) with the coefficients (\ref{CD-ABCDnew}%
) up to the fourth order in perturbation. By substituting the resulting
expression into Eq.~(\ref{CD-Lyap}) and expanding the logarithm within the
same approximation, after the averaging over $\Theta_n$ with uniform
distribution, we arrive at final expression

\begin{eqnarray}  \label{CD-LyapMixed}
\frac{1}{l_{\xi}}=\frac{\zeta\sigma^4}{4} \frac{[(2\varphi^2-\sin^2\varphi)%
\cos\varphi-\varphi\sin\varphi]^2}{\varphi^2-\sin^2\varphi}.
\end{eqnarray}
Here the constant

\begin{equation*}  \label{CD-Zeta-New}
\zeta=\frac{\langle\delta(n)^4\rangle-\langle\delta(n)^2\rangle^2}{%
\langle\delta^2\rangle^2}
\end{equation*}
is specified by the form of distribution of $\delta_{\alpha,\beta}(n)$. For
Gaussian and flat distributions we have $\zeta=0,-6/5$, respectively.

Equation (\ref{CD-LyapMixed}) determines the asymptotics for large $%
\lambda\gg \max(\sigma,1)$,

\begin{equation*}  \label{CD-LyapMixedOmega}
\frac{1}{l_{\xi}}\equiv\gamma\approx\frac{2^4}{3^3 5^2}(\zeta+2)\sigma^4 k^8
,
\end{equation*}
that results in a quite surprising wavelength dependence of the localization
length, $l_{\xi}\propto\lambda^8$. Thus, the dependence $l_{\xi}\propto%
\lambda^6$, numerically found for large $\lambda$ in Refs.~\cite%
{Asatryan07,Asatryan10a} and confirmed later in \cite{} should be regarded
as the intermediate one, apparently emerging due to not sufficiently large
lengths $N$ over which the average of $\gamma$ is performed.


\section{Localization in Complex Media}

\label{sec:novel} \numberwithin{equation}{section}


\subsection{Nonreciprocal Transmission in Magnetoactive Optical Structures}

\label{subsec:nonreciprocal}

In this Subsection we present the results of analytical and numerical study
of the Anderson localization of light propagating through random
magnetoactive layered structures. We demonstrate that an interplay between
strong localization and magnetooptical effects produces a number of
non-reciprocity features in the transmission characteristics.

Magnetooptical effects and nonreciprocity are widely exploited in modern
optics and applied physics~\cite{Zvezdin,Potton}. In particular,
magnetoactive periodic structures are currently attracting growing attention
\cite{Lyubchanskii,Inoue}. The main phenomena of interest are the enhanced
Faraday effect on resonances~\cite{InoueAraiFujii} and one-way propagation
(nonreciprocal transmission) \cite%
{FigotinVitebsky,YuWangFan,Khanikaev,KhanikaevSteel} employed for the
concept of optical insulators. The resonant Faraday effect has also been
shown in connection with the localization of light in random layered
structures~\cite{InoueFujii}.

Here we examine the transmission properties of one-dimensional random
layered structures with magneto-optical materials. We employ
short-wavelength approximation, where the localization is strong, and
consider both Faraday and Voigt geometries. In the Faraday geometry,
magneto-optical correction to the localization length $l$ results to a
significant broadband non-reciprocity and polarization selectivity in the
typical, exponentially small transmission. In the Voigt geometry, averaging
over random phases suppresses the magneto-optical effect, in contrast to the
case of periodic structures where it can be quite pronounced~\cite%
{FigotinVitebsky,KhanikaevSteel}. At the same time, in both the geometries
we reveal the nonreciprocal frequency shifts of narrow transmission
resonances, corresponding to the excited localized states inside the
structure \cite{Frisch,BBF,Bliokh-2,we-PRL-1}. This offers efficient
unidirectional propagation at the given resonant frequency.

Consider the light transmission through the long stack composed with
magnetooptical materials in the short-wavelength approximations. In the
localized regime, we can neglect in Eq. (\ref{total transf matr}) the
external interface transfer matrices ${\hat F}^{0\alpha}$, ${\hat F}^{\beta
0}$ just replacing the exact matrix $\hat{T}$ by the truncated matrix $\hat{T%
}^{\prime}$

\begin{equation}  \label{truncated}
\hat{T}^{\prime}={\hat F}_{N}{\hat S}_{N}{\hat F}_{N-1}{\hat S}_{N-1}{\hat F}%
_{N-2}~...~{\hat F}_{2} {\hat S}_{2}{\hat F}_{1}{\hat S}_{1}.
\end{equation}
Then, if the wavelength within the $k$-th layer is much shorter than the
variance of the layer thickness \cite{BerryKlein}, then the phases $\varphi_k
$ modulo $2\pi$ in the propagation matrices ${\hat S}_{j}$ (\ref{space}) are
independent and nearly uniformly distributed in the range $(0,2\pi)$. In
this approximation, the transmittance corresponding to the transfer matrix (%
\ref{truncated}) after averaging over all phases $\varphi_k$, is reduced to
the product of the transmittances of separate layers \cite{Baluni} and,
furthermore, to the product of transmittances of the interfaces only \cite%
{BerryKlein}

\begin{equation}  \label{factorization}
\ln \left({\mathcal{T}}\right)\approx \sum_{j=1}^{2N} \ln
\tau_j~,~~\tau_j=1/|({\hat F}_j)_{11}|^2.
\end{equation}
Substitution of Eq.~(\ref{factorization}) into Eq.~(\ref{LL-0}) in the limit
$N\to\infty$ yields the simple expression for the localization length

\begin{equation}  \label{transm decrement-1}
\frac{1}{l}=\approx\frac{1}{2} {\ln\left|\left({\hat F}^{\alpha\beta}%
\right)_{11}\left({\hat F}^{\beta\alpha}\right)_{11}\right|}.
\end{equation}
in the short-wavelength approximation.

This result can be easily extended to any number of alternating layers. For
instance, considering a random structure consisting of three types of
alternating layers, `$\alpha$', `$\beta$', and `$\gamma$', one has

\begin{equation*}  \label{3-layer}
\frac{1}{l}=\frac{1}{3} {\ln\left|\left({\hat F}^{\alpha\beta}\right)_{11}%
\left({\hat F}^{\beta\gamma}\right)_{11}\left({\hat F}^{\gamma\alpha}%
\right)_{11}\right|}.
\end{equation*}

\begin{figure}[tbh]
\centering
\includegraphics[width=0.9\columnwidth]{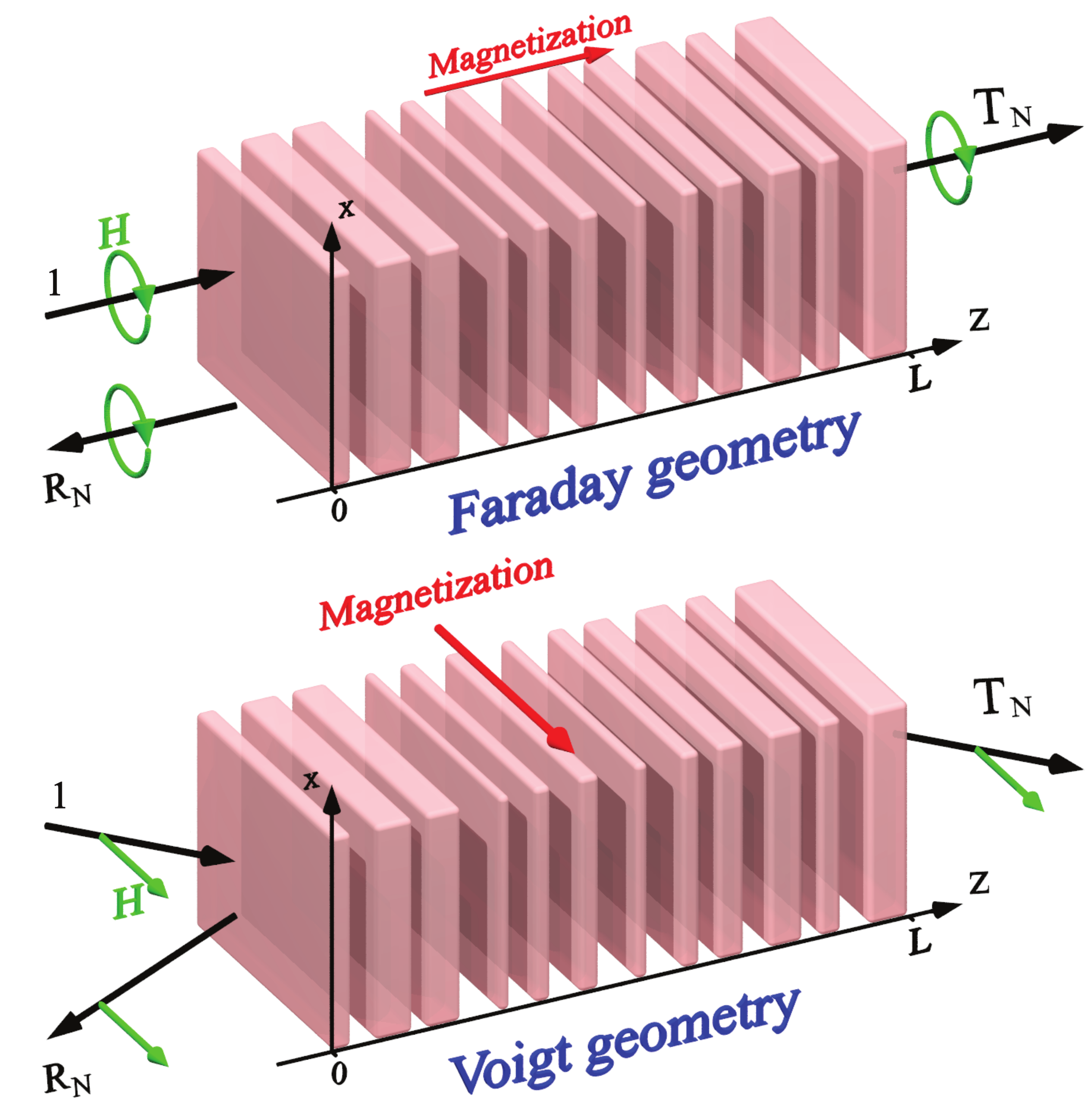} 
\caption{(Ref.~[\onlinecite{Bliokh-mag}], color online.) Schematic picture
of the wave transmission and reflection from a random-layered structure
consisting of two types of alternating layers `$\protect\alpha$' (here --
magnetoactive material) and `$\protect\beta$' (here -- air) with random
widths. Magnetization of the medium, wave polarizations and directions of
propagation are shown for the Faraday and Voigt geometries.}
\label{Mag-stack}
\end{figure}

Transmission through a one-dimensional lossless linear medium is always
reciprocal if there is only one (but propagating in two directions) mode in
the system. Indeed, while the forward transmission of the wave incident from
the left on the medium is described by the $2\times 2$ transfer matrix ${%
\hat T}$ with transmission coefficient $T$ and transmittance ${\mathcal{T}}$%
, the backward transmission of the reciprocal wave incident from the right
is characterized by the inverse transfer matrix ${\hat T}^{-1}$ with the
same transmission coefficient and transmittance\cite{Baluni,BerryKlein}.

If the system possesses two or more uncoupled modes labeled by index $%
\varsigma$, the waves are marked by the propagation direction $\upsilon$ and
mode indices: $h^{\upsilon,\varsigma}$. Still, the forward and backward
propagation of each mode $\varsigma$ through the system with incident waves
of types $(+,\varsigma)$ and $(-,\varsigma)$ are described by the $2\times 2$
transfer matrices ${\hat T}^{\varsigma}$ and $({\hat T}^{\varsigma})^{-1}$
characterized by the same transmittance $\mathcal{T}^{\varsigma}$. However, the wave
reciprocal to $(+,\varsigma)$ is determined by the \textit{time-reversal
operation} which changes $\upsilon \to -\upsilon$ (because of the $\mathbf{k}%
\to -\mathbf{k}$ transformation) but can also affect $\varsigma$ \cite%
{Potton}. In particular, if the time reversal operation changes the sign of
the mode index: $\varsigma \to -\varsigma$, then the \textit{reciprocal}
wave will be $(-,-\varsigma)$ rather than the \textit{backward} wave of the
same mode, $(-,\varsigma)$. Accordingly, the transmittances of the mutually
reciprocal waves through the system, $\mathcal{T}^\vs$ and $\mathcal{T}%
^{-\varsigma}$, can be different. This signals \textit{nonreciprocity} in
the system.

Non-reciprocity in the system under consideration originates from the
difference between the modes $\varsigma$ and $-\varsigma$, and does not
depend explicitly on the direction of incidence $\upsilon$. Therefore, in
practice, it is sufficient to compare only \textit{forward} transmissions of
the modes $\pm\varsigma$, described by the transfer matrices ${\hat T}%
^{\pm\varsigma}$ and transmittances $\mathcal{T}^{\pm\varsigma}$.

There are two main geometries typical for magneto-optical problems \cite%
{Zvezdin}: the Faraday geometry, where the magnetization is collinear with
the direction of propagation of the wave, and the Voigt (or Cotton-Mouton)
geometry, where the magnetization is orthogonal to the direction of
propagation of the wave (see Fig.\ref{Mag-stack}). Below we study  the
averaged transmission decrement and individual transmission resonances in
both geometries and show that propagation of light in disordered
magnetoactive layered media offers nonreciprocal transmission.

In the Faraday geometry both magnetization and the wave vector are directed
across the layers, i.e., along the $z$-axis (see Fig.\ref{Mag-stack}). We
assume that the magnetic tensor is equal to one and the magneto-optical
effects are described exclusively by the dielectric tensor which in the
Faraday geometry has the form\cite{Zvezdin}

\begin{equation*}  \label{vareps}
\hat{\varepsilon}= \left\Vert
\begin{array}{ccc}
\varepsilon & -iQ & 0 \\
iQ & \varepsilon & 0 \\
0 & 0 & \varepsilon \\
&  &
\end{array}
\right\Vert.
\end{equation*}

The eigenmodes of the problem are circularly polarized waves of magnetic $%
\mathbf{H}$

\begin{equation}  \label{magnetic}
\mathbf{H}^{\upsilon,\varsigma}=\frac{H^{\upsilon,\varsigma}}{\sqrt{2}}
\left(
\begin{array}{c}
1 \\
i\varsigma \\
0 \\
\end{array}
\right){\text{e}}^{ i(\upsilon kz-\omega t)}, \ \ \ \upsilon,\varsigma=\pm 1,
\end{equation}
and electric $\mathbf{E}$

\begin{equation}  \label{electric}
\mathbf{E}^{\upsilon,\varsigma}= i\upsilon\varsigma \frac{k_0}{k}\mathbf{H}%
^{\upsilon,\varsigma}.
\end{equation}
fields. Here $H^{\upsilon,\varsigma}$ ($E^{\upsilon,\varsigma}$) are the
wave amplitudes, whereas $k$ is the propagation constant affected by the
magnetization parameter $q$ and depending on $\varsigma $.

\begin{eqnarray}  \label{wavenumbers}
k=nk_{0}\sqrt{1+\varsigma q}, \ \ \ n=\sqrt{\varepsilon},  \notag \\
k_{0}=\frac{\omega}{c}, \ \ \ \ \ \ q=\frac{Q}{\varepsilon}.
\end{eqnarray}
In the linear approximation in $q$, $k\simeq nk_{0}(1+\varsigma q/2)$.

Parameter $\varsigma$ is the mode index which determines the direction of
rotation of the wave field. In this manner, the product $\upsilon\varsigma$
represent the helicity

\begin{equation*}  \label{helicity}
\chi=\upsilon\varsigma,
\end{equation*}
which distinguishes the right-handed ($\chi=+1$) and left-handed ($\chi=-1$)
circular polarizations defined with respect to the direction of propagation
of the wave. Note that the time reversal operation keeps helicity unchanged,
whereas $\varsigma$ changes its sign \cite{Potton}. Thus, the reciprocal
wave is given by $\mathbf{H}^{-\upsilon,-\varsigma}$, precisely as described
above.

The total field in a layer is the sum $\mathbf{H}^{+,\varsigma}+\mathbf{H}%
^{-,\varsigma}$ of the eigenvectors (\ref{magnetic}) with the amplitudes $%
H^{\pm,\varsigma}.$ Consider the wave transformation at the interface
between the media `$a$' and `$b$'. The helicity of the wave flips upon the
reflection and remains unchanged upon transmission. As a result, parameter $%
\varsigma$ remains unchanged, so that there is no coupling between the modes
with $\varsigma=+1$ and $\varsigma=-1$ (see Fig.\ref{Mag-stack}), and these
modes can be studied independently. From now on, for the sake of simplicity,
we omit $\varsigma$ in superscripts and write explicitly only the values of
the direction parameter $\upsilon=\pm 1$.

Using the standard boundary conditions for the wave electric and magnetic
fields at the `$\alpha$'-`$\beta$' interface for the normalized fields

\begin{eqnarray*}  \label{normalized}
\vec{h}=\frac{k_{0}}{k} \left(%
\begin{array}{c}
H^{+} \\
\\
H^{-}%
\end{array}
\right), \ \ \ \ \ \ \ \ \vec{h}_{\alpha}={\hat F}^{\alpha\beta} \ \vec{h}%
_{\beta}
\end{eqnarray*}
with the normalized interface transfer matrix

\begin{equation}  \label{normalized matrix-Faraday}
{\hat F}^{\alpha\beta}=\frac{1}{2\sqrt{k_{\alpha} k_{\beta}}} \left\Vert
\begin{array}{cc}
\displaystyle{k_{\beta}+k_{\alpha}} & \displaystyle{k_{\beta}-k_{\alpha}} \\
\displaystyle{k_{\beta}-k_{\alpha}} & \displaystyle{k_{\beta}+k_{\alpha}} \\
&
\end{array}
\right\Vert,
\end{equation}
where $k_{\alpha,\beta}$ are the wave numbers (\ref{wavenumbers}) in the
corresponding media.

Calculating the localization decrement from Eq. (\ref{transm decrement-1})
with Eqs.~(\ref{wavenumbers}) and (\ref{normalized matrix-Faraday}), we
obtain in the linear approximation in $q$:

\begin{eqnarray}  \label{transm decrement-2}
&&\kappa=2\ln\frac{k_{\alpha}+ k_{\beta}}{2\sqrt{k_\al k_\be}}\simeq
\kappa^{(0)}+\kappa^{(1)},  \notag \\
&&\kappa^{(0)}\!=\!\ln\frac{(n_\al+n_\be)^2}{4n_\al n_\be},  \notag \\
&&\kappa^{(1)}\!=\!\frac{\sigma}{2} (q_\al-q_\be)\frac{n_\al-n_\be}{%
n_\al+n_\be}.
\end{eqnarray}
Thus, the localization decrement acquires the first-order magneto-optical
correction $\kappa^{(1)}$ caused by the Faraday effect. This correction
depends on $\varsigma$, i.e., on the polarization helicity $\chi$ and the
propagation direction $\upsilon$ through $\varsigma=\chi\upsilon$. For the
reciprocal waves with the same $\chi$ and opposite $\upsilon$, $\kappa^{(1)}$
has opposite signs. This signals nonreciprocal localization in a Faraday
random medium. In practice, the nonreciprocal difference in the transmission
decrements (\ref{transm decrement-2}) can be observed by changing sign of
either propagation direction $\upsilon$ (with the helicity fixed), or
polarization $\chi$, or magnetization $q$.

\begin{figure}[t]
\centering
\includegraphics[width=0.8\columnwidth]{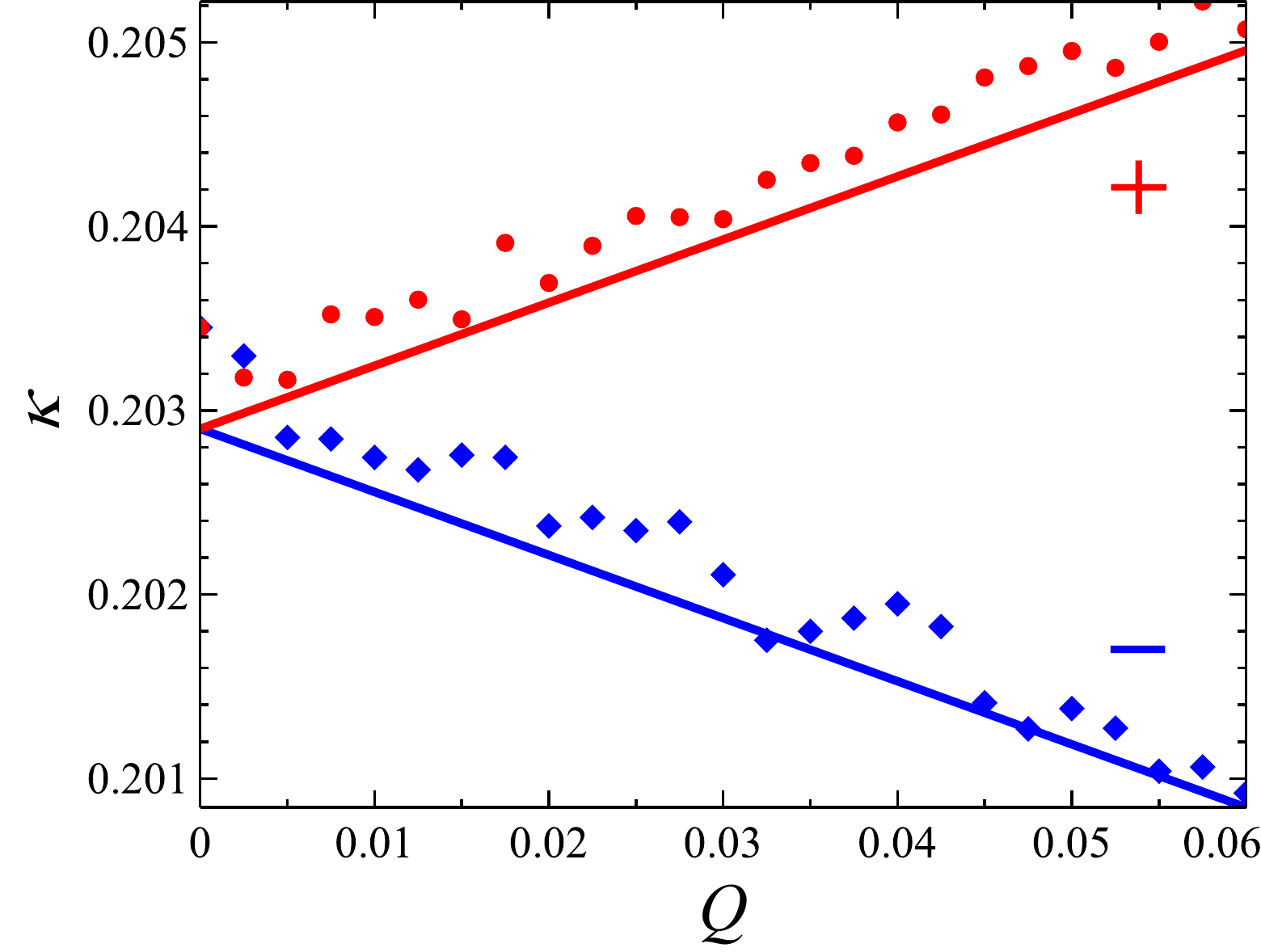}
\caption{(Ref.~[\onlinecite{Bliokh-mag}], color online) Localization
decrement $\protect\kappa$ vs. magneto-optical parameter $Q$ for opposite
modes propagating through a two-component random structure in the Faraday
geometry (see details in the text). The modes with $\protect\varsigma=\pm 1$
correspond to either opposite circular polarizations or propagation
directions. Numerical simulations of exact equations (symbols) and
theoretical formula (\protect\ref{transm decrement-2}) (lines).}
\label{Mag-Fig2}
\end{figure}

Despite the magneto-optical correction to the localization decrement is
small in magnitude, $\kappa^{(1)} \ll \kappa^{(0)}$, it still might result
in a significant difference in the typical transmission spectrum. This
difference is described by an additional factor of $\propto \exp[%
-2N\kappa^{(1)}]$ in transmittance, which is exponential with respect to the
length of the structure. Hence, small correction (\ref{transm decrement-2})
brings about significant broadband nonreciprocity or polarization
selectivity in the typical small transmission when $N\left|\kappa^{(1)}%
\right|\geq 1$.

Fig.~\ref{Mag-Fig2} shows dependence of the localization decrement on the
magnetization parameter $Q=\varepsilon q$ calculated numerically and
compared to analytical result (\ref{transm decrement-2}). Numerical
simulations were performed for the structure containing ${\mathcal{N}}=2N=90$
alternating layers of air ($\varepsilon = 1, Q=0$), and bismuth iron garnet
(BIG), with dielectric constant $\varepsilon = 6.25$ and magneto-optic
parameter reaching $Q=0.06$. The thicknesses of layers were randomly
distributed in the range 50$\div$150 $\mu$m (i.e., $\bar{w}=100\mu$m, $%
d=50\mu$m), whereas the excitation wavelength was 632 nm. The averaging was
performed over $10^5$ realizations of the random sample. One can see
excellent agreement between numerical simulations and analytical results
showing linear splitting of the $\varsigma=1$ and $\varsigma=-1$
localization decrements as a function of the magneto-optic parameter.\newline

In the Voigt geometry, the dielectric tensor is\cite{Zvezdin}

\begin{equation*}  \label{varepsilon-V}
\hat{\varepsilon}=\varepsilon\left\Vert
\begin{array}{ccc}
1 & 0 & iq \\
0 & 1 & 0 \\
-iq & 0 & 1 \\
&  &
\end{array}
\right\Vert.
\end{equation*}
The first-order interaction of the wave with the magnetization occurs only
upon \textit{oblique} propagation of the wave in the $xz$-plane, i.e., when $%
k_{x}=\text{const} \neq 0$ (see Fig.\ref{Mag-stack}).

The eigenmodes of the problem are the TE mode which is uncoupled from the
magnetization, and TM mode with the tangential components

\begin{eqnarray*}  \label{fields}
H_y^{\upsilon,\varsigma} &=& H^{\upsilon,\varsigma} \ \text{e}^{i(\varsigma
x k_{\perp} + \upsilon z k_{\parallel}-\omega t)},  \notag \\
E_x^{\upsilon,\varsigma} &=& A^{\upsilon,\varsigma} H_y^{\upsilon,\varsigma}.
\end{eqnarray*}

Here parameters $\upsilon=\pm 1$ and $\varsigma=\pm 1$ indicate propagation
in the positive and negative $z$ and $x$ directions, respectively, $%
k_{\parallel}=\sqrt{k^2-k_x^2}$, $k_{\perp}=|k_x|$, whereas

\begin{eqnarray}  \label{wavenumbers-1}
A^{\upsilon,\varsigma}&=&-\left(A^{-\upsilon,\varsigma}\right)^{*}=\frac{%
i\varsigma q k_{\perp}+ \upsilon k_{\parallel}}{\varepsilon(1-q^2)k_0}, \\
k&=&nk_{0}\sqrt{1-q^{2}} .  \notag
\end{eqnarray}
In the linear approximation in $q$, $A^{\upsilon,\varsigma}\simeq (\upsilon
k_{\parallel} +i\varsigma q k_{\perp})/(\varepsilon k_0)$ and $k\simeq nk_0$
so that the magnetization affects imaginary parts (i.e., phases) of the
amplitudes $A^{\upsilon,\varsigma}$ and does not affect the propagation
constant, cf. Eqs.~(\ref{electric}) and (\ref{wavenumbers}).

In the Voigt geometry, direction of the transverse wave vector component, $%
\varsigma$, serves as the mode index. The mutually reciprocal waves are $%
H^{\upsilon,\varsigma}$ and $H^{-\upsilon,-\varsigma}$ because the time
reversal transformation reverts the whole wave vector, $\mathbf{k}\mapsto -%
\mathbf{k}$.

The parameter $\varsigma$ is not changed upon reflection and transmission
through the layers, i.e., the modes with $\varsigma=\pm 1$ are uncoupled
from each other. Therefore, for the sake of simplicity, we omit the mode
index in superscripts, and write explicitly only the values of the direction
parameter $\upsilon = \pm 1$.

Using the standard boundary conditions for the wave electric and magnetic
fields at the `$\alpha$'-`$\beta$' interface, for the normalized interface
transfer matrix ${\hat F}^{\alpha \beta}$ we obtain\cite{KhanikaevSteel}

\begin{eqnarray*}  \label{normalized matrix-Voigt}
{\hat F}^{\alpha \beta}=\frac{1}{\sqrt{ 4\mathrm{Re}A_{\alpha}^{+}\mathrm{Re}%
A_{\beta}^{+}}}\times  \notag \\
\left\Vert
\begin{array}{cc}
\displaystyle{A_{\beta}^{+}+A_{\alpha}^{+*}} & \displaystyle{%
A_{\alpha}^{+*}-A_{\beta}^{+*}} \\
\displaystyle{A_{\alpha}^{+}-A_{\beta}^{+}} & \displaystyle{%
A_{\alpha}^{+}+A_{\beta}^{+*}} \\
&
\end{array}
\right\Vert.
\end{eqnarray*}

In contrast to the Faraday geometry, in the Voigt geometry the linear
magneto-optical correction changes only phases of the transmission and
reflection coefficients, whereas corrections to the interface transmittance
start with the terms $\propto q^2$. In short-wave limit, only these
transmittances determine the total transmittance, Eq.~(\ref{factorization}).
Therefore, a short-wavelength transmission through a random multilayered
stack is reciprocal and is not affected by magnetization in the first-order
approximation. In the short-wave limit, this statement remains true for any
number of types of alternating layers. It was verified numerically for the
three-layer system\cite{Bliokh-mag}. At the same time, a periodic structure
with a cell consisting of three different layers (which breaks the mirror
reflection symmetry) can demonstrate significant nonreciprocity \cite%
{YuWangFan,KhanikaevSteel} but beyond the short-wave approximation.


\begin{figure}[t]
\centering
\includegraphics[width=0.83\columnwidth]{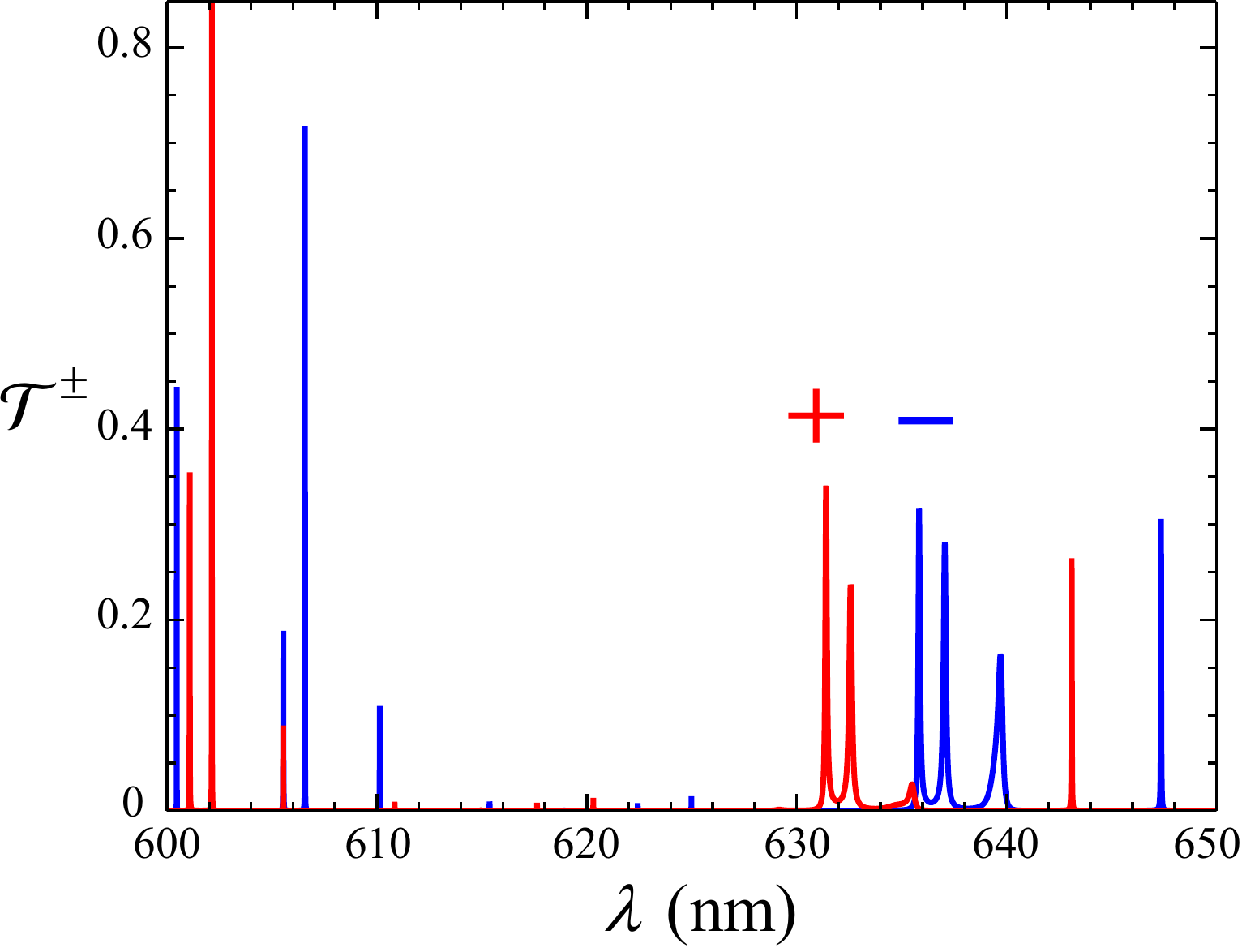}
\caption{(Ref.~[\onlinecite{Bliokh-mag}], color online.) Transmission
spectra of a random magneto-optical sample in the Faraday geometry (see
details in the text) for waves with $\protect\varsigma=\pm 1$. While the
averaged localization decrements are only slightly different (Fig.\protect
\ref{Mag-Fig2}), all individual resonances are shifted significantly as
compared with their widths, Eq.~(\protect\ref{frequency shift-1}).}
\label{Mag-faraday-spectrum}
\end{figure}

Averaged localization decrement is associated with exponential decay of the
incident wave deep into the infinite sample \cite%
{Baluni,Ping91,FG,BerryKlein}. For a finite sample, this is so only for
typical realizations. However, there exist some resonant realizations of the
sample at a given frequency (or, equivalently, resonant frequencies for a
given realization) where transmission is anomalously high and is accompanied
by the accumulation of energy inside the sample. \cite{Lifshits,Frisch,BBF}
Such resonant transmission corresponds to excitation of the Anderson
localized states (quasi-modes) inside the sample. Akin to the resonant
localized states in photonic crystal cavities, the transmission resonances
in random structures are extremely sensitive to small perturbations:
realization\cite{Lifshits}, absorption,\cite{BBF} nonlinearity, \cite%
{Shadrivov} and, as we show here, magnetoactivity.

\begin{figure}[h]
\centering
\includegraphics[width=0.95\columnwidth]{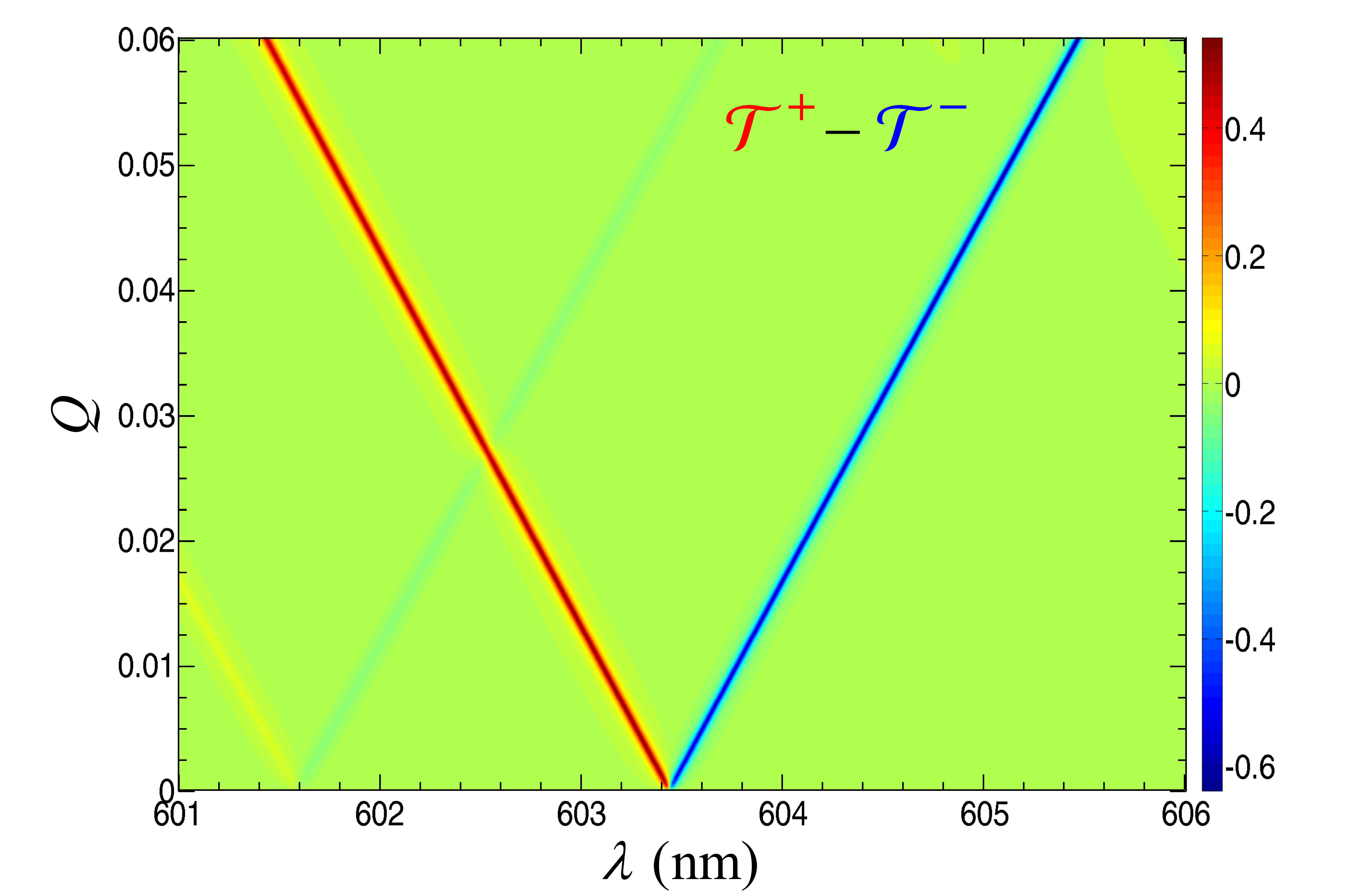}
\caption{(Ref.~[\onlinecite{Bliokh-mag}], color online.) Differential
transmittance, ${\mathcal{T}}^{+}-{\mathcal{T}}^{-}$, for two resonances
from Fig.\protect\ref{Mag-Fig2} as dependent on the value of magneto-optical
parameter $Q$, cf. Eq.~(\protect\ref{frequency shift-1}).}
\label{Mag-faraday-diff-transm}
\end{figure}

Figure~\ref{Mag-faraday-spectrum} shows transmission spectra for two modes $%
\varsigma=\pm 1$ (i.e., either with opposite helicities or propagation
directions) in one realization of a magnetooptical sample in the Faraday
geometry. The parameters of the sample are the same as in Section IIIA with $%
Q=0.06$. One can see strong splitting of the $\varsigma=\pm 1$ transmission
resonances which have exponentially narrow widths\cite{BBF} $%
\propto\kappa\exp (-\kappa N)/2\bar{w}$ . This offers strongly
nonreciprocal, practically unidirectional, propagation or polarization
selectivity in the vicinity of resonant frequencies.

To estimate the splitting of resonances, we note that the wavenumbers in
magnetooptical materials are shifted due to the Faraday effect, Eq.~(\ref%
{wavenumbers}). Hence, the shifts of the resonant wavenumbers of the random
Faraday medium can be estimated by averaging of this shift over different
materials in the structure:

\begin{equation}  \label{frequency shift-1}
\Delta k_{\mathrm{res}} \simeq \varsigma\,\overline{\frac{qnk_0}{2} },
\end{equation}
where $\overline{(...)}$ stands for some average of $(...)$. Using $%
\overline{qn}\sim (q_a n_a + q_b n_b)/2$ for estimation in the two-component
structure, we obtain $\Delta \lambda_{\mathrm{res}} \sim -\varsigma \,3.6$
nm, which agrees with the $\varsigma$-dependent splitting observed in Fig.~%
\ref{Mag-faraday-spectrum}.

Figure~\ref{Mag-faraday-diff-transm} displays the differential transmission
for the waves with $\varsigma=+1$ and $\varsigma$ lying in a narrow
frequency range in Fig.~\ref{Mag-faraday-spectrum}. In agreement with
estimation (\ref{frequency shift-1}), one observes the linear dependence of
the resonance splitting on magnetization.

In the Voigt geometry, the resonances also allow nonreciprocal transmission
and demonstrate splitting of the resonant frequencies. In Fig.~25, the
differential transmission is shown for reciprocal waves with $\varsigma=\pm 1
$ in the vicinity of one resonance for the three-component structure
considered in Section IIIB. The splitting is very small in this case, and $%
\varsigma=+1$ and $\varsigma=-1$ resonances overlap significantly. Because
of this, the differential transmittance in Fig.~25 is tiny, its amplitude
linearly grows with $Q$, whereas the frequency positions of its maximum and
minimum correspond to the width of the original resonance and are
practically unchanged.

\begin{figure}[h]
\label{Voigt-split} \centering
\includegraphics[width=0.95 \columnwidth]{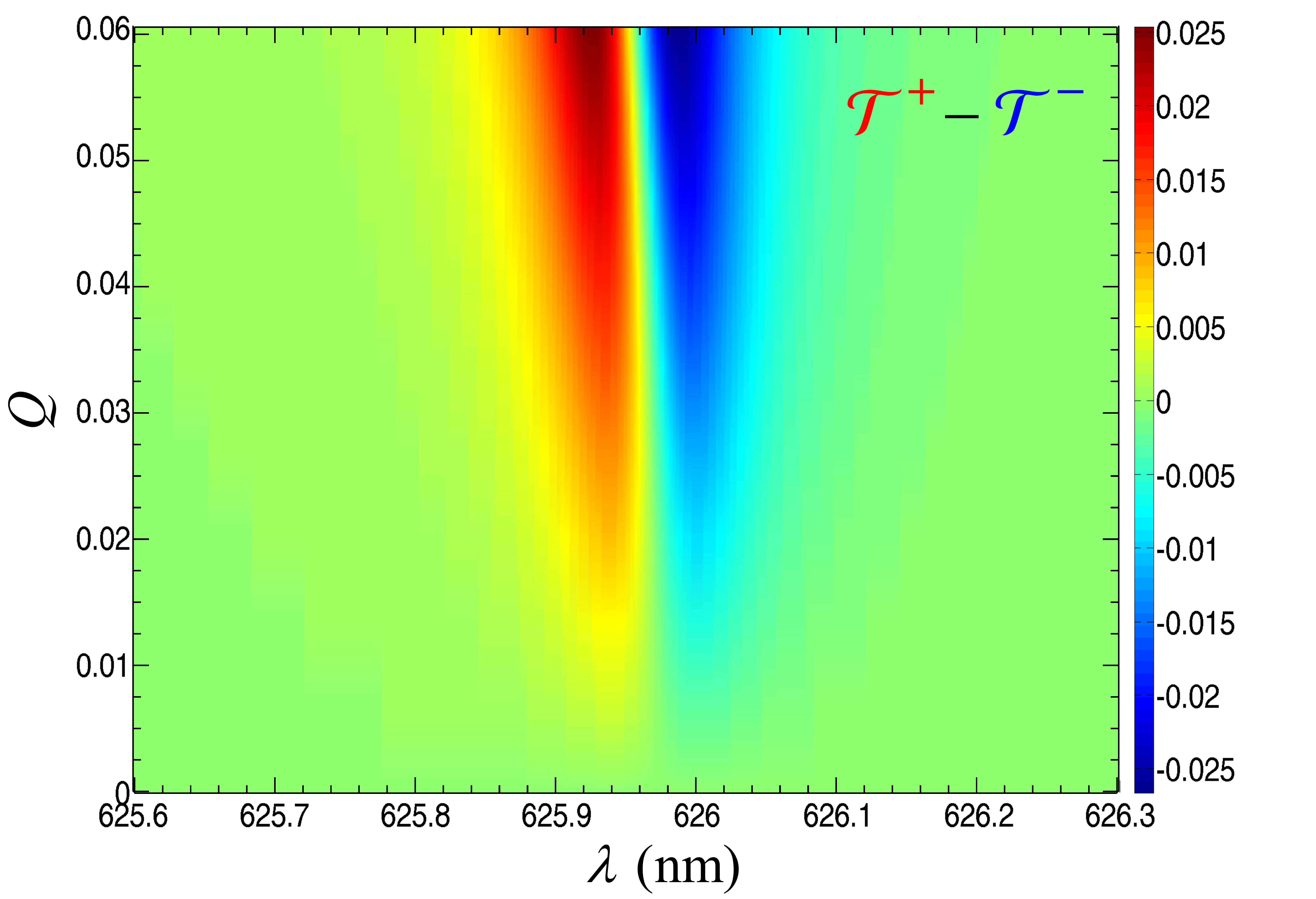}
\caption{(Ref.~[\onlinecite{Bliokh-mag}], color online.) Differential
transmittance, ${\mathcal{T}}^{+}-{\mathcal{T}}^{-}$, for in the vicinity of
a single resonance in the Voigt geometry (see Section IIIB for details) as
dependent on the magneto-optical parameter $Q$.}
\end{figure}

Unlike the wave-number shift in the Faraday geometry, the nonreciprocal
shift of resonant frequencies in the Voight geometry arises from the phases
of the amplitudes $A$, Eq.~(\ref{wavenumbers-1}). These phases are
responsible for the phases of transmission coefficients between the layers
and can be estimated as $\phi \sim q (\varsigma k_{\perp})/(\upsilon
k_{\parallel}) \equiv q \tan\theta$, where $\theta$ is the angle of
propagation with respect to the $z$-axis. The phases accumulated at a layer
effectively shift the wave numbers as $\upsilon \Delta k_{\parallel} =\Delta
k \cos\theta\sim \phi/w$, where $w$ is the thickness of the layer. Averaging
over different materials in the random layered structure, we estimate the
nonreciprocal shift of the resonant wave number:

\begin{equation*}  \label{frequency shift-2}
\Delta k_{\mathrm{res}} \sim \overline{\frac{q\sin\theta}{w\cos^2\theta}}=
\varsigma\,\overline{\frac{q|\sin\theta|}{w\cos^2\theta}}.
\end{equation*}
This shift is $\varsigma$-dependent, i.e., nonreciprocal, and much smaller
than the Faraday-geometry shift (\ref{frequency shift-1}) as $k\bar{w}>kd\gg
2\pi$ in the short-wavelength limit. For the parameters in use, with $Q=0.06$%
, we have $\Delta \lambda_{\mathrm{res}} \sim -\varsigma \,3\cdot 10^{-4}$%
nm, which agrees with the data plotted in Fig.~25.


\subsection{Charge Transport in Disordered Graphene}

\label{subsec:graf} 

Shortly after the discovery of highly unusual physical properties of
graphene it was realized that the electron transport in this material had
many common features with the propagation of light in dielectrics.
Historically, the analogy between Maxwell equations and those used in the
relativistic electron theory has been discussed in different contexts and
for various purposes (see, for example, \cite{BiBi,Zalesny,BFSN,Mehrafarin})
since 1907 when Maxwell equations were reduced~\cite{Silberstein} to an
alternative, more concise form by introducing a complex field $\mathbf{F}=%
\mathbf{E}+i\mathbf{H}$:

\begin{equation}  \label{Silberstein}
c\hat{\boldsymbol{\rho}}\cdot\nabla\Psi=-n\partial\Psi/\partial t,
\end{equation}
where $\Psi$ is the 4-vector with components $-F_x+iF_y$, $F_z$, $F_z$, $%
F_x+iF_y$, $n$ is the refraction index, and the components of the 3-vector $%
\hat{\boldsymbol{\rho}},$ are the Dirac matrices $\hat{\rho_{i}},$ $i=1,2,3$
(Pauli matrices in which the units are replaced by the unit $2\times 2$
matrices).

In the last few yeas, this activity perked up due to the recent developments
in the physics of graphene. Nowadays it is well understood that under some
(rather general) conditions, Dirac equations describing the charge transport
in a graphene superlattice created by applying inhomogeneous external
electric potential could be reduced to Maxwell equations for the propagation
of light in a dielectric medium. To better understand the physics of charge
transport in graphene subject to a coordinate-dependent potential, in what
follows, we compare the results for graphene with those for the propagation
of light in layered dielectric media (for more analogies between quantum and
optical systems, see, e.g. Ref.\cite{Dragoman, analogy}). Additional
analogies, not discussed here, also exist with the transport and
localization of phonons in different kinds of periodic and random
one-dimensional structures \cite{Nori, Nori_1, Nori_2}.

As it was shown above, the light transport of electromagnetic waves in
multilayered media is described in terms of the transfer matrices of two
types. The first type is formed by diagonal matrices $\hat{S}_{j}$
corresponding to the propagation of wave through the $j$-th layer. These
matrices are the same as in Eq.~(\ref{space}) (up to the signs of the
exponent). The second type is formed by the interface transfer matrices $%
\hat{F}_{j,j+1}$ describing transformation of the amplitudes of the
electromagnetic waves at the interface between $j-$th and $(j+1)-$th layers
and having the form

\begin{equation}  \label{matrix3}
\hat{F}_{j,j+1}={\frac{1}{2\cos \theta _{j+1}}}\left\|
\begin{array}{cc}
G_{j,j+1}^{(+)} & G_{j,j+1}^{(-)} \\
G_{j,j+1}^{(-)} & G_{j,j+1}^{(+)}%
\end{array}%
\right\|,
\end{equation}%
where

\begin{equation}  \label{RLmatrix-1}
G_{j,j+1}^{(\pm )}=\cos \theta _{j+1}\pm \cos \theta _{j}\cdot \mathrm{sgn}%
(\nu_{j}\nu_{j+1}){\frac{Z_{j+1}}{Z_{j}}}
\end{equation}%
for $s$-polarized waves and

\begin{equation}
G_{j,j+1}^{(\pm )}={\frac{Z_{j+1}}{Z_{j}}}\cos \theta _{j+1}\pm \cos \theta
_{j}\cdot \mathrm{sgn}(\nu _{j}\nu _{j+1})  \label{RLmatrix-2}
\end{equation}%
for $p$-polarized waves. Here, $\theta _{j}$ is the angle of the propagation
within the $j-$th layer, $Z_{j}$ and $\nu _{j}$ are the impedance and the
refractive index of the $j$th layer defined by Eq.~(\ref{ImpRefInd}). Signs $%
\pm $ correspond, respectively, to $R-$ and $L-$ dielectric layers with
positive and negative refractive indices.

In the case of the charge transport in a graphene superlattice created by a
piecewise-constant electrostatic potential depending on one coordinate $x$
in the plane $(x,y)$ of the graphene layer, the analogues transfer matrix,
which describes the transition through the interface between adjacent
regions with different values of the potential, has the form \cite{BFSN}

\begin{equation}  \label{matrix2}
\hat{\mathcal{F}}_{j,j+1}={\frac{1}{2\cos \theta _{j+1}}}\left\Vert
\begin{array}{cc}
\mathcal{G}_{j,j+1}^{(+)} & \mathcal{G}_{j,j+1}^{(-)} \\
(\mathcal{G}_{j,j+1}^{(-)})^{\ast } & (\mathcal{G}_{j,j+1}^{(+)})^{\ast }%
\end{array}%
\right\Vert ,
\end{equation}%
where

\begin{equation}
\mathcal{G}_{j,j+1}^{(\pm )}=e^{-i\theta _{j+1}}\pm e^{\pm i\theta
_{j}}\cdot \mathrm{sgn}[(\varepsilon -u_{j})(\varepsilon -u_{j+1})],
\label{matrix2-a}
\end{equation}%
Here the $\theta _{j}$ is given by equation $\tan \theta _{j}=\beta /\sqrt{%
(\varepsilon -u_{j})^{2}-\beta ^{2}}$ where $\beta $ is the projection of
the dimensionless momentum on $y$ axis, $\varepsilon$ and $u_j$ are the
dimensionless energy of the charge carrier and the scalar potential of the $j
$-th layer. If $\theta _{j}$ is real, it coincides with the angle of the
propagation of electron within the $j$-th layer.

Comparison of Eqs. (\ref{matrix3}) and (\ref{matrix2}) shows that the role
of the refractive index $\nu $ in graphene is played by the difference $%
\varepsilon -u$. In particular, a layer, in which the potential exceeds the
energy of the particle, $u>\varepsilon $, is similar to a $L$-slab with
negative refractive index (metamaterial), while a layer where $u<\varepsilon
$, is similar to normal material. It is due to this similarity that a
junction of two regions having opposite signs of $u-\varepsilon $ (so-called
\textit{p-n} junction) focuses Dirac electrons in graphene\cite{Cheianov},
in the same way as an interface between left- and right-handed dielectrics
focuses electromagnetic waves\cite{Veselago} .

This analogy is not complete: although the equations are akin, the boundary
conditions are, generally speaking, different. As a result Eq.~(\ref%
{matrix2-a}) (for graphene) does not contain factor $Z_{j+1}/Z_{j}$ which is
present in Eqs.~(\ref{RLmatrix-1}), (\ref{RLmatrix-2}) and determines the
reflection coefficients at the boundary between two dielectrics \cite{born
wolf}. Another important distinction between transfer matrices $\hat{%
\mathcal{F}}$ (graphene) and $\hat{F}$ (electromagnetic waves) is that $\hat{%
\mathcal{F}}$ is a complex-valued matrix, while the $\hat{F}$ is always
real. This is manifestation of the fundamental difference between graphene
wave functions and electromagnetic fields in dielectrics. The graphene wave
functions are complex-valued spinors which describe two different physical
objects: particles (electrons) and antiparticles (holes). The
electromagnetic fields are real that reflects the fact that photons do not
have antiparticles (antiphoton is identical with photon). These distinctions
bring about rather peculiar dissimilarities between the conductivity of
graphene and the transparency of dielectrics.

\begin{figure}[h]
\centering \scalebox{0.9}{\includegraphics{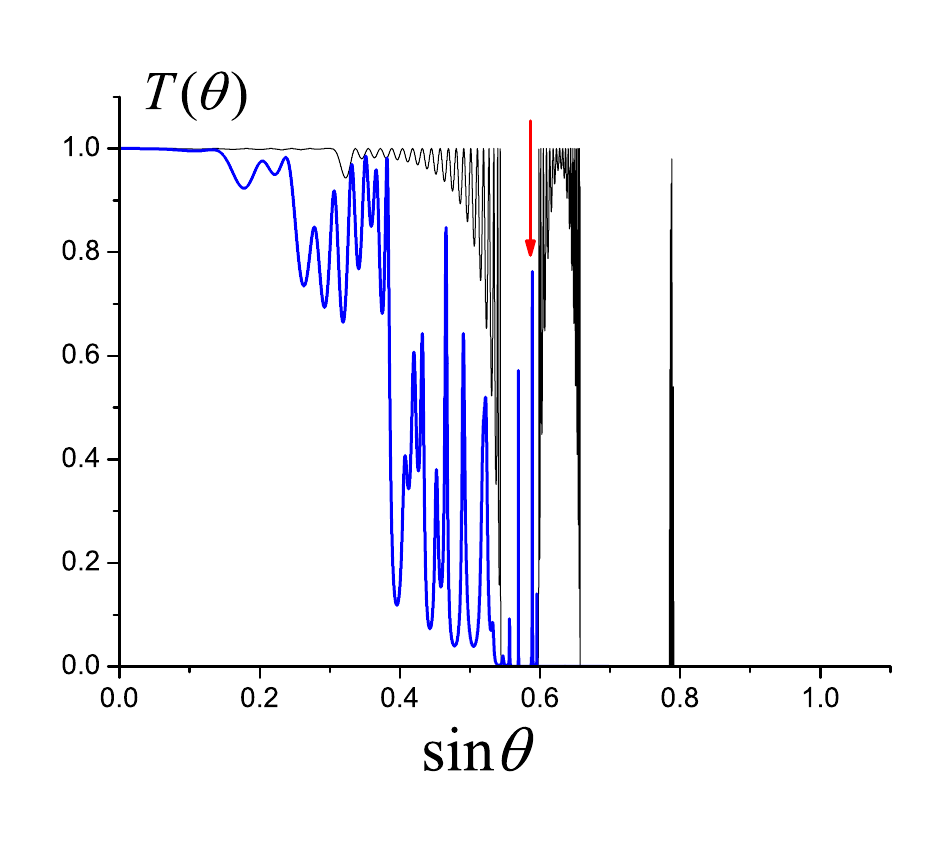}}
\caption{(Ref.~[\onlinecite{BFSN}]) Transmission coefficient $T(\protect%
\theta)$ for periodic (thin black line) and disordered (bold blue line)
graphene.}
\label{FG1}
\end{figure}

However in the particular case of normal incidence $\theta_{j}=\theta_{j+1}=0
$ and equal impedances $Z_{j}=Z_{j+1}$, the transmission of Dirac electrons
through a junction is similar to the transmission of light via an interface
between two media with different refractive indices (but equal impedances).
Such an interface is absolutely transparent to light and therefore both
\textit{p-n} and \textit{p-p} junctions are absolutely transparent to the
Dirac electrons in graphene\cite{Cheianov, Katsnelson1}. This is related to
the absence of backscattering and antilocalization of massless Dirac
fermions caused by their spin properties \cite{anti1,anti2}. This also
explains Klein paradox\cite{Klein} (perfect transmission through a high
potential barrier) in graphene systems, and leads (together with symmetry
and spectral flow arguments) to the surprising conclusion that Dirac
electrons are delocalized in disordered 1D graphene structure, providing a
minimal non-zero overall transmission, which cannot be destroyed by
fluctuations, no matter how strong they are \cite{Titov-1}. However, this
statement (being correct in some sense) should be perceived with a certain
caution. Indeed, many features of Anderson localization can be found in
random graphene systems. It has been shown in \cite{Mehrafarin} that
although the wave functions of normally incident ($\theta =0$) particles are
extended and belong to the continuous part of the spectrum, away from some
vicinity of $\theta =0,$ 1-D random graphene systems manifest all features
of disorder-induced strong localization. In particular, for a long enough,
disordered graphene superlattice the transmission coefficient, $T$, as a
function of the angle of incidence, $\theta $, (or of the energy $E$, if $%
\theta \neq 0$ is fixed) has typical for Anderson localization shape, Fig.~%
\ref{FG1}.

\begin{figure}[htb]
\centering \scalebox{0.5}{\includegraphics{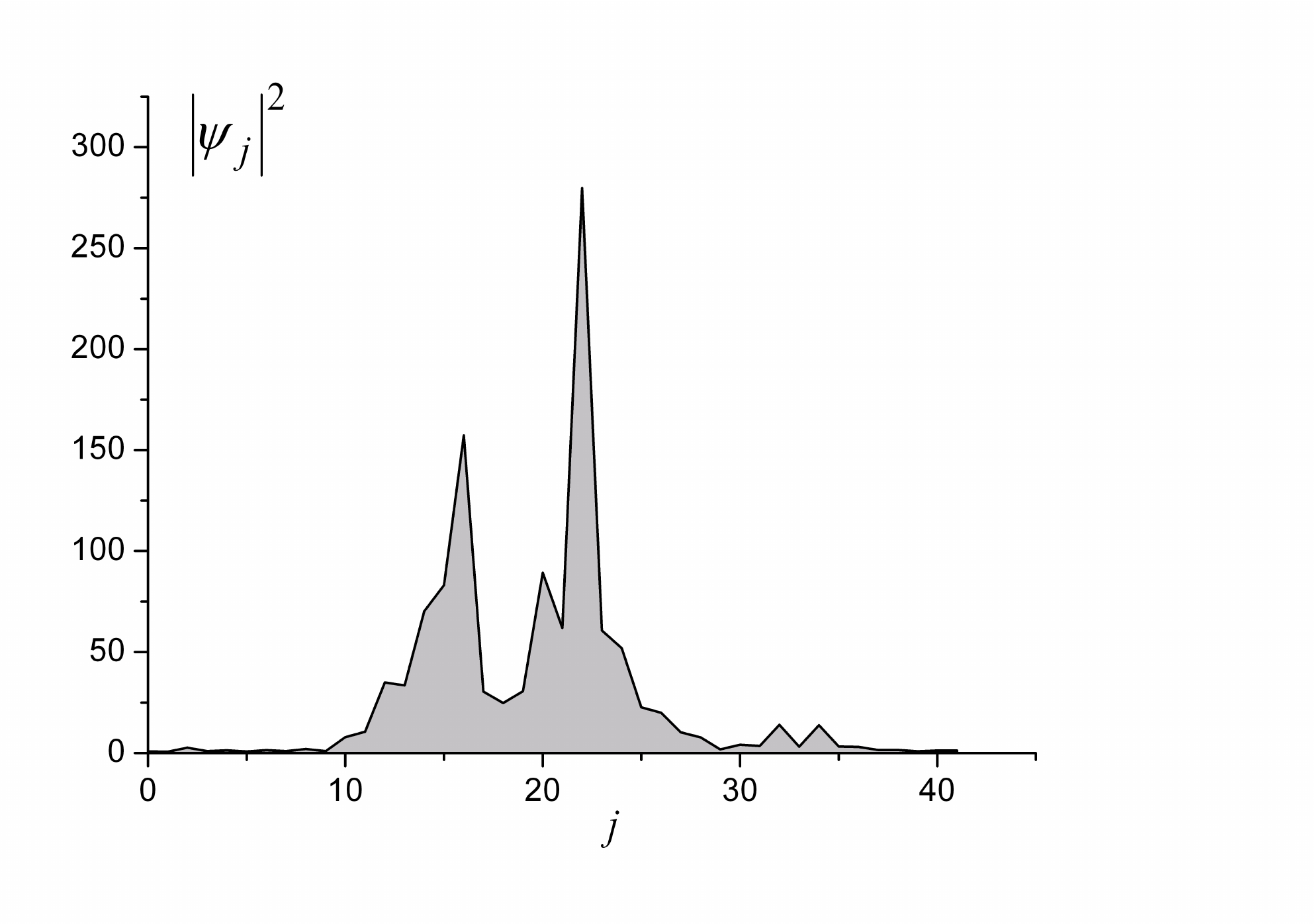}}
\caption{(Ref.~[\onlinecite{BFSN}]) Spatial distribution of the wave
function localized inside the sample for $\protect\theta$ marked by red
arrow in Fig.~\protect\ref{FG1}}
\label{FG2}
\end{figure}

Along with continuous of typical angles (or energies), for which the
transmission is exponentially small, there exists a quasi-discrete random
set of directions where the sample is well transparent, i.e., the
transmission coefficient is close to one. At these angles, the wave
functions are exponentially localized (Fig.~\ref{FG2}), with the Lyapunov
exponent (inverse localization length $l _{\xi}$) being proportional to the
strength of disorder.

Charge transport in a graphene sheet subjected to a disordered electrostatic
potential is determined by the ratio between its values $u(\xi)$ and the
energy $\varepsilon$ of the particle. In particular, in randomly-layered
potential $u_{j}=u_{0}(j)+\Delta u_{j}$ ($j$ is the number of a layer, $%
u_{0}(j)$ is a non-random function, $\Delta u_{j}$ are independent random
variables homogeneously distributed in the interval $[-\delta u,\delta u]$)
it manifests essentially different features in the following three different
systems \cite{Mehrafarin}:

(i) $u_{j}<\varepsilon$, $u_{0}(j)$ is a periodic function. In this case, a
relatively weak disorder drastically changes the transmission spectrum. All
features of the spectrum of the underlying periodic structure are washed
out, and a rather dense (quasi-)discrete angular spectrum appears, with the
corresponding wave functions being localized at random points inside the
sample (disorder-induced resonances). However, there is one fundamental
difference from the usual Anderson localization: in the vicinity of normal
incidence, the transmission spectrum of graphene is continuous with extended
wave functions, and the transmission coefficient is finite ($T=1$ at $\theta
=0)$. It is this range of angles that provides the finite minimal
conductivity, which is proportional to the integral of $T(\theta )$ over all
angles $\theta $.

(ii) $\varepsilon\leq u_{0}(j)=\mathrm{const.}$Under these conditions, the
transmission of the unperturbed system is exponentially small (tunneling)
and, rather unusually, gets enhanced by the fluctuation of the potential.

(iii) $\varepsilon=0$, $u_{0}(j)$ is a periodic set of numbers with
alternating signs. The behavior of the charge carriers in the potential of
this type is most unusual. It is characteristic of two-dimensional Fermions
and have no analogies in electron and light transport. The disorder
obliterates the transmission peaks of the underlying periodic system, makes
much wider the transparency zone around normal angle of incidence, and gives
rise to a new narrow peak in the transmission coefficient, associated with
wave localization in the random potential. Unlike the peaks in the periodic
structure, the wave function of this disorder-induced resonance is
exponentially localized. In distinction to the case (i), the transmission in
(iii) is extremely sensitive to fluctuations of the applied potential:
relative fluctuations $\Delta u/u_{0}=0.05$ reduce the angular width of the
transmission spectrum more than four times.

Propagation of light in analogous L-R and R-R disordered dielectric
structures demonstrates completely different behavior. As the degree of
disorder (variations of the refractive index) grows, the averaged angular
spectra quickly reach their asymptotic ``rectangular'' shape: a constant
transmission in the region where all interfaces between layers are
transparent followed by an abrupt decrease in transmission in the region of
angles where the total internal reflection appears.


\subsection{Bistability of Anderson Localized States in Nonlinear Media}

\label{subsec:nonlin} 


Recent renewed interest to Anderson localization is driven by a series of
experimental demonstrations in optics~\cite{optics,Segev,Silberberg} and
Bose-Einstein condensates~\cite{BEC,BEC1}. One of the important issues risen
in these studies is that the disordered systems can be inherently nonlinear,
so that an intriguing interplay of nonlinearity and disorder could be
studied experimentally.

Nonlinear interaction between the propagating waves and disorder can
significantly change the interference effects, thus fundamentally affecting
localization~\cite{nonlin-loc1,nonlin-loc2,Soven,Azbel}. However, most of
the studies of the localization in random nonlinear media deal with the
ensemble-averaged characteristics of the field, such as the mean field and
intensity, correlation functions, etc. These quantities describe the
averaged, typical behavior of the field, but they do not contain information
about individual localized modes (resonances), which exist in the localized
regime in each realization of the random sample \cite%
{Soven,Azbel,BBF,we-PRL-1,Topolancik}. These modes are randomly located in
both real space and frequency domain and are associated with the exponential
concentration of energy and resonant tunneling. In contrast to regular
resonant cavities, the Anderson modes occur in a statistically-homogeneous
media because of the interference of the multiply scattered random fields.
Although the disorder-induced resonances in linear random samples have been
the subject of studies for decades, the resonance properties of nonlinear
disordered media have not been explored so far.

In this Section we present the study of the effect of nonlinearity on the
Anderson localized states in a one-dimensional random medium~\cite{Shadrivov}%
. As a result of interplay of nonlinearity and disorder, the bistability and
nonreciprocity appear upon resonant wave tunneling and excitation of
disorder-induced localized modes in a manner similar to that for regular
cavity modes. At the same time, weak nonlinearity has practically no effect
on the averaged localization background.


First, let us consider a stationary problem of the transmission of a
monochromatic wave through a one-dimensional random medium with Kerr
nonlinearity. The problem is described by the equation

\begin{equation}  \label{oscillator}
\frac{d^2\psi}{dx^2}+k^{2}\left[n^2 -\chi|\psi|^2 \right]\psi=0~,
\end{equation}
where $\psi$ is wave field, $x$ is coordinate, $k$ is wave number in the
vacuum, $n=n(x)$ is the refractive index of the medium, and $\chi $ is the
Kerr coefficient.

In the linear regime, $\chi|\psi|^2 = 0$, the multiple scattering of the
wave on the random inhomogeneity $n^2(x)$ brings about Anderson
localization. The main signature of the localization is an exponential decay
of the wave intensity, $I=|\psi|^2$, deep into the sample and, thus, an
exponentially small transmission \cite{And,LGP,Ping91,BerryKlein}: $I_{%
\mathrm{out}}^{\mathrm{(typ)}}\sim I_{\mathrm{in}}\exp(-2L/l)\ll 1$. Here $L$
is the length of the sample and $l$ is the localization length which is the
only spatial scale of Anderson localization. Along with the typical wave
transmission, there is an anomalous, resonant transmission, which
accompanies excitation of the Anderson localized states inside the sample
and occurs at random resonant wave numbers $k=k_{\mathrm{res}0}$ \cite%
{Soven,Azbel,BBF,we-PRL-1,Topolancik}. In this case, the intensity
distribution in the sample is characterized by an exponentially localized
high-intensity peak inside the sample, $I_{\mathrm{peak}} \gg I_{\mathrm{in}}
$, and a transmittance much higher than the typical one: $I_{\mathrm{out}}^{%
\mathrm{(res)}}\gg I_{\mathrm{out}}^{\mathrm{(typ)}}$.

Excitation of each localized mode inside the random sample can be associated
with an effective resonator cavity located in the area of field localization
and bounded by two potential barriers with exponentially small
transparencies~\cite{Bliokh-2}. According to this model, the transmittance
spectrum $T(k, I_{out})$ in the vicinity of a resonant wavelength for the
case of weak nonlinearity ($\chi|\psi|^2 \ll 1$) is given in the form~\cite%
{BBF,Genack1,Shadrivov}:

\begin{equation}  \label{nonlinear-transmission}
T(k,I_{\mathrm{out}})\equiv \frac{I_{\mathrm{out}}}{I_{\mathrm{in}}}=\frac{%
T_{\mathrm{res}}}{1+\left[A\chi I_{\mathrm{out}}+\delta\right]^2}~,
\end{equation}
where $T_{res}$ is the transmission coefficient at resonance, and
dimensionless parameters $A$ and $\delta$ characterize, respectively, the
strength of the nonlinear feedback and the detuning from the resonant wave
number:

\begin{equation}  \label{params}
A= \frac{2Q}{\chi} \left.\frac{d \ln k_{\mathrm{res}}}{d I_{\mathrm{out}}}%
\right|_{I_{\mathrm{out}}=0},~ \delta=2Q\left(1-\frac{k}{k_{\mathrm{res}0}}%
\right).
\end{equation}

\begin{figure}[t]
\centering \scalebox{0.42}{\includegraphics{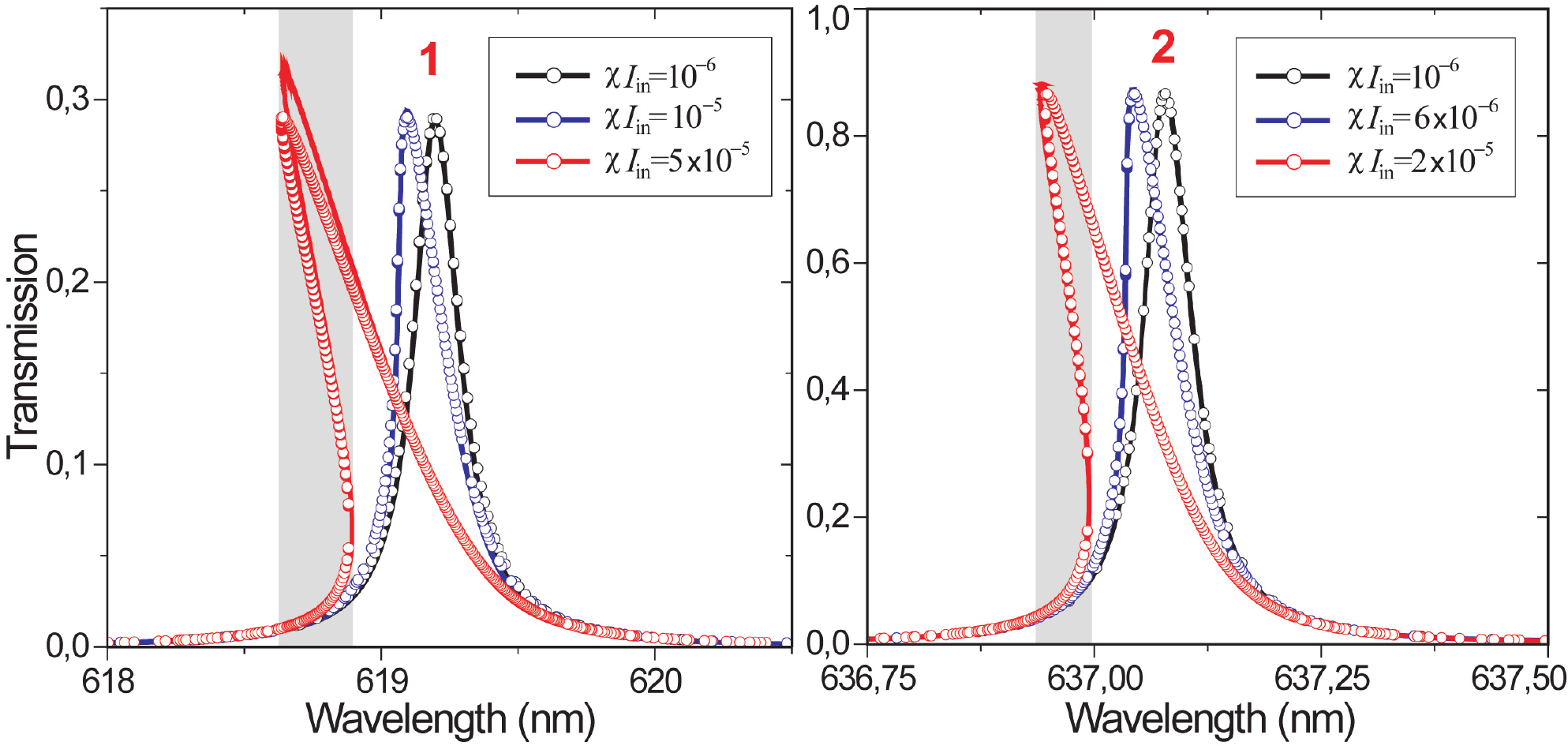}}
\caption{(Ref.~[\onlinecite{Shadrivov}], color online.) Nonlinear
deformations of the transmission spectra of two random resonances at
different intensities of the incident wave. Numerical simulations of the Eq.~%
\protect\ref{oscillator} (curves) and theoretical Eq.~\protect\ref%
{nonlinear-transmission} (symbols) are shown for the case of defocusing
nonlinearity, $\protect\chi>0$. Light-grey stripes indicate three-valued
regions for the high-intensity curves, where only two of them (corresponding
to the lower and upper branches) are stable.}
\label{Bist_Fig2}
\end{figure}

\begin{figure}[t]
\centering \scalebox{0.42}{\includegraphics{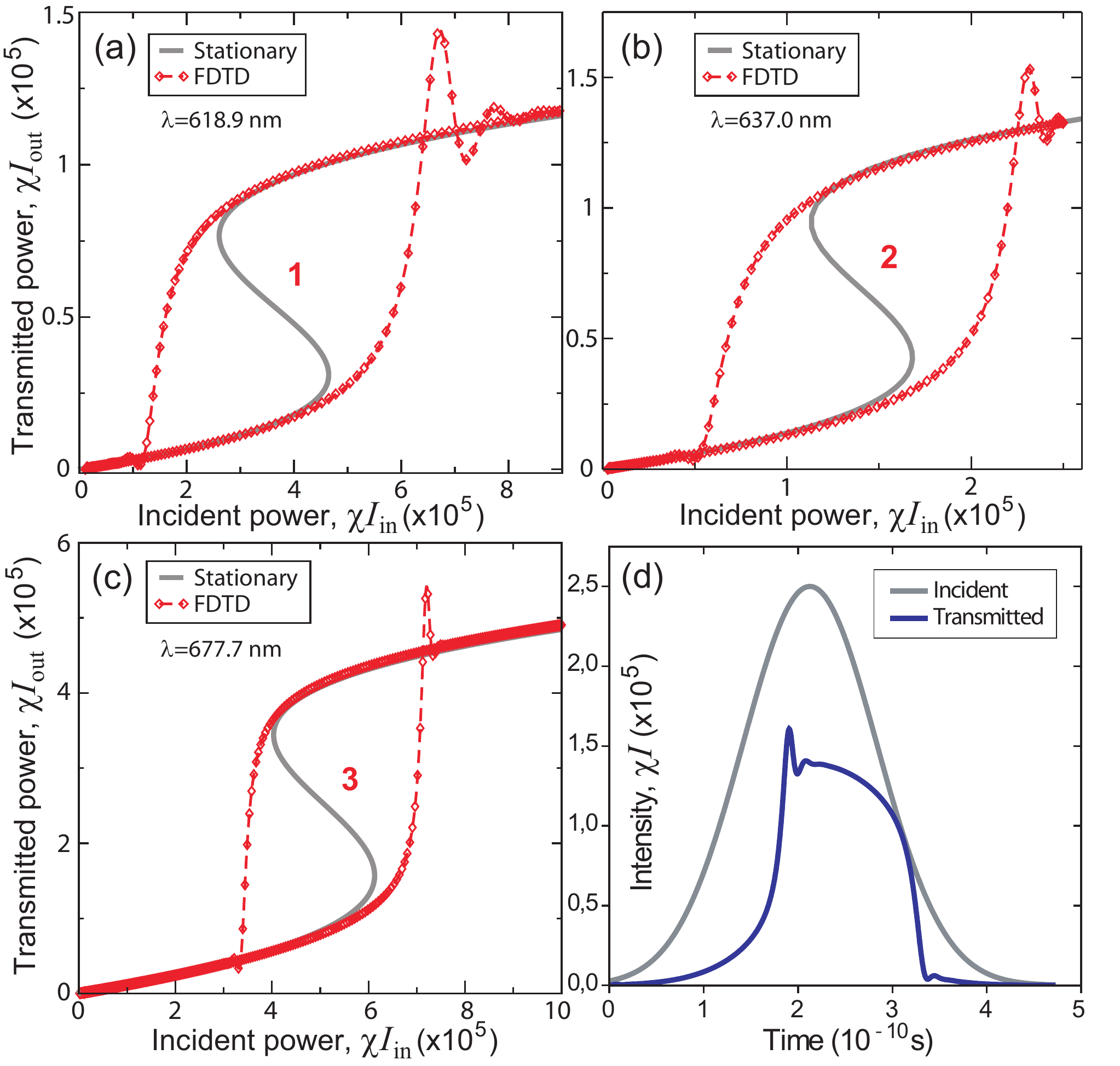}}
\caption{(Ref.~[\onlinecite{Shadrivov}], color online.) Stationary and FDTD
simulations showing hysteresis loops in the output vs. input power
dependence for three different resonances. Panel (d) shows deformation of
the transmitted Gaussian pulse corresponding to the hysteresis switching on
the resonance 2.}
\label{Bist_Fig3}
\end{figure}

Equation (~\ref{nonlinear-transmission}) establishes relation between the
input and output wave intensities, which is given by a cubic equation with
respect to $I_{\mathrm{out}}$. It has a universal form typical for nonlinear
resonators possessing optical bistability \cite{Bistability}. From Eq.~(\ref%
{params}) it follows that in the region of parameters:

\begin{equation*}  \label{eq6}
A\delta < 0~,~~\delta^2 >3~,~~|\chi| I_{\mathrm{in}} > \frac{8}{3\sqrt{3}}%
\frac{1}{|A|T_{\mathrm{res}}}~,
\end{equation*}
the dependence $I_{\mathrm{out}}(I_{\mathrm{in}})$ is of the S-type and the
stationary transmission spectrum $T(k)$ is a three-valued function. In most
cases, one of the solutions is unstable, whereas the other two form a
hysteresis loop in the $I_{\mathrm{out}}(I_{\mathrm{in}})$ dependence (see
Figs.~\ref{Bist_Fig2} and \ref{Bist_Fig3}).

It is important to emphasize two features of the equations (\ref%
{nonlinear-transmission}) and (\ref{params}), describing the nonlinear
resonant transmission through a localized state. First, they have been
derived without any approximations apart from the natural smallness of the
nonlinearity and Lorentzian shape of the spectral line. Second, although the
resonant transmission, the effect of nonlinearity, and bistability owe their
origin to the excitation of the Anderson localized mode inside the sample,
equations (\ref{nonlinear-transmission}) and (\ref{params}) contain only
quantities which can be found via outside measurements~\cite{Shadrivov}.

Figure \ref{Bist_Fig2} shows nonlinear deformations of the resonant
transmission spectra $T(k)$ for different values of $I_{\mathrm{in}}$, which
exhibit transitions to bistability. The analytical dependence $T(k)$ given
by Eqs.~(\ref{nonlinear-transmission}, \ref{params}) with the parameters $T_{%
\mathrm{res}}$, $Q$, and $A$ found from the numerical experiments are in
excellent agreement with the direct numerical solutions of Eq.~(\ref%
{oscillator}) \cite{remark}. In numerical simulations of stationary regime
we used the standard 4-th order Runge-Kutta method. We note, that the
incident field amplitude is a single-valued function of the transmitted
field. Thus, we solve second-order ordinary differential equation Eq.~(\ref%
{oscillator}) using transmitted field value as the boundary conditions for
the equation.

The dimensionless parameters $T_{\mathrm{res}}$, and $Q$ from Eqs. (\ref%
{nonlinear-transmission}, \ref{params}), can also be estimated from a simple
resonator model of the Anderson localized states \cite{BBF,we-PRL-1,Bliokh-2}%
:

\begin{equation}  \label{param-estimates}
T_{\mathrm{res}}=\frac{4T_1 T_2}{\left(T_1+T_2\right)^2}~,~~Q^{-1}\sim\frac{
T_1+T_2}{4k_{\mathrm{res}0}l}~,
\end{equation}
where

\begin{equation*}  \label{barrier-transmissions}
T_1 \sim \exp\left[-2x_{\mathrm{res}}/l\right]~,~~ T_2 \sim \exp\left[%
-2(L-x_{\mathrm{res}})/l\right]~
\end{equation*}
are the transmission coefficients of the two barriers that form the
effective resonator, $x_{\mathrm{res}}$ is the coordinate of the center of
the area of field localization, $l$ is the localization length, and $L$ is
the length of the sample.

Introducing a weak Kerr nonlinearity into the resonator model, one can also
estimate the nonlinear feedback parameter $A$:

\begin{equation}  \label{nonlin-coeff}
A\sim Q/\overline{n^2}T_2~,
\end{equation}
where $\overline{n^2}$ is the mean value of $n^2(x)$.

It is important to note that each disorder-induced resonance is associated
with its own effective cavity, so that the disordered sample can be
considered as a chain of randomly located coupled resonators~\cite{Ref-exp-2}%
.

Equations (\ref{param-estimates}, \ref{nonlin-coeff}) enable one to estimate
the values of the parameters describing the nonlinear resonant wave
tunneling in Eqs.~(\ref{param-estimates}, \ref{nonlin-coeff}) by knowing
only the basic parameters of the localization -- the localization coordinate
and the localization length. In particular, substituting Eqs.~(\ref%
{param-estimates}, \ref{nonlin-coeff}) into Eq.~(\ref{params}) and taking
into account that the most pronounced transmission peaks correspond to the
localized states with $x\simeq L/2$ and $T_1 \sim T_2$, we estimate the
incident power needed for bistability of localized states

\begin{equation}  \label{power}
|\chi| I_{\mathrm{in}} \gtrsim \frac{\exp(-2L/l)}{k_{\mathrm{res}0}l}~.
\end{equation}
For the parameters used in our simulations this gives quite reasonable value
$|\chi|I_{\mathrm{in}}\gtrsim 10^{-5}$. If we increase the length of the
sample, the Q-factors of the resonances grow, and the incident power needed
to observe the bistability becomes smaller.

To demonstrate temporal dynamics upon the bistable resonant tunneling, an
explicit iterative nonlinear finite-difference time-domain (FDTD) scheme was
implemented. For precise modeling of the spectra of narrow high-Q
resonances, fourth-order accurate algorithm was used, both in space and in
time~\cite{fourthorder}, as well as the Mur boundary conditions to simulate
open boundaries and total-field/scattered-field technique for exciting the
incident wave. Sufficient accuracy was achieved by creating a dense spatial
mesh of 300 points per wavelength ($dx=\lambda/300$). To assure stability of
the method in nonlinear regime, the time step was selected as $dt = dx/3c$,
and each simulation ran for $N=2*10^8$ time steps. To compare the results of
the FDTD simulations with the steady-state theory, the transmission of long
Gaussian pulses with central frequencies and amplitudes satisfying
conditions (\ref{power}) was considered, see Fig.~\ref{Bist_Fig3}(d). With
an appropriate choice of the signal frequencies, we observe hysteresis loops
in the $I_{\mathrm{out}}(I_{\mathrm{in}})$ dependencies which are in
excellent agreement with stationary calculations, as shown in Figs.~\ref%
{Bist_Fig3}(a--c). Characteristic transitional oscillations accompany jumps
between two stable branches, and strong reshaping of the transmitted pulse
evidences switching between the two regimes of transmission, Fig.~\ref%
{Bist_Fig3}(d). We note, that different choice of the signal frequencies
near the resonance can lead to various other behaviors of output vs. input
curves, with transmission either increasing, when nonlinear resonance
frequency shifts towards the signal frequency, or decreasing in the opposite
case.

\begin{figure}[t]
\centering \scalebox{0.43}{\includegraphics{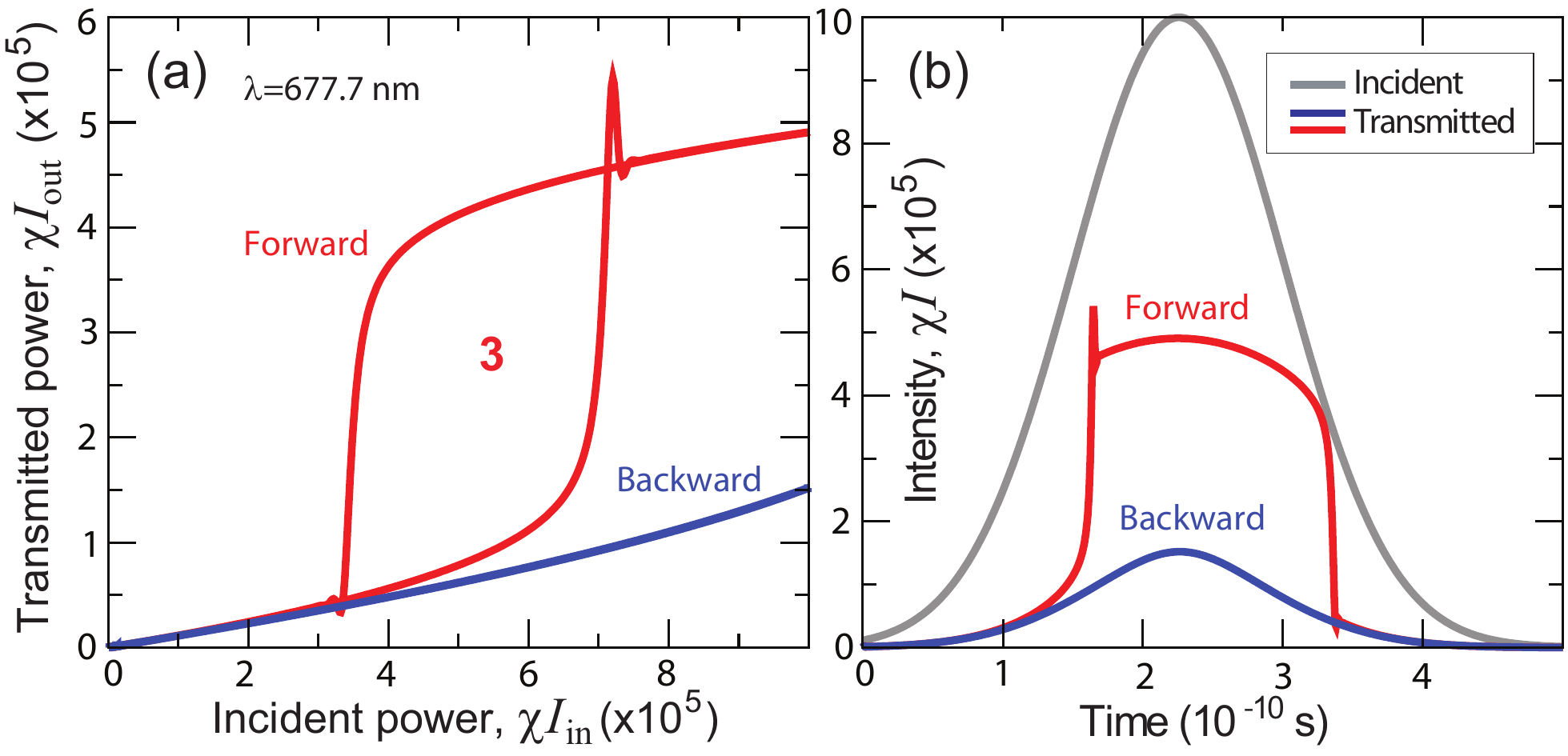}}
\caption{(Ref.~[\onlinecite{Shadrivov}], color online) (a) Non-reciprocal
transmission through the nonlinear disordered structure, showing different
output powers for identical waves incident from different directions. (b)
Corresponding shape of the incident pulse, and pulses transmitted in
different directions.}
\label{Bist_Fig4}
\end{figure}

In addition to the bistability, the resonant wave tunneling through a
nonlinear disordered structure is nonreciprocical. As is known for regular
systems, nonsymmetric nonlinear systems may possess nonreciprocal
transmission properties, resembling the operation of a diode. An all-optical
diode is a device that allows unidirectional propagation of a signal at a
given wavelength, which may become useful for many applications~\cite{diode}%
. A disordered structure is naturally asymmetric in the generic case, and
one may expect a nonreciprocal resonant transmission in the nonlinear case.
To demonstrate this, we modeled propagation of an electromagnetic pulse
impinging the same sample from different sides and monitor the transmission
characteristics. One case of such nonreciprocical resonant transmission is
shown in the panel (a) of Fig.\ref{Bist_Fig4}. We observe considerably
different transmission properties in opposite directions with the maximal
intensity contrast between two directions 7.5:1. Moreover, the threshold of
the bistability is also significantly different for two directions: there is
a range of incident powers, for which the wave incident from one side of the
sample is bistable, while there is no signs of bistability for the incidence
from the other side. Figure~\ref{Bist_Fig4}(b) shows the pulse reshaping for
incidence from opposite sides of the structure.

In this Section we have presented the study of new manifestations of the
interplay between nonlinearity and disorder. It is shown that even weak
nonlinearity affect dramatically the resonant transmission associated with
the excitation of the Anderson localized states leading to bistability and
nonreciprocity. Despite random character of the appearance of Anderson
modes, their behavior and evolution are rather deterministic, and,
therefore, these modes can be used for efficient control of light similar to
regular cavity modes. These results demonstrate that, unlike infinite
systems, the Anderson localization in finite samples is not destroyed by
weak nonlinearity -- instead it exhibits new intriguing features typical for
resonant nonlinear systems.


\section{Conclusions}

\label{sec:Concl}

We have reviewed the transmission and localization wave properties of the
complex disordered structures composed of (i) left-handed metamaterials,
(ii) magneto-active optical materials, (iii) graphene superlattices, and
(iv) nonlinear dielectric media. Interference origin of the wave
localization, together with strong energy concentration, makes Anderson
localization highly sensitive to weak modifications of the material
properties. We have shown that the exotic properties of novel materials can
drastically modify the main features of the wave localization. This brings
about anomalous pronounced dependences of the wave transmittance and
localization length on both wave and material parameters: frequency, angle
of incidence, polarization, magnetization, nonlinearity, etc. As a result,
remarkable phenomena appear, such as anti-(de-)localization, unidirectional
transmission, slow-light propagation, and bistability.

We have described a number of novel features accompanying the wave
localization in complex media, including: (i) dramatic suppression of
localization in mixed stacks with left-handed metamaterials, (ii) Brewster,
zero-$\varepsilon$, and zero-$\mu$ delocalization, and (iii) anomalous
transmission enhancement in periodic metamaterials with only one disordered
electromagnetic characteristics, (iv) nonreciprocal localization and
unidirectional transmission through magneto-active disordered stacks, (v)
angle-dependent transmission resonances in graphene superlattices, and (vi)
bistability and nonreciprocity of transmission resonances in nonlinear
disordered structures.

We believe that presented results significantly extend and enrich theory and
potential application of the wave localization in complex disordered media.
In particular, they provide a theoretical toolbox which can serve for design
of novel optical and electronic devices with unusual transport properties.


\section{Acknowledgments}

\label{sec:ackn} 

We are pleased to dedicate this review paper to 80-th anniversary of
Academician Victor Valentinovich Eremenko and wish him a good health, good
mood, and new scientific achievements.

We also thank our co-authors
especially L.C. Botten, M.A. Byrne, R.C. McPhedran, F. Nori, P. Rajan, and
S. Savel'ev, for fruitful collaboration and discussions of many original
results summarized in this review paper. S.G. is grateful to N.M. Makarov,
P. Markos, and L.A. Pastur for useful comments and helpful discussions.

This
work was partially supported by the European Commission (Marie Curie Action). V.F acknowledges partial support from the Israeli Science Foundation (Grant \# 894/10)




\begin{thebibliography}{999}
\bibitem{remark} A minor discrepancy between the theoretical model and exact
numerical simulations can be seen for high incident power for the first
resonance. This is because at high intensity the shape of the resonance line
changes, which is not taken into account by the model.
\end{thebibliography}
\end{document}